\documentclass[trackchanges]{aastex7}

\usepackage{amsmath}
\usepackage{subcaption}

\newcommand{\vect}[1]{\boldsymbol{#1}}

\newcommand{\vxi}{\boldsymbol{\xi}}

\newcommand{\Ma}{\mathcal{M}_\ast}

\newcommand{\qtr}{q_{\rm tr}^{(1)}}
\newcommand{\qtrr}{q_{\rm tr}^{(2)}}

\newcommand{\hang}[1]{\textcolor{black}{#1}}

\begin{document}

\title{Dynamical tide modified Roche limit in eccentric, asynchronous binaries}

\author[orcid=0000-0002-6011-6190, gname='Hang', sname='Yu']{Hang Yu}
\affiliation{eXtreme Gravity Institute, Department of Physics, Montana State University,
Bozeman, Montana 59717, USA}
\email{hang.yu2@montana.edu}

\author[orcid=0000-0002-8239-0174, gname='Shu Yan', sname='Lau']{Shu Yan Lau}
\affiliation{eXtreme Gravity Institute, Department of Physics, Montana State University,
Bozeman, Montana 59717, USA}
\email{shuyan.lau@montana.edu}

\author[orcid=0009-0002-8956-7849]{Ethan Mckeever}
\email{ethanmckeever@montana.edu}
\affiliation{eXtreme Gravity Institute, Department of Physics, Montana State University,
Bozeman, Montana 59717, USA}

\author[orcid=0000-0001-5611-1349]{Phil Arras}
\email{pla7y@virginia.edu}
\affiliation{Department of Astronomy, University of Virginia, P.O. Box 400325, Charlottesville, VA 22904, USA}

\author[orcid=0000-0001-9194-2084]{Nevin N. Weinberg}
\email{nevin@uta.edu}
\affiliation{Department of Physics, University of Texas at Arlington, Arlington, TX 76019, USA}

\correspondingauthor{Hang Yu}
\email{hang.yu2@montana.edu}
\correspondingauthor{Shu Yan Lau}
\email{shuyan.lau@montana.edu}

\begin{abstract}

The Roche limit, or the threshold separation within which a celestial object (the donor) $M$ cannot remain in a stable configuration due to a companion's tidal field, has been well established when $M$ is in hydrostatic equilibrium and has synchronous rotation in a circular orbit. However, limited analyses exist considering corrections to the Roche limit due to hydrodynamical effects. We fill in the gap by providing a general theoretical framework involving nonlinear hydrodynamics. We consider both exact nonlinear equations derived from an affine model describing incompressible ellipsoids and series-expanded ones that can be calculated for realistic stars and planets.  
Our formulation addresses the Roche problem in generic orbits and synchronization levels of $M$, and fully accounts for the history-dependent hydrodynamical effects. We show that as the orbital eccentricity increases, fluid instability is more likely to develop at the pericenter due to the increased dynamical tide that accumulates over multiple orbits. When $M$ moves in a highly eccentric orbit (with eccentricity around 0.9) and the damping of the fluid is small, the threshold pericenter separation at which mass loss from $M$ can occur can be at least 30\% higher than the value predicted for a circular orbit with hydrostatic equilibrium. If only a single passage is considered, however, the threshold separation is 20\% smaller than the static limit.
The nonlinear interaction at each pericenter passage can also trigger a diffusive fluid evolution inside $M$ even with moderate eccentricities, complementing previous studies of diffusive tides caused by random propagation phases. Our work has broad implications for interacting binaries in eccentric orbits, including migrating gaseous exoplanets, repeated partial tidal disruption events, and more.

\end{abstract}

\keywords{ \uat{Roche limit}{1404}, \uat{Hydrodynamics}{1963}, \uat{Exoplanet dynamics}{490}, \uat{Exoplanet tides}{497}, \uat{Tidal disruption}{1696}, \uat{X-ray transient sources}{1852}}


\section{Introduction}
\setcounter{footnote}{0} 

The Roche model has long been the foundational framework for studying mass loss and transfer in interacting binary systems. Of particular importance is the Roche limit, a critical separation within which a donor object (with mass $M$ and radius $R$) can no longer hold itself but will lose mass due to the tidal force of the companion (treated as a point particle with mass $M'$ here). Many applications of the Roche model rely on the fit by \citet{Eggleton:83}, which assumes that the donor is in hydrostatic equilibrium--a valid approximation for evolved circular binaries but one that may break down in eccentric systems, particularly when eccentricity is high and the donor has not reached synchronization.

Highly eccentric binaries can form in various environments. One such example lies in the realm of exoplanets. The formation of hot Jupiters (HJs) and hot Neptunes (HNs) at separations as close as $0.05\,{\rm AU}$ to their host stars is one of the longest outstanding puzzles in exoplanetary science \citep{Mayor:95, Dawson:18}. High-eccentricity migration of HJ and HN is a promising explanation \citep{Fabrycky:07}: A giant planet initially forms with a semimajor axis of $\mathcal{O}({\rm AU})$ and is dynamically excited onto a highly eccentric orbit. It then migrates inward due to tidal dissipation while the orbit circularizes. This picture naturally reproduces some key characteristics of HJs \citep{Dawson:18}, including their highly inclined, sometimes retrograde orbits. While qualitatively successful, this channel still has large uncertainties on a quantitative level. For example, the pericenter separation required for inward migration can be dangerously close to tidal disruption of the planet, and a small variation in the disruption threshold can have a significant impact on the formation efficiency of HJs and HNs \citep{Yu:24b}. 

Another example is found in the nucleus of a galaxy that hosts a massive black hole (MBH). In such regions, a stellar-mass binary may be drawn close to the MBH and disrupted by its tidal forces through the Hills mechanism, which ejects one star while leaving the other bound to the MBH on an eccentric orbit \citep{Hills:88}. If the bound star approaches the MBH closely enough, it may itself be disrupted, resulting in a tidal disruption event (TDE; \citealt{Rees:88, Komossa:15}). 
At slightly greater separations, the star may survive a full disruption by the MBH and repeat mass transfer over multiple orbits, causing repeated partial TDEs (rpTDEs; \citealt{Hayasaki:13, Coughlin:2019, Cufari:22, Liu:24, Bandopadhyay:24, Bandopadhyay:25, Yao:25}). Recently, a new class of recurring X-ray bursts, known as quasi-periodic eruptions (QPEs), has been observed and is associated with low-mass MBHs~\citep{Miniutti:19}. The mechanism driving QPEs remains uncertain, yet similar to the production of TDEs \citep{Maguire:20}, several proposed models involve a star or white dwarf (WD) in an eccentric orbit around an MBH, undergoing periodic mass transfer episodes that power the QPEs~\citep{King:20, King:22, Wang:22, Linial:23}. 
The recent work of \cite{Lau:25} demonstrates that both can be the result of a tidally-induced mass loss at different pericenter distances. 
For orbits with moderately high eccentricities ($0.90<e<0.99$) where the orbital period is short compared to the tidal dissipation timescale, the tide inside the WD builds up \hang{diffusively (i.e., as a random-walk process)} over multiple orbits. The tidal energy eventually approaches the stellar binding energy and is released via a mass ejection as the extended tidal bulge breaks. Neither the WD nor the orbit will be disrupted by a single break, allowing the process to repeat several times. This may lead to rpTDEs with time intervals of a few years, consistent with the observations (e.g., \citealt{Campana:15, Wevers:23, Miniutti:23b, Somalwar:25}).
Meanwhile, the model also shows the possibility of QPEs being sourced by tidal stripping at closer pericenter separations, where the tidal field is strong enough such that a single passage is sufficient to induce mass transfer, though this mechanism is challenged by certain observed features in some QPEs, like the alternating long-short recurrence time and brightness of bursts \citep{Miniutti:19, Giustini:20, Arcodia:22, Zhou:24}. 
The proposed mechanism by \citet{Lau:25} accounts for the dynamical nature of the tide over multiple orbits, improving previous studies of \citet{King:22, Wang:22} based on the static Roche model that does not apply in eccentric orbits. Nonetheless, the threshold distance for the tidally induced mass transfer, either via multiple orbit buildup or a single passage, depends on a phenomenological parametrization that is not derived from fundamental hydrodynamics.

Despite the importance of mass loss in eccentric binaries, illustrated by the examples above, theoretical advancements beyond the classic Roche model are still limited. Some pioneering work by \citet{Sepinsky:07a} extended the Roche model to non-circular orbits with non-synchronized donors, identifying significant corrections to the classic fit of \citet{Eggleton:83}; however, their work was still based on a quasi-static approximation. In reality, the donor also exhibits a dynamical response, corresponding to oscillations of the eigenmodes (mainly the fundamental modes, or f-modes) excited at each pericenter passage~\citep{Press:77}. This component is known as the dynamical tide. 
The analysis of \citet{Diener:95}, based on an affine model developed by \citet{Carter:83}, accounted for the dynamical tide but focused on the hydrodynamical stability in a single passage (though they considered the effects due to varying the ellipsoid's orientation). 
Unlike the equilibrium component, which depends solely on the instantaneous orbital configuration, the dynamical tide can accumulate over multiple orbits, allowing its amplitude to grow to a scale comparable to the unperturbed radius \citep{Mardling:95, Ivanov:04, Wu:18, Vick:18, Yu:21}. Such an accumulation is expected to play a critical role in the dynamics of mass transfer.

The numerical study by \citet{Guillochon:11} examining the threshold separation at which a proto-HJ is disrupted by its host star strongly motivates the need to include DT in modeling mass transfer. 
According to the fit of \citet{Eggleton:83}, Roche lobe overflow should only occur when the planet's pericenter separation $D_p$ reaches $D_p {\lesssim} 2.0 R_t$ with $R_t{=}(M'/M)^{1/3} R$ the tidal radius. However, \citet{Guillochon:11} found that with high orbital eccentricity ($e{=}0.9$), a planet is disrupted at a much greater pericenter separation of $D_p \simeq 2.7 R_t$, $35\%$ greater than the separation from Eggleton's fit. The discrepancy is further amplified when the results of \citet{Sepinsky:07a} are used, which predict a Roche lobe $\sim10\%$ greater than Eggleton's fit (mainly due to the lack of centrifugal force), requiring an even smaller separation for the overflow to occur. Similarly, the critical single-passage separation of $D_p\simeq 1.7 R_t$ from \citet{Diener:95} also cannot explain the numerical results of \citet{Guillochon:11}. 

In this work, we will show that the dynamical tide is the needed ingredient to reconcile the discrepancy. For this, we will consider the nonlinear hydrodynamical equations following both the affine model of \citet{Diener:95} (see also \citealt{Carter:83, Carter:85}) and those derived by \citet{VanHoolst:94} following an order-by-order expansion (see also \citealt{Weinberg:12, Yu:23a}). Our model is fully dynamical and makes no (quasi-)static assumptions. Consequently, it applies to arbitrary eccentricities of the orbit and asynchronicities of the donor (though we still restrict our analysis to aligned spins). On the other hand, when the perturbed fluid is in hydrostatic equilibrium, our model also reproduces the classic results obtained by \citet{Chandrasekhar:63} (see also \citealt{Chandrasekhar:87} for a comprehensive summary) on homogeneous ellipsoids. 

The rest of the paper is organized as follows. We will first describe a toy model in \S \ref{sec:toy_model} that captures the essence of our model and illustrates the significance of the dynamical tides. The nonlinear hydrodynamical model is introduced in \S \ref{sec:nl_hydro}, after which we will demonstrate its consistency with the analysis of \citet{Chandrasekhar:63} in \S \ref{sec:instab_st_tide} under the static limit. A theoretical investigation on instantaneous fluid stability at the pericenter follows in \S \ref{sec:instab_eq_tide}. Then, in \S \ref{sec:instab_dyn_tide_w_num}, a numerical experiment is conducted to address the build-up of the instability throughout multiple orbits. How the model applies to different astrophysical settings is described in \S \ref{sec:app}. Lastly, we conclude in \S \ref{sec:conclusion}. 

Throughout the paper, we will use natural units $G=M=R=1$.

\subsection{A toy model}
\label{sec:toy_model}

\hang{
The appearance of the Roche limit can be schematically shown with a toy model:
\begin{equation}
    \ddot{\xi} = \omega_f^2 (\underbrace{-\xi}_{\text{linear oscillator}} + \underbrace{I \epsilon}_{\text{linear tide}} + \underbrace{\kappa \xi^2}_{\text{combined nonlinear corrections}} ),
    \label{eq:toy}
\end{equation}
where $\xi \in \mathbb{R}$ denotes the fluid displacement due to a large-scale perturbation (i.e., the f-mode of the donor whose eigenfrequency is denoted by $\omega_f$). The first term on the right hand side (RHS) describes the linear restoration of the fluid to the equilibrium, the $I\epsilon$ term the linear tidal drive, with $I>0$ a coupling constant, $\epsilon = M'/D^3$ the instantaneous tidal field exerted by a companion (the accretor) with mass $M'$, and $D$ the instantaneous binary separation. Lastly, the $\kappa \xi^2$ term represents a combination of nonlinear effects in both the internal fluid response and the external tidal drive. Theoretical calculation predicts $\kappa>0$ when $I>0$ (see Appendix \ref{appx:eom_affine}).  
}
For the linear problem (ignoring the $\kappa \xi^2$ term), a static solution ($\ddot{\xi}=0$) can always be arranged by setting $\xi = I \epsilon$. With the nonlinear $\kappa \xi^2$ term, the RHS of Eq. (\ref{eq:toy}) must take the opposite (same) sign as $\xi$ for the system to be oscillatory (grow exponentially). We immediately see that with $\kappa \xi^2\geq0$, the nonlinear term can lead to instability when $\xi>0$, and an increasing $\xi$ will make the system more vulnerable to instability. 
\hang{This is consistent with the standard Roche lobe overflow picture where mass loss occurs mainly at the L1 and L2 Lagrangian points, corresponding to locations of maximum stretch of the donor.}
If we assume the hydrostatic limit and write $\xi\sim e^{\sigma t}$, at the instability threshold ($\sigma=0$), we need
\begin{equation}
     \kappa\xi^2 -\xi + I\epsilon = 0,
\end{equation}
which is formally a quadratic function of $\xi$, and a real root exists only if its discriminant is non-negative
\begin{equation}
    1-4\kappa I \epsilon \geq 0 \text{ or }\epsilon \leq \frac{1}{4\kappa I} \text{ for stability}. 
    \label{eq:toy_stability_st}
\end{equation}
We thus obtain that $1/(4\kappa I)$ is the maximum perturbation allowed. This is the essence of the static Roche limit found in \citet{Chandrasekhar:63}. See later in \S \ref{sec:instab_st_tide}.
\hang{Alternatively, Eq. (\ref{eq:toy}) can also be viewed as a particle with displacement $\xi$ in an effective potential $U_{\rm eff}(\xi)$, with $\ddot{\xi} = -d U_{\rm eff}/d\xi$ and
\begin{equation}
    U_{\rm eff}(\xi) = \underbrace{\frac{1}{2}\omega_f^2 \xi^2 - \frac{1}{3} \kappa \omega^2_f \xi^3}_{\text{internal energy + self-gravity}} \underbrace{- I\epsilon\xi.}_{\text{external interaction}}  
\end{equation}
As the donor is sufficiently deformed, the total energy of the internal fluid can decrease (see Appendix \ref{appx:eom_affine}), creating a local maximum in the effective potential. Going beyond the local maximum will then lead to a runaway in $\xi$ \citep{Wu:98}.}\footnote{\hang{In numerical experiments to be presented in \S \ref{sec:instab_dyn_tide_w_num}, we confirm the presence of the instability by solving the fully nonlinear equations derived in Appendix \ref{appx:eom_affine}. While previous studies (e.g., \citealt{Wu:98}) suggested that the four-wave interaction (the $\propto \xi^4$ term) in the effective potential could prevent the instability, the stabilization effect is eventually taken over by even higher-order corrections. When considering the f-modes, the full stability threshold is similar to that obtained at the three-wave order, as presented in the toy model here.}
}

To see the effect of the dynamical effects, we first consider $\xi \sim \cos(\varpi t) e^{\sigma t}$ with $\varpi^2\lesssim\omega_f^2$. Plugging it back to Eq. (\ref{eq:toy}) and consider the threshold case, $\sigma=0$, we have
\begin{equation}
    \omega_f^2 \kappa \xi^2 - (\omega_f^2 - \varpi^2)\xi  + \omega_f^2 I\epsilon=0,
\end{equation}
which is again a quadratic function of $\xi$. It has a real solution only if 
\begin{equation}
    (\omega_f^2 - \varpi^2)^2 - 4 \omega_f^4 \kappa I \epsilon \geq 0 \text{ or } \epsilon \leq \frac{(\omega_f^2-\varpi^2)^2}{4\omega_f^4 \kappa I}\text{ for stability}. 
    \label{eq:toy_stability_dyn}
\end{equation}
Compared to the static case of Eq. (\ref{eq:toy_stability_st}), including the dynamical effect makes instability more likely to occur as the maximum $\epsilon$ is smaller by a factor of $(\omega_f^2-\varpi^2)^2/\omega_f^4$. This model will be relevant for the discussions in \S \ref{sec:instab_eq_tide} about finite-frequency corrections to the equilibrium tide, \hang{relevant mainly to asynchronous donors in mildly eccentric binaries.} 

Another way to incorporate the dynamical effect is to consider $\xi=\xi_{\rm dyn} \cos(\varpi t) + \eta e^{\sigma t}$ with $\xi_{\rm dyn} \in \mathbb{R}$ treated as a known quantity. Eq. (\ref{eq:toy}) is now treated as a quadratic function of $\eta$, and a solution with $\sigma=0$ can be arranged, by again examining the discriminant, if
\begin{equation}
    \epsilon \leq \frac{1 - 4(\varpi^2/\omega_f^2)\kappa\xi_{\rm dyn} }{4\kappa I } \text{ for stability.}
    \label{eq:toy_stability_dyn2}
\end{equation}
The $\xi_{\rm dyn}$ term reduces (increases) the maximum $\epsilon$ a stable configuration can allow if $\xi_{\rm dyn} >0$  ($\xi_{\rm dyn} <0$). If a positive $\xi_{\rm dyn}$ has triggered some exponential growth in $\eta$, suddenly changing its sign will only make $\eta$ oscillatory but not damp away the growth. Therefore, for a sufficiently large $\xi_{\rm dyn}$ with a randomly fluctuating sign, the net effect will again make the instability more likely to occur. This model is relevant to the dynamical tide \hang{in a donor moving in a highly eccentric orbit} in \S \ref{sec:instab_dyn_tide_w_num}. 

The goal of our study is to quantify the enhanced hydrodynamical instability caused by the dynamical tide through analytical and numerical investigations. 

\section{Nonlinear hydrodynamics}
\label{sec:nl_hydro}

We focus mainly on a homogeneous, incompressible fluid object with a polytropic index $n=0$ and an adiabatic index $\Gamma\to\infty$ for the theoretical discussion, though our numerical investigations later in \S \ref{sec:instab_dyn_tide_w_num} will be extended to touch upon $n=1$ polytropes. 
This choice is for simplicity and to assist a direct comparison with the classical results obtained by \cite{Chandrasekhar:63, Chandrasekhar:87}. 
Denote the Lagrangian displacement of a mass element $dM$ as $\vxi=\vect{r} - \vect{x}$ with $\vect{r}$ and $\vect{x}$ the perturbed and original position vectors, respectively. We further decompose $\vxi$ into eigenmodes in the configuration space, as
\begin{equation}
    \vxi(t, \vect{x}) = \sum_m q_m(t) \vxi_m(\vect{x}),
    \label{eq:vxi_expand}
\end{equation}
For the Roche limit calculation, it is sufficient to consider only the quadrupolar ($l=2$) fundamental modes (f-modes). 
We drop the $l\geq 3$ perturbations, restrict our analysis to aligned spins when rotation is incorporated, and ignore the inertial modes \citep{Ho:99, Lai:06, Braviner:15, Xu:17}. 
Therefore, we label the modes only with their azimuthal quantum numbers $m$. 
The eigenfunctions $\vxi_m$ are computed from a non-rotating background as (e.g., \citealt{Poisson:14})
\begin{subequations}
\begin{align}
    &\vxi_{2}=\sqrt{\frac{2\pi}{3\omega_f^2}}\nabla(x_r^2Y_{22})=\frac{5}{4}\left[(x+iy)\vect{e}_x + (ix-y)\vect{e_y}\right], 
    \\
    &\vxi_{0} = \sqrt{\frac{2\pi}{3\omega_f^2}}\nabla(x_r^2Y_{20}) = -\frac{5}{2\sqrt{6}}[x\vect{e}_x+y\vect{e_y}-2z\vect{e}_z], \\
    &\vxi_{-2}=\sqrt{\frac{2\pi}{3\omega_f^2}}\nabla(x_r^2Y_{2,-2})=\frac{5}{4}\left[(x-iy)\vect{e}_x + (-ix-y)\vect{e_y}\right]. 
\end{align}
\label{eq:vxi}
\end{subequations}
and we have used $\omega_f^2=4/5$ as the (non-rotating) eigenfrequency under the linear theory.
The modes are normalized so that $\omega_f^2\langle \vxi_{m'}, \vxi_m\rangle = \omega_f^2 \int \vxi_{m'}^\ast\cdot\vxi_{m} \rho d^3x= \delta_{mm'}$. A mode with $|q_m|=1$, therefore, has a potential energy of unity (also approximately the magnitude of the binding energy of $M$). 
Note that the mode amplitudes $q_m$ are complex and can be further decomposed as $q_m= v_m e^{im\phi_q}$ with $v_m\in {\mathbb{R}}$ and $m\phi_q$ describing the phase.  
The $m=\pm 1$ modes are not excited by the tide and are dropped from the discussion. 

Using the modal expansion, $\vxi(t, \vect{x})$ is fully determined once $q_m(t)$ is obtained. First, we ignore rotation. The equations of motion of $q_m$ are given by
\begin{align}
    \ddot{q}_m = - \omega_f^2 q_m + \omega_f^2 K_m, 
    \label{eq:ddq_tide_only}
\end{align}
where 
\begin{align}
    K_2 &= \frac{3}{4} \epsilon e^{-2i\Phi} + q_2\left( -\frac{95\sqrt{6}}{84}q_0 + \frac{5}{8} \epsilon \right) - \frac{5\sqrt{6}}{16} \epsilon q_0 e^{-2i\Phi}, \\
    K_0 &= -\frac{\sqrt{6}}{4} \epsilon -  \frac{95\sqrt{6}}{168}( 2q_2 q_{-2}- q_0^2)  - \frac{5\epsilon}{16} (\sqrt{6}q_2 e^{2i\Phi} +  \sqrt{6}q_{-2}e^{-2i\Phi} + 2q_0),
\end{align}
with $\epsilon=M'/D^3$. We use $D$ and $\Phi$ to denote orbital separation and phase. 
The derivation follows a variational principle and is presented in detail in Appendix \ref{appx:eom_affine}. Note that in a homogeneous ellipsoid, the perturbation satisfies the affine model of \citet{Carter:83, Carter:85, Diener:95}, where the perturbed position $\vect{r}$ is related to the unperturbed position $\vect{x}$ via 
\begin{equation}
    r_i(t, \vect{x})=q_{ij}(t) x_j,
    \label{eq:affine_def}
\end{equation}
where the mapping between $q_{ij}(t)$ and $q_m(t)$ is given in Eq. (\ref{eq:r_vs_x}). The affine model allows the exact nonlinear form of $K_m$ to be obtained (available in Appendix \ref{appx:eom_affine}, Eqs. \ref{eq:Pim}-\ref{eq:Rm}, and allows for arbitrary values of $n$ and $\Gamma$ with the caveat that Eq. \ref{eq:affine_def} may not be accurate beyond $n=1/(\Gamma-1)=0$). Nonetheless, we expand the results to the three-wave order (that is, we count $q_m\sim \epsilon$ and include terms to $\mathcal{O}(\epsilon^3)$ in the energy variations and $\mathcal{O}(\epsilon^2)$ in the equation of motion). It is already sufficient to reproduce the Roche limit at this order. More importantly, such expanded equations can also be obtained following \citet{VanHoolst:94}, which takes a general set of linear eigenmodes as input to compute their nonlinear couplings and applies to generic compressible fluids (realistic stars and planets). This is advantageous over the affine model, which only self-consistently describes idealized homogeneous ellipsoids. Therefore, our study bridges the two different formulations of nonlinear hydrodynamics. The exact agreement between the two formulations at the three-wave order for homogeneous ellipsoids is presented in Appendix \ref{appx:grav_E}. 

We pause here and briefly describe how the nonlinear term leads to instability. 
As the eigenfunctions $\vxi_m \propto e^{im\phi}$ (Eq. \ref{eq:vxi}), we multiply both sides of Eq. (\ref{eq:ddq_tide_only}) by $e^{im\phi}$ and set $\phi=\phi_q$. Effectively, this corresponds to evaluating the fluid displacement along the major axis of the perturbed ellipsoid. 
At the linear order and under the static limit $\ddot{q}_m=0$, we have 
\begin{equation}
    v_2^{\rm (st)} = \frac{3}{4} \epsilon >0, \text{ and } v_0^{(\rm st)}=-\frac{\sqrt{6}}{4}\epsilon<0,
\end{equation}
with $v_m = (q_m e^{im\phi})_{\phi=\phi_q}$.  Treating $(-v_0)\sim v_2\sim \xi$, we see that Eq. (\ref{eq:ddq_tide_only}) takes the same form as the toy model of Eq. (\ref{eq:toy}), and the nonlinear terms in $K_m$ tend to destabilize the system when $\xi>0$. Thus, instability is most likely to occur when $\xi$ reaches its maximum along the major axis. In the equilibrium limit (without mode resonances; following the convention of \citealt{Yu:24a}), where the tidal bulge follows the companion, we have $\phi_q=\Phi$, matching the intuition that mass transfer first occurs at the L1 Lagrangian point. At the quadrupolar order, the L2 point ($\phi=\Phi+\pi$) destabilizes at the same time. 

Before solving for the instability threshold quantitatively, we now extend the equations of motion to include rotation. In this analysis, we focus on the case where the background spin is uniform and the spin axis is aligned with the orbital angular momentum (along $\vect{e}_z$). The equations of motion are now described in a frame that corotates with $M$ with three modifications. 

The first is a simple Doppler effect, which corresponds to a phase shift $\Phi_C=\Phi - \omega t$. The mode amplitude changes accordingly, as $q_{mC} = q_m e^{im\omega t}$ while keeping $q_me^{im\Phi} = q_{mC}e^{im\Phi_C}$. Quantities with a subscript $C$ are evaluated in the corotating frame. Since the analyses later will be in the corotating frame as the default, we will drop the subscript $C$ in the mode amplitude $q_{m}$ from this point onward. For the orbital phase, however, we will still make the distinction between the inertial frame $\Phi$ and the corotating frame $\Phi_C$. 

The second correction comes from the centrifugal force whose detailed incorporation is presented in Appendix \ref{appx:eom_affine}. Since the force can be derived from a specific centrifugal potential $d\mathcal{R}/dM = - \omega^2(r^2-z^2)/2$, its impact on the equations of motion can be obtained through a variational principle by considering the energy $\mathcal{R}$. When doing so, it is important to introduce an $l=0$ mode whose amplitude we denote as $q_{\rm tr}$ (a perturbation of the trace of $q_{ij}$ in \citealt{Diener:95}) and allow it to be coupled to the $q_m$ mode at the three-wave level to properly capture the correction at $\mathcal{O}(\epsilon^2)$. The same result can also be obtained from the formulation of \citet{VanHoolst:94} by splitting $d\mathcal{R}/dM = (d\mathcal{R}/dM)_{l=0} + (d\mathcal{R}/dM)_{l=2} = - \omega^2 r^2/3 - \omega^2 (x^2+y^2-2z^2)/6$, and adding $(d\mathcal{R}/dM)_{l=0}$ to the background (monopolar) gravitational potential and $(d\mathcal{R}/dM)_{l=2}$ to the first-order, quadrupolar Eulerian potential perturbation (same as the tidal interaction potential). Note that our approach treats the centrifugal potential as a (nonlinear) drive of the modes defined by a non-rotating background, instead of redefining a new set of eigenmodes of a non-spherical background where the centrifugal perturbation is absorbed. In other words, the centrifugal potential is treated on an equal footing as the tidal potential. 

Lastly, to incorporate the Coriolis force to the lowest order in $\omega$, we split the corotating frame mode amplitude $q_m$ as
\begin{equation}
    q_m=c_{m,+} + c_{m,-} = c_{m,+} + c_{-m,+}^\ast,
    \label{eq:qm_decom_into_cmpm}
\end{equation}
with
\begin{subequations}
\begin{align}
    &\dot{c}_{m,+}=-i\omega_{f,+} c_{m,+} + \frac{i\omega_f K_m}{2}, \\
    &\dot{c}_{m,-}=i\omega_{f,-} c_{m,-} - \frac{i\omega_f K_m}{2}.
\end{align}
\label{eq:dcdt}
\end{subequations}
This splitting is based on the phase-space expansion of \citet{Schenk:02}, and the Coriolis force is absorbed as a frequency shift \citep{Lai:21, Yu:24a, Yu:25a},
\begin{align}
    \omega_{f,\pm}=\omega_f \pm\frac{\langle\vxi_m, -i\vect{\omega}\times\vxi_m \rangle}{\langle\vxi_m, \vxi_m \rangle}=\omega_f \mp \frac{m}{l} \omega. 
    \label{eq:omega_f_pm}
\end{align}
Higher-order corrections from the Coriolis force are ignored (but see, e.g., \citealt{Dewberry:22}
).  
The corresponding second-order equation can be obtained by differentiating $\dot{c}_{m,\pm}$ one more time and eliminating the first-order derivatives with the equations of motion, leading to
\begin{equation}
    \ddot{q}_m = -\omega_f^2 q_m + m \omega_f\omega (c_{m,+}-c_{m,-}) + \omega_f^2 K_m.
\end{equation}
Compared to Eq. (\ref{eq:ddq_tide_only}), there is an additional term $\propto (c_{m,+} - c_{m,-})$ due to the Coriolis effect. While this in principle represents a new degree of freedom (as prograde and retrograde modes are split due to rotation), under the static limit $c_{m,\pm}\simeq[1\pm(m/l) \omega/\omega_f] (K_m/2)$, we have
\begin{equation}
    m \omega_f\omega (c_{m,+}-c_{m,-}) \simeq \frac{(m\omega)^2}{l} K_m (\simeq \frac{3}{2}\omega^2 \epsilon e^{-2i\Phi_C} \equiv C_2^{\rm (st)},\ \text{for }m=2)
    \label{eq:dcm_ad}
\end{equation}
which is $\mathcal{O}(\omega^2 \epsilon)\sim\mathcal{O}(\epsilon^2)$. Since this is already as small as the smallest terms in Eqs. (\ref{eq:ddq_tide_only}), we will approximate this term as $C_2^{\rm (st)}$ from Eq. (\ref{eq:dcm_ad}) in our analytical discussion. When doing numerical experiments, we will solve directly the equations of $c_{m,\pm}$ (Eq. \ref{eq:dcdt}) to account for the additional degree of freedom.

Summarizing, the equations of motion we will solve in \S \ref{sec:instab_st_tide} and \S \ref{sec:instab_eq_tide} are 
\begin{equation}
    \ddot{q}_m = - \omega_f^2 q_m + \omega_f^2 K_m + C_m^{(\rm st)},
    \label{eq:ddq}
\end{equation}
where 
\begin{align}
    K_2 &= \frac{3}{4} \epsilon e^{-2i\Phi_C} + q_2\left( -\frac{95\sqrt{6}}{84}q_0 + \frac{5}{8} \epsilon + \frac{5}{4} \omega^2 \right) - \frac{5\sqrt{6}}{16} \epsilon q_0 e^{-2i\Phi_C}, \label{eq:K2_3m}
    \\
    K_0&=-\frac{\sqrt{6}}{4} \epsilon - \frac{\sqrt{6}}{6} \omega^2- \frac{95\sqrt{6}}{168}( 2q_2 q_{-2}- q_0^2)  - \frac{5\epsilon}{16} (\sqrt{6}q_2 e^{2i\Phi_C} +  \sqrt{6}q_{-2}e^{-2i\Phi_C} + 2q_0) + \frac{5}{4}\omega^2 q_0, 
    \label{eq:K0_3m}
\end{align}
and $C_2^{(\rm st)}$ is given by Eq. (\ref{eq:dcm_ad}) while $C_0^{\rm (st)}=0$. We will also consider 
\begin{equation}
    \ddot{q}_m e^{im\phi_q} = - \omega_f^2 v_m + \omega_f^2 \bar{K}_m + \bar{C}_m^{\rm (st)},
    \label{eq:ddv}
\end{equation}
where $\bar{K}_m=K_m e^{im\phi_q}$ and $\bar{C}_m^{\rm (st)} = C_m^{\rm (st)} e^{im\phi_q}$. In the equilibrium limit (as considered in \S \ref{sec:instab_st_tide} and \ref{sec:instab_eq_tide}) without free oscillations of the mode due to resonances, $\phi_a=\Phi_C$, and one just needs to replace $q_m$ with $v_m$ and drop the phases following $\epsilon$ in $K_m$ to obtain $\bar{K}_m$. Eq. (\ref{eq:ddv}) is real valued. 

\subsection{Consistency with previous analyses}
\label{sec:instab_st_tide}

\begin{figure}
    \centering
    \includegraphics[width=0.5\linewidth]{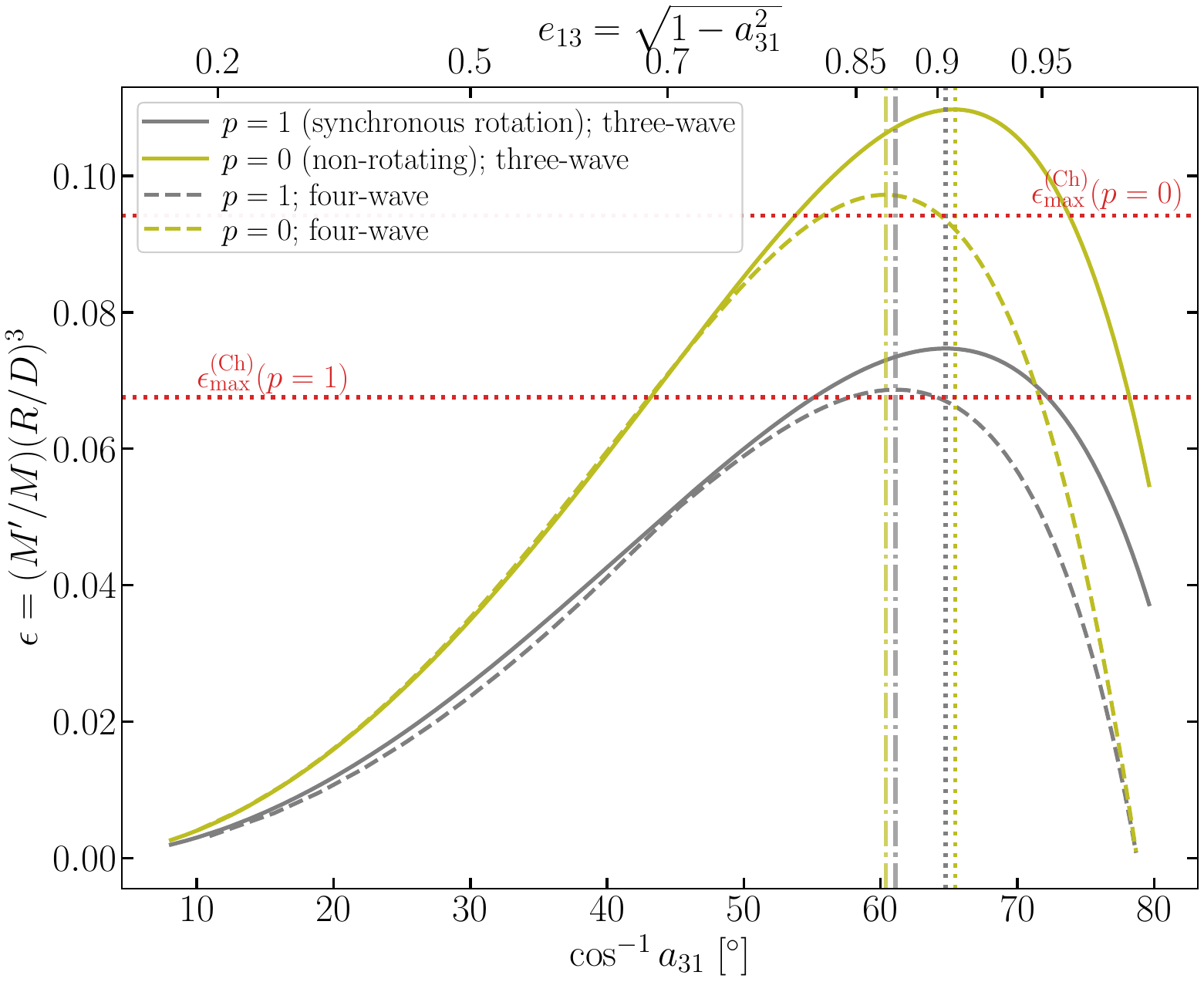}
    \caption{Equilibrium sequences under the static limit. Curves in gray correspond to synchronized donors (with $p=1$), while those in yellow are for the non-rotating case ($p=0$). The solid and dashed lines are evaluated at the three-wave and four-wave orders, respectively. Vertical lines mark the locations where $\epsilon$ reaches the maxima. The classic results of \citet{Chandrasekhar:63} are shown in the red horizontal lines. }
    \label{fig:config_stat_tide}
\end{figure}

First, we consider the case where $M$ moves in a circular orbit with synchronous rotation, so the left-hand sides of Eq. (\ref{eq:ddq}) or (\ref{eq:ddv}) vanish. Our goal is to establish the consistency between our formulation based on an order-by-order expansion with the classic analyses of \citet{Chandrasekhar:63} under this static limit. 
We write the spin frequency $\omega$ as $\omega^2=p \epsilon$, with $p=1+M/M'$ for synchronized rotation in a circular orbit. Note that our $p$ differs from the one in \citet{Chandrasekhar:63} by 1, as $p=p^{\rm (Ch)} + 1$. The superscript ``(Ch)'' will be used when we quote notations or results from \cite{Chandrasekhar:63}. In our analysis, we will focus on $M'\gg M$ (as for planet migration or TDEs), so a synchronous donor has $p=1$. 

Inspired by \citet{Chandrasekhar:63}, we note that the perturbed fluid can be described by an ellipsoid with $\bar{x}^2/a_1^2 + \bar{y}^2/a_2^2 + \bar{z}^2/a_3^2=1$ at the surface. The barred coordinates are aligned with the axes of the ellipsoid. In the limit with circular orbit and synchronous rotation, the barred frame coincides with the corotating frame. The axes are given by (Eqs. \ref{eq:a123_full} and \ref{eq:qtr_incomp})
\begin{subequations}
\begin{align}
    a_1 &= 1 + \frac{5}{2}v_2 - \frac{5\sqrt{6}}{12} v_0, \\
    a_2 &= 1 - \frac{5}{2}v_2 - \frac{5\sqrt{6}}{12} v_0, \\
    a_3 &= 1 + \frac{5\sqrt{6}}{6} v_0. 
\end{align}
\end{subequations}
For a given $a_{31}=a_3/a_1$, it defines a linear relation between $v_0$ and $v_2$,
\begin{equation}
    v_0 = \frac{ \sqrt{6}[5a_{31}v_2 +2(a_{31}-1)] }{5(2 + a_{31})},
    \label{eq:v0_vs_v2_given_a31}
\end{equation}
Using this relation to eliminate $v_0$, Eq. (\ref{eq:ddv}) now formally become two quadratic equations of $v_2$,
\begin{align}
    &X_{22} v_2^2 + X_{21}v_2 + X_{20}=X_{02} v_2^2 + X_{01}v_2 + X_{00}=0.
    \label{eq:v2_quad_form}
\end{align}
where the coefficients $X_{mn}$ are functions of $(\epsilon, a_{31})$. 
The first subscript indicates the azimuthal number $m$ of the corresponding equation (of $\ddot{q}_2$ or $\ddot{q}_0$), while the second indicates the power in $v_2$. 
The expressions of the coefficients $X_{mn}$ are straightforward to obtain but lengthy, so we report their values in Appendix \ref{appx:coeff_poly}. 

If there exists a $v_2$ that simultaneously solves the above two polynomials, then the resultant of the two polynomials must vanish \citep{MathWorld:Resultant}. That is, 
\begin{align}
    \det[\boldsymbol{S}(X_{2n}, X_{0n})]&=0, \label{eq:det_syl}\\
    \text{with } \boldsymbol{S}(X_{2n}, X_{0n}) &= 
    \begin{bmatrix}
        X_{22}, & X_{21}, & X_{20}, &0 \\
        0, & X_{22}, & X_{21}, & X_{20} \\
        X_{02}, & X_{01}, & X_{00}, &0 \\
        0, & X_{02}, & X_{01}, & X_{00} 
    \end{bmatrix},
\end{align}
where $\boldsymbol{S}(X_{2n}, X_{0n})$ is also known as the Sylvester matrix, and the resultant is its determinant. For given $a_{31}$, Eq. (\ref{eq:det_syl}) can then be used to determine the value of $\epsilon$, thereby determining an equilibrium sequence. 

Some examples of equilibrium sequences (plotted as $\epsilon$ versus $\cos^{-1} a_{31}$ or $e_{13}=\sqrt{1-a_{31}^2}$) are shown in Fig. \ref{fig:config_stat_tide}. We use the color gray to represent a synchronized donor with $p=1$ (as $M\ll M'$), corresponding to the Roche sequence, while yellow for the non-rotating case with $p=0$ (the Jeans sequence; note $\ddot{q}_m$ are ignored in this calculation). As the ellipsoid becomes more deformed (towards the right of the figure), the $\epsilon$ that leads to an equilibrium does not increase monotonically. Instead, it reaches a maximum, $\epsilon_{\rm max}$, at around $\cos^{-1} a_{31} \simeq 60^\circ-65^\circ$. This means for $\epsilon>\epsilon_{\rm max}$, no equilibrium configuration of the ellipsoid can be arranged, indicating hydrodynamical instability. We can also translate $\epsilon>\epsilon_{\rm max}$ to $(D/R_t) < (1/\epsilon_{\rm max})^{1/3}$. In other words, when the binary separation $D$ measured in the tidal radius $R_t$ is less than $(1/\epsilon_{\rm max})^{1/3}$, the donor $M$ cannot remain in equilibrium, and hydrodynamic instability occurs, defining the Roche limit. 

In our calculation, we find that it is sufficient to capture the classic results of \citet{Chandrasekhar:63} with percent-level accuracy using only three-wave interactions (solid lines; Eqs. \ref{eq:K2_3m} and \ref{eq:K0_3m}). For the synchronous ($p=1$) case, we find $\epsilon_{\rm max}\simeq 0.0747$ at $\cos^{-1} a_{31} \simeq 64.8^\circ$, meaning the threshold separation is at $ D^{(\rm th)}=(1/\epsilon_{\rm max})^{1/3} R_t \simeq 2.38 R_t$, matching the classic Roche limit (indicated by the horizontal dotted lines) of $2.44 R_t$ to $2.5\%$. A similar level of accuracy is also obtained for the non-rotating $p=0$ case, with $\epsilon_{\rm max}\simeq 0.110$, $(1/\epsilon_{\rm max})^{1/3}\simeq 2.09$, achieved at $\cos^{-1} a_{31}\simeq 65.4^\circ$. 

The accuracy can be improved further by including the next order, four-wave interactions. The equations of motion at this order are derived in Appendix \ref{appx:eom_affine}, Eqs. (\ref{eq:K2_4m}) and (\ref{eq:K0_4m}). Still using Eq. (\ref{eq:v0_vs_v2_given_a31}) to eliminate $v_0$ in terms of $v_2$ and $a_{31}$,\footnote{
Strictly, at the four-wave order one should account for corrections to $a_{1,2,3}$ from the monopolar mode, $q_{\rm tr}$; Eqs. (\ref{eq:a123_full}) and (\ref{eq:qtr_incomp}). This introduces nonlinear corrections to Eq. (\ref{eq:v0_vs_v2_given_a31}), which we ignore here. 
}
we again obtain two polynomial equations of $v_2$, 
\begin{equation}
    \sum_{n=0}^3 W_{mn} v_2^n=0,
    \label{eq:v2_cubic_form}
\end{equation}
by setting $\ddot{q}_m=0$, with $m=2$ and $0$. The resultant (now functions of $W_{mn}$) needs to vanish as before for the two equations to hold simultaneously. We can thus find the equilibrium sequences at the four-wave order, shown in the dashed lines in Fig. \ref{fig:config_stat_tide}. The result now matches those from \citet{Chandrasekhar:63} almost exactly. For the $p=1$ case shown in gray, we have $\epsilon_{\rm max} \simeq 0.0687$ at $\cos^{-1} a_{31}=61.1^\circ$, corresponding to a Roche limit of $D^{(\rm th)}=(1/\epsilon_{\rm max})^{1/3} R_t=2.44 R_t$. 

While the main result of this Subsection is to reproduce the classic ones by \citet{Chandrasekhar:63} on homogeneous ellipsoids, it is significant for two reasons. First, it validates our formulation based on which $\ddot{q}_m\neq 0$ corrections will be introduced. Second, it bridges the analyses of \citet{Chandrasekhar:63} on idealized ellipsoids and \citet{VanHoolst:94} on general stars and planets (see also \citealt{Weinberg:12, Weinberg:16, Yu:23a}). In general, exact nonlinear expressions are challenging to find, but an order-by-order expansion as in \citet{VanHoolst:94} is straightforward to compute (though lengthy). Our result provides a framework for how one can start from \citet{VanHoolst:94} to determine the Roche limit of a realistic star/planet theoretically.

\subsection{Finite-frequency corrections in asynchronous, low eccentricity systems}
\label{sec:instab_eq_tide}

In this Subsection, we will extend the stability analysis to analytically consider finite-frequency corrections due to the equilibrium tide (following the convention of \citealt{Yu:24a}), where the perturbation varies with the instantaneous tidal forcing. In other words, the tidal bulge still points towards the companion.  
The component that varies at a mode's natural frequency and depends on the excitation history of the mode (what we call the dynamical tide) will be deferred to the numerical investigation in \S \ref{sec:instab_dyn_tide_w_num} and ignored in the analysis here. 
This means that the prediction derived here will be accurate for binaries with low eccentricity $e\lesssim 0.3$. 
For more eccentric binaries, the dynamical tides can exhibit complicated, often diffusive, evolutions that may trigger instabilities at separations greater than the prediction of this Subsection by tens of percent. 

To proceed, we write $q_m=(1+z_m) q_{m}^{(0)}$ where $q_{m}^{(0)}\simeq I_m \epsilon e^{-i m\Phi_C}$ represents an approximate solution of the perturbation, with $I_2=3/4$ and $I_0=-\sqrt{6}/4$ (Eqs. \ref{eq:K2_3m} and \ref{eq:K0_3m}). We treat $z_m$ as a low-frequency variable (whose time derivatives will be ignored) that we will use to determine the equilibrium structure of the donor. 
Note Eq. (\ref{eq:ddv}) still holds as long as we incorporate the time derivatives on the left-hand side due to $\ddot{q}_m\simeq\ddot{q}_m^{(0)}$. 
In essence, this model corresponds to the toy model described in Eq. (\ref{eq:toy_stability_dyn}). 

The procedure for finding the Roche limit follows identically those described in \S \ref{sec:instab_st_tide}. 
Choosing again a linear relation between $v_0$ and $v_2$ as in Eq. (\ref{eq:v0_vs_v2_given_a31}), Eq. (\ref{eq:ddv}) now becomes two quadratic equations of $z_2$:
\begin{align}
    &Z_{22} z_2^2 + Z_{21} z_2 + Z_{20} =Z_{02} z_2^2 + Z_{01} z_2 + Z_{00} = 0.
    \label{eq:z2_quad_form}
\end{align}
The coefficients, now depending on $(\epsilon, a_{31})$ as well as $q_{m}^{(0)}$ and their derivatives, are given in Appendix \ref{appx:coeff_poly}. 
Requiring the resultant of the two polynomial equations in Eq. (\ref{eq:z2_quad_form}) to vanish allows one to find $\epsilon$ given $a_{31}$. For a given orbit and synchronization level, the maximum $\epsilon$ in the equilibrium sequence then determines the corresponding Roche limit.
As the tidal deformation is the strongest at the pericenter, we will evaluate all the quantities entering $Z_{mn}$ there. Consequently, the analysis considers the instantaneous hydrodynamical stability at each pericenter passage. 

To find $\ddot{q}_{m}^{(0)}$, we note
\begin{align}
    &\dot{\Phi}_p^2 = \varpi_p^2= \frac{(M+M')(1+e)}{D_p^3} = \epsilon(1+q)(1+e),
\end{align}
at the pericenter (denoted with the subscript $p$) and $q=M/M'(=0 \text{ in this work by default})$. 
A simple approximation of $q_{m}^{(0)}\simeq I_m \epsilon e^{- i m \Phi_c}$ with $\Phi_c=\Phi - \omega t$ leads to
\begin{align}
    &[\ddot{q}_{m}^{(0)}]_{p} e^{im\Phi_c}\simeq I_m \epsilon \left[-m^2(\varpi_p-\omega)^2 - 3\left(\varpi_p^2-\frac{1+M'}{D_p^3}\right)\right] = -I_m\epsilon^2\left\{m^2\left[\sqrt{(1+q)(1+e)} - \sqrt{p}\right]^2 + 3e(1+q)\right\}.
    \label{eq:ddqm_apprx}
\end{align}
We have evaluated the result at the pericenter and assumed an unperturbed Keplerian orbit. 

\begin{figure}
    \centering
    \includegraphics[width=0.5\linewidth]{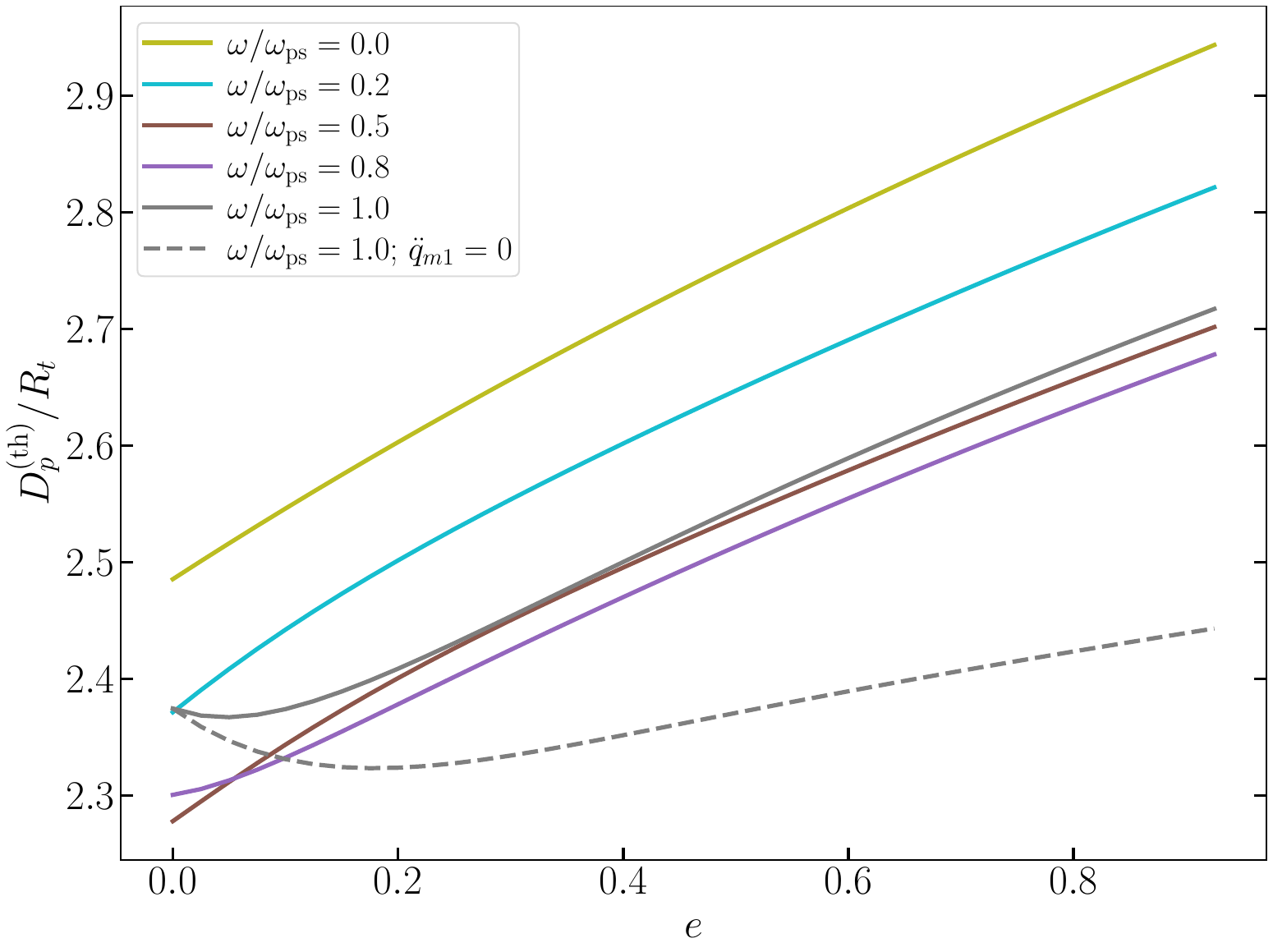}
    \caption{The threshold pericenter separation at which hydrodynamical instability occurs as a function of the orbital eccentricity. The plot is for the equilibrium tide model and ignores the dynamical tide. 
    Different colors represent different levels of spin synchronization (gray for pseudo-synchronized rotation; yellow for zero spin).  }
    \label{fig:Dp_th_eq_tide}
\end{figure}

For practical usefulness, we show in Fig. \ref{fig:Dp_th_eq_tide} the threshold pericenter separation, $D_p^{(\rm th)}/R_t= (1/\epsilon_{\rm max})^{1/3}$, below which hydrodynamical instability can occur. Different colors represent different spin rates relative to the pseudo-synchronization values $\omega_{\rm ps}$ in \citet{Hut:81}, 
\begin{equation}
    \frac{\omega_{\rm ps} P_{\rm orb}}{2\pi} = \frac{1 + \frac{15}{2}e^2 + \frac{45}{8}e^4 + \frac{5}{16}e^6}{(1+3e^2+\frac{3}{8}e^4)(1-e^2)^{3/2}},
\end{equation}
where $P_{\rm orb}$ is the orbital period.
Also shown as a reference in the gray-dashed line is the threshold $D_p^{\rm (th)}$ under the static tide limit (obtained by setting $\ddot{q}_m=0$). At a fixed eccentricity, the nonrotating configuration has the largest $D_p^{(\rm th)}$ (most prone to instability) as it has the largest $|\ddot{q}_{m}^{(0)}|$ (Eq. \ref{eq:ddqm_apprx}) among the configurations we consider. As the rotational rate increases, $D_p^{(\rm th)}$ first decreases due to the reduction of the $|\ddot{q}_{m}^{(0)}|$ term, and then increases as the centrifugal potential dominates. The minimum $D^{\rm (th)}_p$ is achieved when $\omega/\omega_{\rm ps}\simeq 0.5$ for circular binaries. 
When the orbital eccentricity $e$ increases, $D^{\rm (th)}_p$ increases approximately linearly with $e$, especially for slowly rotating systems. For synchronous systems, there is a decrease in $D_p^{(\rm th)}$ when $e \lesssim 0.1$ due to the decrease in $\omega_{\rm ps}$, which reduces the centrifugal force.

We pause here to discuss further the meaning of Fig. \ref{fig:Dp_th_eq_tide}. As the coefficients in Eq. (\ref{eq:z2_quad_form}) are evaluated at the pericenter, Fig. \ref{fig:Dp_th_eq_tide} should be interpreted as the instantaneous instability threshold at the pericenter. Therefore, a donor $M$ marginally unstable at the pericenter may become stable once it moves away from the pericenter and $\epsilon$ (defined by the instantaneous separation) drops below the instability threshold. As a result, the growth of hydrodynamical instability during a single pericenter passage may not disrupt $M$ sufficiently to cause mass loss. Nonetheless, as long as the fluid dissipation timescale is long compared to the orbital period (which is very well satisfied for the large-scale f-mode when $1-e>0.01$ under the linear theory assuming damping due to turbulent convection; \citealt{Goldreich:77, Wu:05}), the growth of instability at every pericenter passage will accumulate over multiple orbits, eventually disrupting $M$ and triggering mass loss (see also numerical examples to be presented in the next Subsection). For systems with low eccentricities $e\lesssim 0.3$, Fig. \ref{fig:Dp_th_eq_tide} provides a decent approximation of the onset of mass loss as a runaway in mode amplitude occurs within a few ($<10$) orbits. For systems with higher eccentricities, the amount of growth per passage is smaller, given the smaller fraction of time $M$ spends near the pericenter ($2\pi/(\dot{\Phi}_p P_{\rm orb})\propto \sqrt{(1-e)^3/(1+e)}$). This requires more orbits for the perturbation to build up until a runaway eventually occurs. However, the dynamical tide component can also be excited in those systems, significantly lowering the threshold $\epsilon$ for a stable configuration (see Eq. \ref{eq:toy_stability_dyn} and note the dynamical tide plays the role of $\xi_{\rm dyn}$ in the toy model). In the weak damping case, mass loss from $M$ can occur at $D_p$ significantly greater than the prediction of Fig. \ref{fig:Dp_th_eq_tide} for highly eccentric systems. 

\subsection{Dynamical tide induced instability and diffusive tide}
\label{sec:instab_dyn_tide_w_num}

For binaries with significant eccentricities, the truly dynamical component of a mode can be excited at a pericenter passage. Such a component oscillates at a mode's natural frequency (hence phase incoherent with the orbit) and does not decrease in amplitude as the companion moves to the apocenter (assuming dissipation is small). In the next pericenter passage, the existing component of the dynamical tide can interfere with the new tidal kick it receives at the pericenter, hence making this component's evolution history dependent (as opposed to the equilibrium tide that depends only on the instantaneous orbital configuration). 
This dynamical component formally modifies the stability threshold according to the toy model in Eq. (\ref{eq:toy_stability_dyn2}), though its complicated evolution prohibits a simple analytical study as we have done in \S \ref{sec:instab_st_tide} and \ref{sec:instab_eq_tide}. 
Nonetheless, the hydrodynamical equations are sufficiently generic for arbitrary rotations and orbital configurations. We will use a numerical approach to investigate the associated hydrodynamical stability.  

In this Subsection, we will solve directly the phase-space expanded equations of $c_{m,\pm}$, Eqs. (\ref{eq:dcdt}), so that the Coriolis effect can be accounted for naturally. For the driving terms ($K_m$), we will consider both the version expanded to the three-wave order, Eqs. (\ref{eq:K2_3m}) and (\ref{eq:K0_3m}), and the exact nonlinear version derived in Appendix \ref{appx:eom_affine} following the affine model of \citet{Diener:95} (see also \citealt{Carter:83, Carter:85}). The monopolar mode $q_{\rm tr}$ is treated fully dynamically, where a differential equation of $\ddot{q}_{\rm tr}$ is solved instead of reducing $q_{\rm tr}$ under the static limit with Eqs. (\ref{eq:qtr_incomp}). 
Note that the affine model in \citet{Diener:95} allows for generic values of polytropic index $n$ and adiabatic index $\Gamma$. Thus, our numerical experiments will consider both a homogeneous ellipsoid with $(n,\ \Gamma)=(0,\ 20)$ as before\footnote{
Numerically, $\Gamma\to \infty$ is hard to implement. We therefore use a finite $\Gamma=20 \gg 1$ as an approximation.
}, 
and a polytrope with $(n,\ \Gamma)=(1,\ 2)$ that serves as a good approximation of a buoyancy-neutral Jovian planet.  However, a caveat for the $n=1$ model is that the starting point of the affine model (Eq. \ref{eq:affine_def}) is not self-consistently satisfied as the Lagrangian displacement $\vxi$ is no longer linear in position. We fix the companion's mass to be $M'=10^3$ and treat it as a point particle. When computing the orbital motion, we ignore tidal back-reaction and simply model it as Keplerian, with 
\begin{equation}
    \ddot{D} = D\dot{\Phi}^2 - \frac{M+M'}{D^2}; \quad \ddot{\Phi} = -2\frac{\dot{D}\dot{\Phi}}{D}. 
\end{equation}
The initial conditions are set at the apocenter, with the modes computed with their linear equilibrium tide solution. The set of differential equations is evolved using the \texttt{solve\_ivp} function in \texttt{SciPy} \citep{Scipy} with an explicit Runge-Kutta method of order 8 \citep{Hairer:93}.


\begin{figure}
    \centering
    \includegraphics[width=0.9\linewidth]{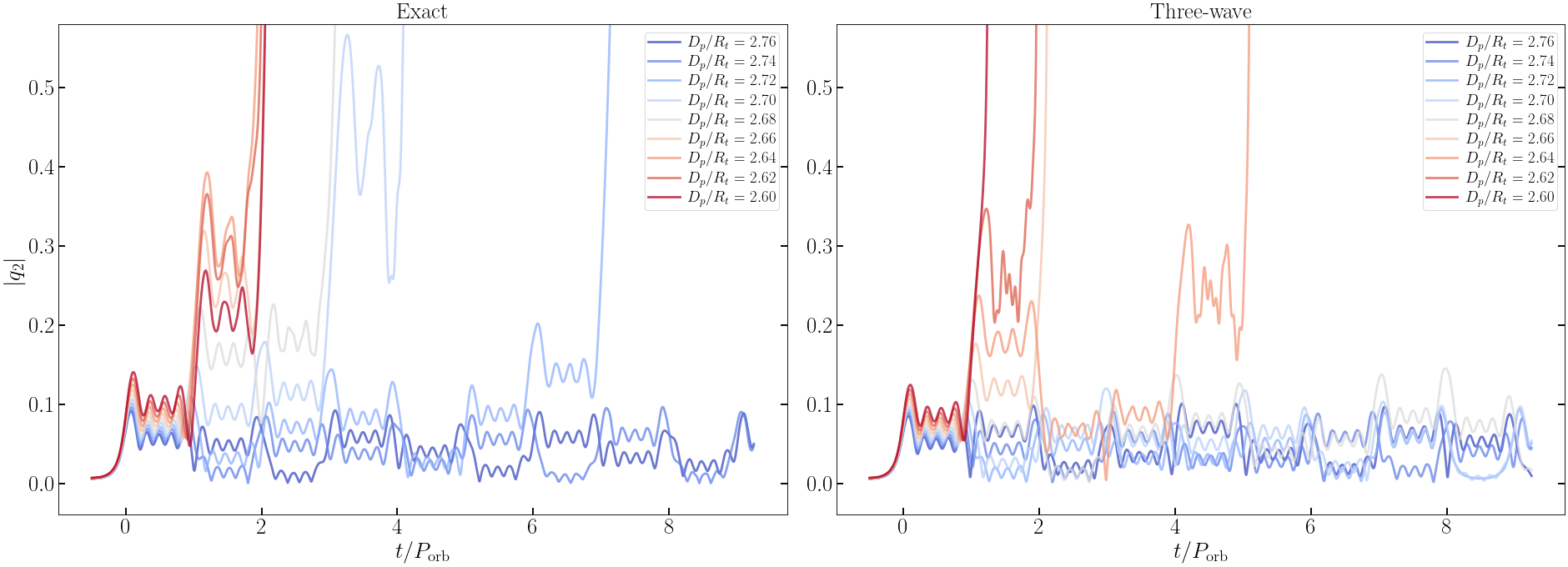}
    \caption{Numerical evolution of the $q_2$ mode for a non-rotating, homogeneous donor star in an eccentric orbit with $e=0.3$. 
    The left panel uses the exact nonlinear equations, while the right one uses the expanded version to the three-wave order.
    Runaway of the mode amplitude happens when $D_p/R_t \lesssim 2.72$ (left) or $D_p /R_t \lesssim 2.66$. This matches Fig. \ref{fig:Dp_th_eq_tide} where the threshold separation is predicted to be $D_p^{\rm (th)}\simeq 2.66 R_t$. }
    \label{fig:num_q2_e3_ns}
\end{figure}

In the first experiment shown in Fig. \ref{fig:num_q2_e3_ns}, we validate the results of \S \ref{sec:instab_eq_tide} by considering the numerical evolution of $|q_2| = |c_{2,+} + c_{-2,+}^\ast|$ for a non-rotating ($p=0$), homogeneous ($n=0$) star in a moderately eccentric orbit with $e=0.3$. We evolve the system using both the exact nonlinear equations (left panel) and the three-wave-expanded ones (right panel). Qualitatively, the two sets of equations yield similar results, with the exact version allowing the runaway to occur at a slightly further separation $D_p/R_t\simeq 2.72$. This is consistent with Fig. \ref{fig:config_stat_tide}, which shows that the three-wave equations slightly underestimates $D_p^{(\rm th)} \propto (1/\epsilon_{\rm max})^{1/3}$. 

The numerical result of Fig. \ref{fig:num_q2_e3_ns} also matches the theoretical prediction of Fig. \ref{fig:Dp_th_eq_tide}, which predicts $D_p^{(\rm th)}(e=0.3, p=0)\simeq 2.66 R_t$ at the three-wave order. However, a subtle difference is worth emphasizing. Fig. \ref{fig:Dp_th_eq_tide} predicts the instantaneous stability at the pericenter. Fig. \ref{fig:num_q2_e3_ns} shows that the instability may not directly lead to a runaway in the mode amplitude\footnote{In the following, we will use the words ``instability'' and ``runaway'' with different meanings. ``Instability'' is said to occur when a mode experiences exponential growth instantaneously (typically near the pericenter). If the pericenter passage timescale is much shorter than the growth timescale, the mode may become oscillatory, and the donor may remain bounded by its self-gravity once it moves away from the pericenter. In contrast, we use ``runaway'' when a mode's amplitude grows significantly to exceed order unity within a single pericenter passage. A runaway in the mode amplitude will require the donor to significantly restructure itself to remove the energy in the oscillation, which likely results in mass loss.   } 
in a single passage. As the donor moves away from the pericenter and $\epsilon$ decreases, the mode stabilizes and becomes oscillatory. However, in the next pericenter passage, the growth resumes, eventually leading to a divergence in the mode amplitude, potentially triggering mass loss from the donor.

\begin{figure}
    \centering
    \includegraphics[width=0.9\linewidth]{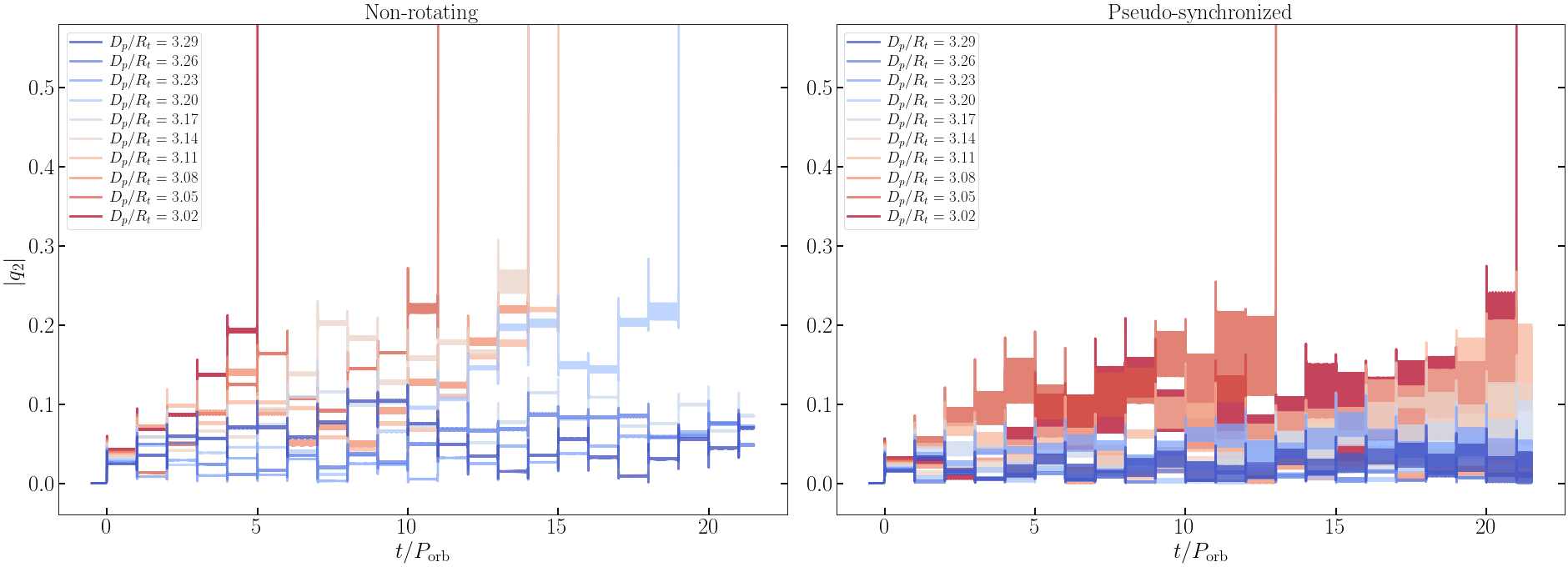}
    \caption{Numerical evolution of the $q_2$ mode for a homogeneous donor in an eccentric orbit with $e=0.9$. The donor is assumed to be non-rotating in the left panel and pseudo-synchronized in the right. Over multiple cycles' accumulation, runways in mode amplitude may happen at pericenter separation much greater than the prediction of Fig. \ref{fig:Dp_th_eq_tide}.  }
    \label{fig:num_q2_e9_ns_vs_ps}
\end{figure}

The accumulation of the dynamical tide becomes more significant in systems with higher eccentricities, with one such example shown in Fig. \ref{fig:num_q2_e9_ns_vs_ps} for a donor moving in a highly eccentric ($e=0.9$) orbit. We still consider a homogeneous donor but vary its rotation rates (left for the non-rotating case and right for the pseudo-synchronized case). Both cases are evolved using the three-wave equations of motion. As the hydrodynamical instability grows only momentarily at each pericenter passage, an eventual runaway in the mode amplitude requires multiple orbits to occur.
On the other hand, this accumulation allows runaway to occur at pericenter separations much greater than the prediction of Fig. \ref{fig:Dp_th_eq_tide}. For the non-rotating case, the left panel of Fig. \ref{fig:num_q2_e9_ns_vs_ps} shows that the runaway can kick in at a separation as large as $D_p/R_t\simeq 3.20$, whereas without the dynamical tide, the separation needs to be $D_p \lesssim 2.93 R_t$ based on Fig. \ref{fig:Dp_th_eq_tide}. Rotation suppresses the dynamical tide slightly \citep{Lai:97}, requiring a smaller $D_p/R_t \simeq 3.05$ for the runaway to occur within 20 orbits. We also note a larger variation in $|q_2|$ away from the pericenter. This is due to the beat of the $c_{2,+}$ (prograde) and $c_{-2, +}$ (retrograde) modes entering $q_2$. Note from Eq. (\ref{eq:omega_f_pm}) that the two modes have different natural frequencies when the background is rotating.  

Note the mode evolution largely resembles the \hang{diffusive tide (also known as the chaotic tide)} studied in, e.g., \citet{Mardling:95, Ivanov:04, Ivanov:07, Wu:18, Vick:18, Yu:21, Yu:22a}. 
We will elaborate on this point further in \S \ref{sec:app_chaotic_tide}.

\begin{figure}
    \centering
    \includegraphics[width=0.9\linewidth]{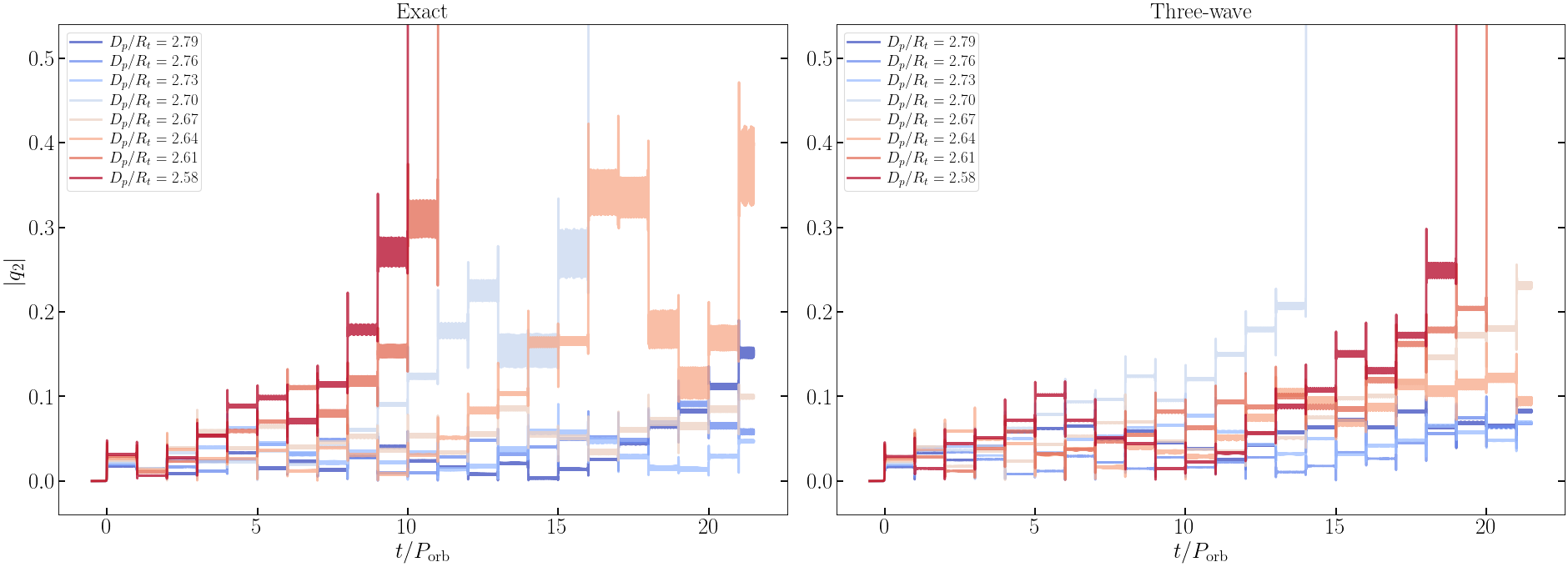}
    \caption{Numerical evolution of the $q_2$ mode for a non-rotating donor with $(n, \Gamma)=(1,2)$ in an eccentric orbit with $e=0.9$. Both the exact (left) and three-wave (right) nonlinear equations are considered. Runaways in mode amplitudes are observed when $D_p/R_t\lesssim 2.7$, matching nicely the predictions of \citet{Guillochon:11}. The threshold $D_p/R_t$ can be even greater if the system is integrated for more orbits. 
    }
    \label{fig:num_q2_e9_n_1}
\end{figure}

As our theoretical study is largely motivated by the numerical simulations of \citet{Guillochon:11}, we consider in Fig. \ref{fig:num_q2_e9_n_1} the evolution of a donor with $n=1$ and $\Gamma=2$ (a decent approximation for gaseous giant planets) moving in an eccentric orbit with $e=0.9$. In the affine model, the linear tidal overlap ($\propto I_m$) of an $n>0$ model is smaller than that of the $n=0$ model by a factor of $(5-n) \Ma$ with $\Ma=(\int x^2 dM)/3$; see Appendix \ref{appx:eom_affine}. For $n=1$, $(5-n)\Ma\simeq 0.523$. As a result, the $n=1$ model requires a $D_p/R_t$ smaller than the $n=0$ model by a factor of $\sim 1/[(5-n)\Ma]^{1/3} \simeq 1.24$ for the runaway to occur. Other than this overall scaling difference, the evolution of the $n=1$ model largely resembles the $n=0$ model, and the predictions of the exact nonlinear equations (left) can be largely captured by the three-wave equations (right). Remarkably, our results show nice consistency with the simulations of \citet{Guillochon:11}. Although the donor is not significantly disrupted in a single passage, the perturbation accumulates diffusively, eventually leading to a runaway in the mode amplitude in $\mathcal{O}(10)$ orbits when $D_p/R_t\lesssim 2.7 R_t$. 

\begin{figure}
    \centering
    \includegraphics[width=0.9\linewidth]{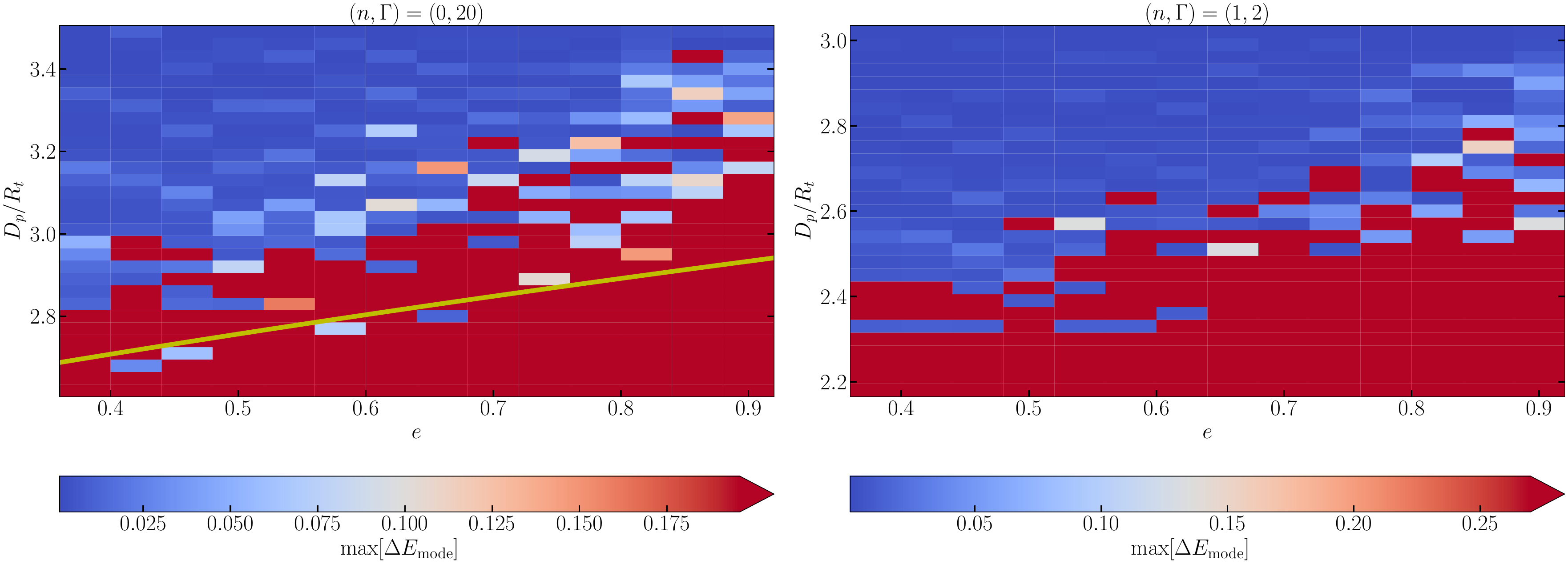}
    \caption{Maximum change in the mode energy for a non-rotating donor over 22 orbits. The left panel considers a homogeneous fluid ($n=0$) while $n=1$ in the right panel. The maximum energy is saturated at values where runaways in the mode amplitudes are observed. As a comparison, the equilibrium tide prediction (yellow curve from Fig. \ref{fig:Dp_th_eq_tide}) is shown again in yellow. } 
    \label{fig:e_Dp_map_num}
\end{figure}

In Fig. \ref{fig:e_Dp_map_num}, we show a survey of the maximum change in the mode energy $\Delta E_{\rm mode}$ reached in 22 orbits versus the orbital eccentricity $e$ and pericenter separation $D_p/R_t$. The mode energy is defined as $\Delta E_{\rm mode}= \Delta T + \Delta U + \Delta \Omega$, corresponding to the sum of the kinetic energy $\Delta T$ (Eq. \ref{eq:Delta_T}), internal energy $\Delta U$ (Eq. \ref{eq:Delta_U}), and self-gravity $\Delta \Omega$ (Eq. \ref{eq:Delta_Omega}). At the leading order and for $n=0$, $\Delta E_{\rm mode} \simeq (1/2)\sum_m(\dot{q}_m^2/\omega_f^2 + |q_m|^2)$ in our normalization. 
The exact nonlinear equations from the affine model are used in the hydrodynamical evolution and the energy calculation (see Appendix \ref{appx:eom_affine}; \citealt{Carter:85, Diener:95}). The donor is assumed to be non-rotating. In the left and right panels, we use respectively $(n, \Gamma)=(0, 20)$ and $(n, \Gamma)=(1, 2)$. In the color scale, we saturate at values where runaways in the mode amplitudes are observed. We empirically determine these saturation values by plotting $\Delta E_{\rm mode}$ as a function of $|q_2|$, which leads to two clearly separate sectors. For smaller values $|q_2|$, we observe a nearly quadratic relation $\Delta E \propto |q_2|^2$, corresponding to bounded oscillations. The other sector has $|q_2|$ piled at unity (a criterion we use to terminate the integration), and at this fixed $|q_2|$, $\Delta E_{\rm mode}$ is multi-valued. The maximum $\Delta E_{\rm mode}$ from the first branch is taken to set the maximum in the color bar. 

We note an approximately linear increase in the threshold $D_p/R_t$ with increasing orbital eccentricity $e$. This is the most critical modification of the earlier prediction of \citet{Sepinsky:07a}, which mainly considered the equipotential and found an expanded Roche lobe in eccentric orbits (harder for mass loss to occur; opposite of our results). Our study, therefore, highlights the significance of accounting for the hydrodynamical response of the fluid. 
Compared to the prediction assuming instantaneous equilibrium tide in Fig. \ref{fig:Dp_th_eq_tide}, the accumulation of the dynamical tide over multiple orbits makes the runaways occur at even greater values of $D_p/R_t$. 
We also note that at fixed $e$, the energy $\Delta E_{\rm mode}$ does not vary monotonically from red to blue as $D_p$ increases. Instead, a random alternation of red and blue is observed near the threshold, indicating the stochastic nature of the evolution (see \S \ref{sec:app_chaotic_tide} for more discussions). 
We further emphasize that Fig. \ref{fig:e_Dp_map_num} serves only as a conservative estimation of the Roche limit, as the integration lasts only 22 orbits. 
Figs. \ref{fig:num_q2_e9_ns_vs_ps} and \ref{fig:num_q2_e9_n_1} suggest that more systems may run away given more orbits, as the evolution is diffusive. 
However, integrating the system for longer would require a more careful treatment of the numerical scheme beyond brute-force integration of the differential equations to improve both efficiency and accuracy. This is left for future investigations.

\begin{figure}
    \centering
    \includegraphics[width=0.9\linewidth]{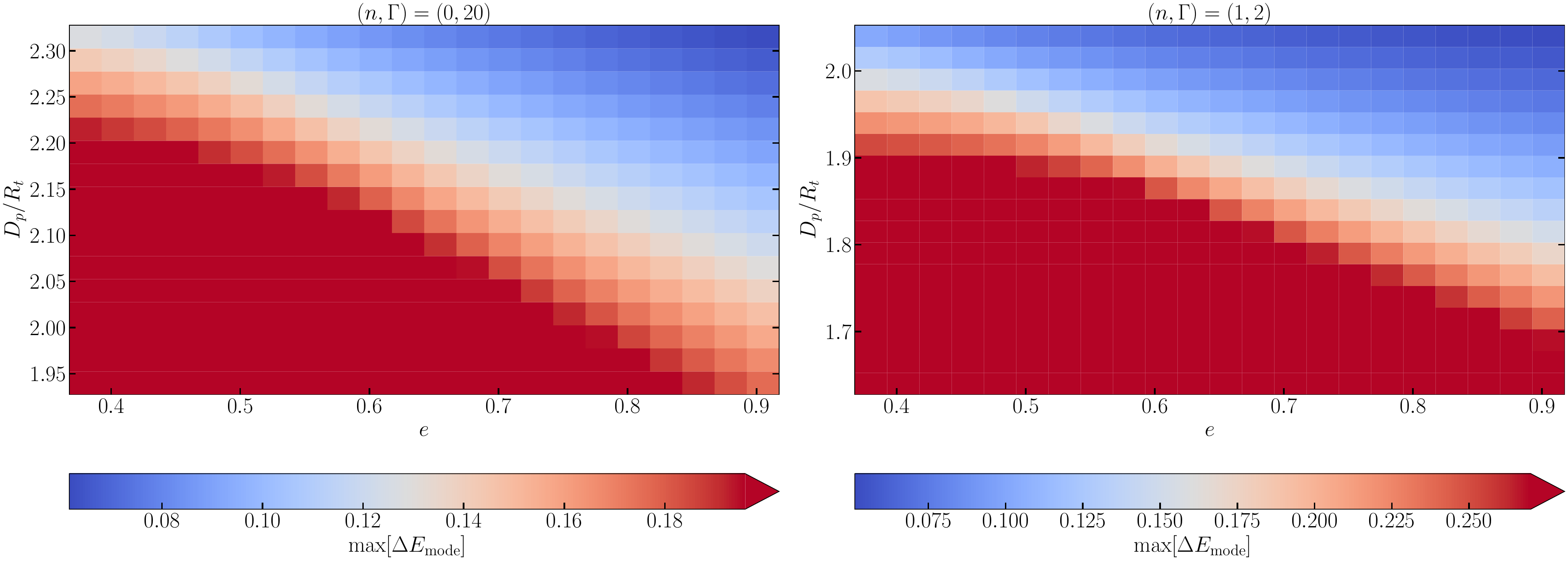}
    \caption{Similar to Fig. \ref{fig:e_Dp_map_num} but shows the mode energy after only a single pericenter passage. } 
    \label{fig:e_Dp_map_num_one_pass}
\end{figure}

So far, the discussions have ignored damping. If there exists some efficient damping mechanism that could remove all the energy in the dynamical tide within one orbit, we would need to consider the hydrodynamical stability in a single passage. Fig. \ref{fig:e_Dp_map_num_one_pass} shows the energy change acquired after a single passage (still assuming non-rotating $M$). When plotting $\Delta E_{\rm mode}$ versus $|q_2|$, we do not see a clear split into two sectors, and $|q_2|$ does not grow exponentially. Instead, at locations where bounded oscillations terminate in the multi-orbit case, we note $\Delta E_{\rm mode}$ becomes multi-valued at a given $|q_2|$. Such branching suggests instability to additional perturbations (and runaways occur in the next orbit), so we still saturate the color bar as before. We've also checked with three-wave equations of motion (which predict runaways in a single passage) and found consistent thresholds in $D_p/R_t$ as in Fig. \ref{fig:e_Dp_map_num_one_pass} (which also approximately match \citealt{Diener:95}). 

This time, we note a decrease in $\Delta E_{\rm mode}$ with eccentricity.  This is because the tidal interaction is significant only for a fraction of $2\pi/(\dot{\Phi}_p P_{\rm orb})\propto \sqrt{(1-e)^3/(1+e)}$ of the orbit. Although the mode becomes instantaneously unstable near the pericenter as long as $D_p/R_t$ is below the prediction of Fig. \ref{fig:Dp_th_eq_tide}, the time for the instability to grow can be too short compared to the instability growth timescale, thereby preventing a runaway that could lead to mass loss. The threshold $D_p/R_t$ can differ by order unity when comparing Fig. \ref{fig:e_Dp_map_num} and \ref{fig:e_Dp_map_num_one_pass} due to the dynamical tide. Using the linear damping calculation of \citet{Goldreich:77} with a quality factor of $\mathcal{O}(10^{13})$, it requires $1-e \lesssim 10^{-8}$ to suppress the dynamical tide within a single orbit. Even if the quality factor is reduced to $10^{6}$ \citep{Wu:05}, damping is significant only when $1-e < 10^{-3}$. For systems less eccentric, Fig. \ref{fig:e_Dp_map_num} should be a more accurate representation of the Roche limit. 
\hang{As a caveat, the discussion considers only linear damping. \citet{Kumar:96} suggested that nonlinear damping can be highly efficient with a quality factor of $\mathcal{O}(100)$. However, their calculation focused on main sequence stars with masses $0.5-0.75\, M_\odot$ and was based on order of magnitude estimations. Whether nonlinear damping can be as strong as claimed in \citet{Kumar:96} for donors of different types remain to be examined in details by future studies. }

\section{Applications}
\label{sec:app}
After describing the general theoretical results in \S \ref{sec:nl_hydro}, we now describe the applications of the model in various astrophysical settings. 

\subsection{Relation to the diffusive tide}
\label{sec:app_chaotic_tide}

It is interesting to note that the numerical mode evolution studied in \S \ref{sec:instab_dyn_tide_w_num} largely resembles the diffusive (or chaotic) tide studied in, e.g., \citet{Mardling:95, Ivanov:04, Ivanov:07, Wu:18, Vick:18, Yu:21, Yu:22a, Lau:25}. 
In fact, our work presents a new mechanism to trigger the diffusive tide. 
In all previous analyses, the diffusive behavior ultimately originates from a random phase accumulation away from the pericenter. Such a random phase could be produced by fluctuations either in the orbital periods due to tidal back-reactions \citep{Mardling:95, Ivanov:04, Wu:18, Vick:18} or gravitational-wave radiation \citep{Ivanov:07, Lau:25}, or in the mode frequencies due to nonlinear anharmonicities \citep{Yu:21, Yu:22a}. A limitation of this mechanism is that the diffusive evolution can hardly be maintained once the orbital eccentricity drops below $e\simeq 0.9$ and the orbital period is not long enough for the random phase to accumulate. For example, for $M'=10^3$, $e=0.7$, and $D_p/R_t=2.645$, we have $P_{\rm orb} = 164.5$ (deliberately chosen to be far away from linear resonance) and $E_{\rm orb} = - MM'(1-e)/(2D_p)=-5.7$ (all in natural units). After a single passage, the mode acquires an energy $\Delta E_{\rm mode}\simeq 1.7 \times10^{-3}$ for the $(n, \Gamma)=(1, 2)$ model. The one-pass perturbation of the mode phase from the tidal back-reaction is $2 \Delta \phi_q = \omega_f \Delta P_{\rm orb}\simeq (-3/2) \omega_f P_{\rm orb} (\Delta E_{\rm mode}/E_{\rm orb})\simeq 0.091\,{\rm rad}$, much smaller than the threshold value of $2\Delta \phi_q \gtrsim 1\,{\rm rad}$ derived in \citet{Vick:18}. Moreover, back-reaction is turned off in our simulations. The anharmonicity-induced phase perturbation differs from the back-reaction phase by a factor of $\mathcal{O}(|E_{\rm orb}|)$ and typically has a greater numerical value \citep{Yu:21}, yet $2\Delta \phi_q = P_{\rm orb}\Delta \omega_f \sim \omega_f P_{\rm orb}\Delta E_{\rm mode} \simeq 0.34\,{\rm rad} $ is still smaller than unity to trigger the diffusive growth. Nonetheless, our Fig. \ref{fig:e_Dp_map_num} shows that this system indeed grows diffusively and eventually runs away. 

Crucially, all studies referenced above used the linear theory of \citet{Press:77, Lai:97} to compute the tidal kick each mode receives at the pericenter. In this work, we incorporate nonlinear hydrodynamics at each pericenter passage, which may allow the mode to grow momentarily due to instability. If the mode's phase at pericenter is periodic, then the growth will repeat after a few orbits and accumulate. This allows the mode to reach an amplitude large enough to trigger the diffusive tide. The growth then continues stochastically until the mode eventually runs away. 

We emphasize that the instability at the pericenter enters at the three-wave order (as we have demonstrated in previous sections both theoretically and numerically), a lower-order effect compared to the nonlinear anharmonicity of \citet{Yu:21}, which fundamentally enters at the four-wave order\footnote{While \citet{Yu:21} also considered three-wave interaction between f- and p-modes, the p-modes are not directly sourced by the tide. Therefore, to produce the anharmonicity, it requires nonlinearly sourcing the p-modes first and then letting the p-mode back-react on the f-mode; see Eq. (\ref{eq:pg_toy}). Since the three-wave coupling is used twice, it effectively becomes a four-wave interaction. The direct four-wave interaction was incorrectly ignored in \citet{Yu:21}. There is also a numerical error in \citet{Yu:21} that overestimates the coupling strength between f- and p-modes.}.
It is also distinct from the nonlinear kick of \citet{Arras:23}. Importantly, the correction to the tidal kick in this work is nonlinear in the fluid displacement $\vxi$, which enters the modes' equations of motion at $\mathcal{O}(|q_m|^2)\sim \mathcal{O}(M'^2/D_p^6)$; see, e.g., Eq. (\ref{eq:K2_3m}). In \citet{Arras:23}, however, the correction is nonlinear in $(\Delta D_p/D_p)$, where $\Delta D_p$ is the change in the pericenter separation due to tidal back-reaction. In other words, the tidal kick in \citet{Arras:23} is evaluated along a tidally perturbed orbit instead of the background Keplerian one. While ignored in this study, one can show that $(\Delta D_p/D_p) \propto (D_p/M')|q_m|^2\sim (M'/D_p^5)$ (see eq. 13 of \citealt{Arras:23}; note that the second term is smaller than the first one by $\mathcal{O}[(R_t/D_p)^{3/2}]$). When evaluating $\epsilon$ at $(D_p + \Delta D_p)$ instead of $D_p$, the correction to the tidal forcing is $\Delta \epsilon\propto (M'/D_p^3)(\Delta D_p/D_p) \sim (M'^2/D_p^8)$, which is smaller than the nonlinear-in-$\vxi$ correction by $(1/D_p^2)$. 
Since the new correction considered here is of lower order than those studied previously, the diffusive (chaotic) tide it leads to may occur over an expanded parameter space (especially for low eccentricities) than predicted by linear tidal kicks at the pericenter.

\begin{figure}
    \centering
    \includegraphics[width=0.5\linewidth]{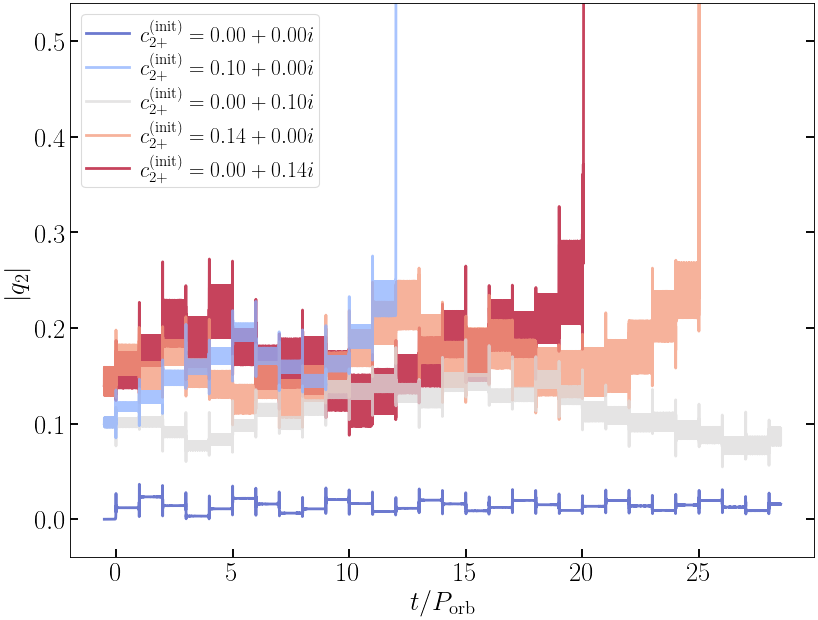}
    \caption{Numerical evolution of the $q_2$ mode for a non-rotating, $n=1$ donor in an $e=0.9$ orbit with $D_p/R_t=2.9$. Different colors represent different initial amplitudes given to the mode. A large initial amplitude can be the consequence of the dynamical tide's diffusive, long-term evolution, and it allows runaways in mode amplitude to occur at $D_p/R_t$ greater than Fig. \ref{fig:e_Dp_map_num}. }
    \label{fig:num_q2_vs_init}
\end{figure}

In turn, for highly eccentric systems, the diffusive tide can make the threshold pericenter separation greater than the prediction of Fig. \ref{fig:e_Dp_map_num}. Note that in Fig. \ref{fig:e_Dp_map_num}, the modes are initiated by their leading-order equilibrium tide solution at the apocenter. For a fixed $D_p$, a large eccentricity essentially suppresses the initial amplitude by a factor of $[(1+e)/(1-e)]^3$. Over merely 22 orbits, the diffusive growth is limited as, on average, the mode energy increases linearly with the number of passages \citep{Wu:18, Vick:18}. One can nonetheless capture the long-term growth history of the modes by giving them different initial conditions. One such example is shown in Fig. \ref{fig:num_q2_vs_init}. We consider in this example the non-rotating $n=1$ model with $e=0.9$ and $D_p/R_t=2.9$. Starting with the equilibrium amplitude, the mode amplitude remains oscillatory within 22 orbits (the bluest curve; also Fig. \ref{fig:e_Dp_map_num}). However, if a significant initial amplitude is used (indicated by curves with different colors), the mode may run away at this large separation. As the diffusive growth is almost unbounded (until reaching near equipartition with the orbit), the threshold $D_p/R_t$ for the mode's runaway can be as large as that for the diffusive tide to occur. 

In fact, the runaways in our simulation correspond to the catastrophic wave breakings during the diffusive evolution postulated in \citet{Wu:18}. When a mode's amplitude runs away, it cannot be supported by the background and has to break somehow. However, our model does not describe the consequence of such a break. 
One possibility, as suggested in \citet{MacLeod:22}, would be shock-heating the surface layer with some potential mass ejection from the donor. 
It further suggests that energy dissipation in highly eccentric systems can be very efficient due to not a continuous dissipation as in the linear theory but discrete epochs of diffusive growth followed by wave breakings as pointed out by \citet{Wu:18}.
Meanwhile, mass loss is likely nonconservative based on \citet{Guillochon:11}. 
More detailed analysis (theoretical or numerical) of the saturation of the instability is left for future investigations.

\subsection{High eccentricity migration of exoplanets}
\label{sec:app_HJs}

Related to specific astrophysical settings, the first application we consider is the formation of HJs via the high-eccentricity migration channel (see, e.g., \citealt{Dawson:18} for a review). In particular, our model provides a theoretical explanation for why \citet{Guillochon:11} observed mass loss from a migrating Jovian planet at a pericenter separation as large as $D_p\simeq 2.7 R_t$, $35\%$ greater than the Roche-lobe-overfilling separation of $D_p\simeq 2.0 R_t$ using the fit by \citet{Eggleton:83}. 
Extrapolating the low-eccentricity Roche lobe calculation of \citet{Sepinsky:07a} will increase the discrepancy even further.  
We highlight that our Fig. \ref{fig:num_q2_e9_n_1} shows remarkable similarities to the simulations of \citet{Guillochon:11}; see, e.g., their fig. 3. Specifically, there should be no significant mass loss from $M$ (the planet) in a single passage. Instead, the perturbation accumulates over orbits, and significant mass loss from $M$ eventually occurs after $\mathcal{O}(10)$ orbits. This consistency validates our analysis.

More importantly, our theoretical model (in terms of simple ordinary differential equations; Eq. \ref{eq:ddq} or \ref{eq:dcdt}) allows the more expensive three-dimensional hydrodynamical simulations of \citet{Guillochon:11} to be faithfully extrapolated to different systems based on each system's unique configuration. 
In particular, the commonly adopted disruption threshold of $2.7 R_t$ (e.g., \citealt{Anderson:16, Vick:19}) is only true for the specific system studied in \citet{Guillochon:11} over $\mathcal{O}(10)$ passages.
More orbits of integration may increase the threshold further due to the diffusive build-up of the mode (captured by a non-equilibrium initial condition shown in Fig. \ref{fig:num_q2_vs_init}). 
\hang{Recent studies by \citet{Yu:24b, Weldon:25} adopted a (potentially more realistic) soft disruption prescription where the planetary mass loss per orbit is parameterized as an exponentially decaying function of $D_p/R_t$, yet the parameters are still calibrated to \citet{Guillochon:11} and held constant throughout the evolution. 
}
If either the orbital eccentricity or the planet's rotational rate varies, a different disruption threshold \hang{(or a different set of mass loss prescriptions)} should apply; see Figs. \ref{fig:e_Dp_map_num} and \ref{fig:num_q2_e9_ns_vs_ps}. The dependence of $D_p^{\rm (th)}$ on these properties may indicate interesting and unanticipated correlations between the current HJ distribution and progenitor configurations. 

Our Fig. \ref{fig:num_q2_vs_init} further suggests that the Roche limit for highly eccentric systems can be as large as the threshold separation to trigger the diffusive tide at about $3.8 R_t$ \citep{Wu:18} for a typical proto-HJ. Interestingly, this distance coincides with the location where a pile-up of HJs is observed (with a current period of about 3 days; \citealt{Wu:18, Dawson:18}). On the other hand, HJ systems with separations shorter than this could imply a high spin rate of the proto-HJ (young system), as the thresholds for both diffusive tide (e.g., fig. 4 of \citealt{Yu:21}) and Roche limit (Fig. \ref{fig:num_q2_e9_ns_vs_ps}) are suppressed by rotation. 

Our Fig. \ref{fig:e_Dp_map_num} may also have implications for formation channels other than the standard high-eccentricity migration. For example, a planet may have disk-migrated before experiencing a high-eccentricity migration. In such a hybrid scenario, the eccentricity excitation required in the second stage is only moderate, with $e<0.9$ (e.g., figs. 5 and 10 of \citealt{Yu:24b}). 
This two-stage migration is suggested by \citet{Guillochon:11} based on observations available at the time and supported by the recent investigation of \citet{Yu:24b} on the formation of the WASP-107 system (their section 4.3). Both studies found that the initial apocenter distance of the planet of the high-eccentricity migration stage is within the snow line. 
Our Fig. \ref{fig:e_Dp_map_num} suggests that the HJ/HN formation efficiency for such a system with $e<0.9$ may be higher than originally expected based on a fixed $D_p^{(\rm th)}=2.7 R_t$ derived from $e=0.9$. An increase in the predicted HJ formation rate will improve the agreement between theory and observation (see, e.g., table 2 of \citealt{Dawson:18}). 

As discussed above in \S \ref{sec:app_chaotic_tide}, the nonlinear tidal interactions leading to the Roche limit can also help trigger and maintain the diffusive tide. For highly eccentric systems with $e>0.9$, an expanded parameter space for the diffusive tide is crucial for explaining the lack of super-eccentric proto-HJs \cite{Socrates:12, Dawson:15}. The efficient energy dissipation due to the diffusive tide naturally explains the paucity, yet the original study of \citet{Wu:18} applies only to those with current periods less than 3 days, a small fraction of the missing population \citep{Dawson:18}. Expanding the threshold for the diffusive tide is therefore crucial for reducing the tension with observations. 
For those with $e<0.9$, the propagation phase of a mode itself is not sufficiently random, and the nonlinear instability at the pericenter is likely what maintains the diffusive tide and drives the orbital evolution. It is important to note that there is no consensus on the dissipation mechanism in moderately eccentric systems after the quench of the diffusive tide. The predicted dissipation rate of f-modes from first-principle calculations \citep{Goldreich:77} is orders of magnitude below what is needed for a planet to circularize without the diffusive tide (but see, e.g., \citealt{Wu:05, Ogilvie:14}).  

We plan to follow up on the points discussed above with a population synthesis study utilizing a disruption threshold described in this work instead of fixing it to $D_p^{(\rm th)}=2.7 R_t$, to explore new features and correlations in the current distribution.

\subsection{rpTDEs and QPEs}
\label{sec:app_TDEs}

Repeating X-ray bursts at galactic centers provide valuable information about the properties of the central MBH and its surrounding environments. A rpTDE corresponds to a star partially disrupted by the tidal force of the MBH, with the remnant continuing to orbit and getting disrupted again in the future. Compared to a full TDE, rpTDEs have a larger $D_p/R_t$, and a different mass fallback rate \cite{Coughlin:2019}. 
Through a hydrodynamic simulation, \cite{Ryu:2020} finds that a main sequence star undergoes a partially disrupting instead of a full TDE for $2 \gtrsim D_p/R_t \gtrsim 0.5$ for a single passage in the Newtonian limit. \cite{Rosswog:2009} performs a simulation of the tidal disruption of a WD orbiting a $10^3~M_\odot$ intermediate mass black hole, where mass loss occurs at $1.1 R_t$ in a single passage. 
Unlike full TDEs that are typically assumed to happen on nearly parabolic orbits, rpTDEs are likely to have more moderate \hang{but still significant} eccentricities with $1-e\sim 0.1-0.01$ \citep{Hayasaki:13, Guillochon:13, Cufari:22, Liu:24, Bandopadhyay:24, Bandopadhyay:25}. 
\hang{The long orbital period limits previous numerical simulations to binaries with $D_p/R_t\lesssim 2$ in order for mass loss to occur in \emph{a single pericenter passage.}}
However, the orbital period is not long enough to damp out the dynamical tide (unless the quality factor is significantly lower than $10^6$ in which case the study of \citealt{Yao:25} applies; see the discussions at the end of \S \ref{sec:instab_dyn_tide_w_num}), allowing it to build up over multiple orbits and potentially change the single-passage predication significantly. 
Similar to the case with Jovian planets, our study provides a \hang{new}, theoretical model to describe the disruption, accounting for the accumulation history of the dynamical tide over \emph{multiple orbits.} \hang{In our picture,} the dynamical tide builds up diffusively until it runs away and ejects mass to power a burst. The stochastic growth can be a promising mechanism to produce rpTDEs with irregular recurrence times \citep{Lau:25, Lau:26}, which has been observed in, e.g., J0456-20 \citep{Liu:24} and HLX-1 \citep{Godet:14, Webb:23} \hang{that previous models requiring mass loss at every pericenter passage struggle to reproduce. For example, the simulations in \citet{Liu:24} require a fractional mass loss of almost 90\% to reproduce the $\sim30\%$ fluctuations in the recurrence times of J0456-20 from $\sim 300 \,{\rm days}$ to $\sim 200\,{\rm days}$. Furthermore, this amount of mass loss is also in tension with the inferred eruption luminosity.}

Moreover, the threshold pericenter separation $D_p^{(\rm th)}$ for the disruption to marginally occur may increase by $\gtrsim 60\%$ with the buildup \hang{compared to the single-passage predictions (e.g., \citealt{Guillochon:13, Ryu:2020}};  comparing Figs. \ref{fig:e_Dp_map_num} and \ref{fig:e_Dp_map_num_one_pass} at $e=0.9$). Since $D_p^{\rm (th)}$ sets the loss cone radius, an increase in $D_p^{(\rm th)}$ increases the rpTDE event rate \citep{Bortolas:23}. If we assume an isotropic distribution of a star's velocity vector at a fixed distance for simplicity, the increase in loss cone event rates is proportional to the increase in its radius \citep{Merritt:13}. 

Our results also complement some previous models on QPEs.
In \cite{King:22, Wang:22}, it is proposed that a highly eccentric WD-MBH binary can simultaneously produce TDEs and QPEs. The predicted tidal stripping based on Eggleton's formula is strictly periodic with $D_p < 2 R_t$. The QPEs produced via this mechanism would therefore be highly regular, which contradicts the observed fluctuations in the recurrence time of bursts. The stochastic nature of the mode provides a natural explanation for these fluctuations. The diffusive evolution also means that the tidal disruption does not repeat every orbit, as the mean mode energy builds up diffusively over time. Its corresponding impact on the orbital evolution and mass transfer has been studied in \cite{Lau:25}. A significant shortcoming within this work is that although the parametrized mass loss formalism is flexible enough to model the case with dynamical tide, the range of the parameters is prescribed to match the observations and previous hydrodynamic simulation results in \cite{Guillochon:11}. In other words, it still lacks a direct connection to the first-principles calculation considering the hydrodynamics under an external tidal field.
This paper provides a theoretical description of the development from mode oscillations into a runaway process that can potentially lead to mass loss.

For a deep encounter ($D_p/R_t < 0.5$) of a WD with the MBH, a thermonuclear runaway can be triggered by the tidal compression \cite{Luminet:89, Rosswog:09, MacLeod:16, Tanikawa:18}, depending on its composition. Based on the model by \citet{Carter:83}, the star undergoes a phase in which it becomes flattened in the orbital plane and the interior is strongly compressed. \cite{Luminet:89} shows that it can trigger nucleosynthesis processes similar to that at the core of a main-sequence star. The energy release can be comparable to that of Type 1a supernovae \cite{Rosswog:09}. This scenario is favored for lower mass MBHs, of $\sim 10^{3} M_\odot$, where the WD orbit can penetrate deeper without getting directly swallowed. Whether a WD ignites or loses mass and powers rpTDE depends on how $\Delta E_{\rm mode}$ from a runaway is released. If $\Delta E_{\rm mode}$ is mostly used to eject mass, then rpTDE is the outcome. If, however, it is mostly converted into heat in the core, then the WD may be ignited in a thermonuclear explosion. Determining the outcome is beyond the scope of the current model, but an interesting question to follow up on in the future.

\section{Conclusion and discussion}
\label{sec:conclusion}

In this work, we studied the nonlinear hydrodynamical stability of a donor $M$ in a compact orbit around a point-particle companion $M'$. 
Our study considers both the exact nonlinear equations derived from the affine model \citep{Diener:95}, which self-consistently apply to homogeneous, incompressible ellipsoids, and the expanded equations following \citet{VanHoolst:94}, which can be readily extended to realistic stars and planets. Our study extends the classic Roche limit derived by \citet{Chandrasekhar:63} of a donor in hydrostatic equilibrium with circular orbit and synchronous rotation to the fully dynamical regime, allowing for arbitrary orbital eccentricities and donor rotation rates (though restricted to aligned spins). 
Schematically, the finite-frequency corrections to the equilibrium tide (accurate for low eccentricities; \S \ref{sec:instab_eq_tide}) can be captured by the toy model in Eq. (\ref{eq:toy_stability_dyn}). The dynamical tide excited in more eccentric systems (\S \ref{sec:instab_dyn_tide_w_num}) corresponds to the toy model of Eq. (\ref{eq:toy_stability_dyn2}).
More specifically, the Roche limit for moderately eccentric systems ($e\lesssim 0.3$) is described by our Fig. \ref{fig:Dp_th_eq_tide}. While a small asynchronicity tends to decrease $D_p^{(\rm th)}/R_t$ (i.e., making the donor more tightly bound) by reducing the centrifugal force, eccentricity can dominate and make $D_p^{(\rm th)}/R_t$ greater (donor more subject to mass loss) than the prediction of \citet{Chandrasekhar:63} by the finite-frequency corrections. 
For a more eccentric binary with $e \sim 0.9$,  the fluid perturbation can grow diffusively due to nonlinear interactions at the pericenter. This eventually leads to a runaway in mode amplitude at a pericenter separation $\gtrsim 30\%$ greater than the Roche limit of circular orbits with synchronous rotations (Figs. \ref{fig:e_Dp_map_num} vs. \ref{fig:config_stat_tide}). Our theoretical result explains the numerical simulation of \citet{Guillochon:11} on the high-eccentricity migration of gaseous planets, and allows the result to be physically extrapolated across possible parameter space and to other similar systems, e.g., rpTDEs. This study also provides a new mechanism to trigger and maintain the diffusive (chaotic) tide that is of lower order in $(1/D_p)$ than all previously considered mechanisms (\S \ref{sec:app_chaotic_tide}). 

We mainly focused on a homogeneous, incompressible ellipsoid to assist a direct comparison with \citet{Chandrasekhar:63}. This simple model is also sufficient to illustrate the crucial role of dynamical effects in the Roche problem and its corrections to the commonly used fit of \citet{Eggleton:83}. We established that predictions based on the exact nonlinear equations of an affine ellipsoid can be largely reproduced by the expanded equations to the three-wave order, or $\mathcal{O}(\xi^3)$ in $\Delta E_{\rm mode}$. Including the next order improves the accuracy, but does not modify the qualitative features (Fig. \ref{fig:config_stat_tide}). 
While the affine model of \citet{Diener:95} allows for different values of $n$ and $\Gamma$, the starting point, Eq. (\ref{eq:affine_def}), is not self-consistently satisfied by a general fluid. 
Nonetheless, the expanded equations of \citet{VanHoolst:94} will apply generically; see also \citet{Weinberg:12, Weinberg:16}. This allows the theoretical study of the Roche limit to be extended to realistic astrophysical objects. It also allows other modes, such as g-modes (present if $\Gamma > 1+1/n$), or those with $l\geq 3$ (important for binaries with $M'\sim 1$) to be included in the problem. Based on the affine model, an increasing $n$ reduces the single-passage $D_p^{(\rm th)}/R_t$ \citep{Diener:95}. It would be interesting to revisit this prediction quantitatively with generic shapes of $\vxi$ (non-affine) under \citep{VanHoolst:94}  and consider the multi-orbit effects. 

In fact, the set of equations required to derive the Roche limit has been used in \citet{Yu:23a} to study the frequency shift of the f-mode due to the companion's tidal field in coalescing binary neutron stars (see also \citealt{Pitre:25} for its relativistic extension). The two effects share the same physical origin. 
\hang{In particular, our analysis also predicts an effective frequency shift of the $m=2$ f-mode given by (cf. Eq. \ref{eq:K2_3m} but allowing for arbitrary $n$ using results from the Appendix \ref{appx:eom_affine})
\begin{align}
    \frac{\Delta (\omega_f^2)}{\omega_f^2} = [\frac{95\sqrt{6}}{84}q_0 -\frac{5}{8} (5-n) \Ma ] \simeq \frac{65}{28}(5-n) \Ma \epsilon,
\end{align}
where the linear, static limit of $q_0\simeq -\sqrt{6}(5-n)\Ma/4$ is used. This agrees with eq. 16 of \citet{Yu:23a} in both the functional form and the numerical value, validating the affine approximation. 
}

The analysis of \citet{Yu:23a} also provides further support for why we can truncate the analysis at the three-wave order for the Roche limit. Specifically, in this work, the interacting $m=2$ and $m=0$ modes are both directly linearly driven by the tide, allowing them to directly interact at the three-wave order. This is of the same order as the frequency shift of \citet{Yu:23a} and lower order than the anharmonicity and pg-instability \citep{Weinberg:13, Weinberg:16, Venumadhav:14}, which fundamentally originate from four-wave interaction ($\mathcal{O}(\xi^3)$ in the equation of motion). In the pg-instability, the p-mode is not linearly forced by the tide, but instead nonlinearly driven through the three-wave channel by an f-mode and a g-mode, schematically written as $q_p\simeq \kappa q_f q_g$ with the subscript indicating the mode (see appendix A of \citealt{Weinberg:13}).  The p-mode then back-reacts on the g-mode, as 
\begin{align}
    &\ddot{q}_g + \omega_g^2 q_g = \omega_g^2\kappa q_f^\ast q_p^\ast = \omega_g^2\kappa^2 q_f^\ast q_f q_g, \nonumber \\
    \text{ or } &
    \ddot{q}_g + \omega_g^2 (1- \kappa q_f^\ast q_f) q_g = \ddot{q}_g + \omega_{g, {\rm eff}}^2 q_g=0,
    \label{eq:pg_toy}
\end{align}
causing the instability when $\omega_{g, {\rm eff}}^2<0$ or
$\kappa q_fq_f^\ast > 1$. Because the three-wave interaction is used twice with the p-mode acting only as a mediator, the pg-three-wave interaction comes in the same order as a direct four-wave interaction between two f-modes and two g-modes, corresponding to a forcing term of the form $\zeta q_f^\ast q_f q_g$ in the equation above. See also similar discussions about the nonlinear anharmonicity studied in \citet{Yu:21} and
\citet{Kwon:24, Kwon:25} caused by effective four-wave interactions. 
In this work, however, the three-wave interaction is only used once because the $m=0$ mode has a linearly forced component ($q_0\sim q_2$ here instead of $q_0\sim \epsilon q_2$ as in the pg-instability), making it a lower-order effect than the anharmonicity. This can also be seen from the left panel of fig. B1 in \citet{Yu:23a}; compare specifically the gray (equivalent to the instability of this work) and the yellow (equivalent to the pg-instability/anharmonicity) curves. This theoretical argument is reflected in Fig. \ref{fig:config_stat_tide}, and is numerically justified by Figs. \ref{fig:num_q2_e3_ns} and \ref{fig:num_q2_e9_n_1}. Including the four-wave interaction (Eqs. \ref{eq:K2_4m} and \ref{eq:K0_4m}) modifies the threshold $D_p^{\rm (th)}/R_t$ by only a few percent (Fig. \ref{fig:config_stat_tide}). 

Similar to \citet{Chandrasekhar:63}, our study does not directly address the evolution of a fluid element after a mode's amplitude runs away. Two crucial questions remain to be addressed. The first is how much mass is removed from $M$ and at what rate. Whereas in circular or low-eccentricity cases, a steady stream of mass flow from the donor to the accretor may form \citep{Shu:81}, in the highly eccentric cases, the mass loss should occur as discrete bursts whenever the wave breaks (after the mode amplitude runs away). An upper bound on the mass loss $\Delta M$ can be set based on the energy released, $\Delta E_{\rm mode}\sim 0.1$. If most of the energy is converted to heat \citep{MacLeod:22}, then $\Delta M$ from the wave breaking may be much smaller than $\Delta E_{\rm mode}$. Nonetheless, a large amount of heat may inflate the donor, making it more likely to lose mass in the subsequent evolution, which quickly disrupts the donor \citep{Yu:24b}. 
On the other hand, if the donor's density increases as it loses mass, it may survive the mass loss \citep{Liu:13}. One such example could be planets in the HN desert \citep{Armstrong:20}. The second question is how a mass element lost from the donor affects the orbit. The numerical simulation of \citet{Guillochon:11} suggests that the total mass and angular momentum of the system are likely to be non-conserved when the mass loss happens with a highly eccentric orbit. One may study the effects on the orbital dynamics following, e.g., \citet{Huang:63, Sepinsky:09, Hamers:19, Lau:25}. Alternatively, since the exterior potential of a perturbed, oscillatory star is simply $\propto Y_{2m}/D^3$, it is straightforward to find a test particle's trajectory under a Roche potential modified by the quadrupolar potential of the donor (and potentially the accretor). The trajectory will determine the fate of the ejected mass and its impact on the orbit. Of particular interest are the changes in the binary's total mass and orbital angular momentum.  
We plan to study these in follow-up investigations.

\begin{acknowledgments}
We thank \hang{Eric Coughlin, Yanqin Wu, Brad Hansen, Smandar Naoz, Grant Weldon, } Fei Dai and Zhen Pan for useful discussions during the preparation of this manuscript. 
This work is supported by Montana NASA EPSCoR Research Infrastructure Development under award No. 80NSSC22M0042 and NSF grant No. PHY-2308415. 
\end{acknowledgments}

\software{\texttt{NumPy} \citep{Numpy},  
          \texttt{SciPy} \citep{Scipy}, 
          \texttt{Matplotlib} \citep{Matplotlib},
          \texttt{Numba} \citep{Numba},
          \texttt{Mathematica} \citep{Mathematica}.
          }


\appendix

\section{Equations of motion of the affine model}
\label{appx:eom_affine}

Write
\begin{equation}
    \vect{r}=\vect{x} + \vxi(\vect{x}, t)=\vect{x} + \sum_mq_m(t)\vxi_m(\vect{x}),
\end{equation}
where $\vect{x}$ and $\vect{r}$ are the original and perturbed locations of a mass element. 
The Lagrangian displacement is denoted by $\vect{\xi}$, which is further decomposed into eigenmodes with eigenfunctions $\vxi_m$ according to Eqs. (\ref{eq:vxi_expand}) and (\ref{eq:vxi}). As argued in the main text, inertial modes and $m=\pm 1$ f-modes are both ignored in this work. 
Because $\xi_m$ is linear in the original position, perturbations of a homogeneous ellipsoid are therefore affine as in \citet{Carter:83, Carter:85, Diener:95}. We additionally introduce a trace term $\vxi_{\rm tr}=q_{\rm tr}\vect{x}$, corresponding to a monopolar (radial with $l=0$) perturbation. While a homogeneous ellipsoid has no radial modes in the linear problem, we find $q_{\rm tr}$ is needed to account for rotation and nonlinear interactions starting at the four-wave order. Note that the trace ($l=0$) mode is normalized differently from the $l=2$ ones; $q_{\rm tr}=1$ corresponds to a surface displacement of unity.  
This allows us to rewrite
\begin{align}
    \begin{bmatrix}
    r_1\\
    r_2\\
    r_3
    \end{bmatrix} = 
    \begin{bmatrix}
        1 + \frac{5}{12}[3 (q_2 + q_{-2}) -\sqrt{6}q_0] + q_{\rm tr}, 
        & i\frac{5}{4}(q_2-q_{-2}), &0 \\
        i\frac{5}{4}(q_2-q_{-2}), &
        1 - \frac{5}{12}[3 (q_2 + q_{-2}) -\sqrt{6}q_0] + q_{\rm tr}, &0 \\
        0, &0, & 1 + \frac{5}{\sqrt{6}}q_0 + q_{\rm tr}
    \end{bmatrix}
    \begin{bmatrix}
        x_1\\
        x_2\\
        x_3
    \end{bmatrix},
    \label{eq:r_vs_x}
\end{align}
which satisfies the form of an affine model $r_i(t, \vect{x})=q_{ij}(t) x_j$, with Einstein summation assumed and the indices standing for coordinates in Cartesian. 
We further use $\vect{q}$ to denote the matrix whose components are $q_{ij}$. The determinant of the matrix will be useful later and is explicitly given
\begin{align}
    \det[\vect{q}] = 1 + 3 q_{\rm tr} - \left(\frac{25}{4}q_2q_{-2} + \frac{25}{8}q_0^2 - 3 q_{\rm tr}^2\right) 
    - \left(\frac{125\sqrt{6}}{24}q_2q_{-2}q_0 - \frac{125\sqrt{6} }{144} q_0^3 + \frac{25}{4}q_2q_{-2}q_{\rm tr} +  \frac{25}{8}q_0^2q_{\rm tr} - q_{\rm tr}^3\right).
    \label{eq:det_qij}
\end{align}

Note that $\vect{q}$ is diagonalized as 
\begin{align}
    \vect{q} = \vect{R}\vect{a}\vect{R}^T=
    \begin{bmatrix}
    \cos \phi_q, & -\sin\phi_q, &0\\
    \sin\phi_q, & \cos\phi_q, & 0\\
    0, & 0, & 1
    \end{bmatrix}
    \begin{bmatrix}
    a_1& 0&0\\
    0 & a_2 & 0\\
    0, & 0, & a_3
    \end{bmatrix}
    \begin{bmatrix}
    \cos \phi_q, & \sin\phi_q, &0\\
    -\sin\phi_q, & \cos\phi_q, & 0\\
    0, & 0, & 1
    \end{bmatrix}
\end{align}
where the angle $m\phi_q$ describes the phase of the mode $q_m$ and 
\begin{subequations}
\begin{align}
    a_1 &= 1 + \frac{5}{2}|q_2| - \frac{5\sqrt{6}}{12} q_0 + q_{\rm tr}, \\
    a_2 &= 1 - \frac{5}{2}|q_2| - \frac{5\sqrt{6}}{12} q_0 + q_{\rm tr}, \\
    a_3 &= 1 + \frac{5\sqrt{6}}{6} q_0 + q_{\rm tr}. 
\end{align}
\label{eq:a123_full}
\end{subequations}
It shows the affine model describes an ellipsoid with semiaxes $a_1$, $a_2$, $a_3$. 
Note that $a_1>a_2$ and $a_1>a_3$. When the dynamical tide dominates, we further have $a_2\geq a_3$ in general, though it is also possible to have $a_2<a_3$ occasionally. 
We further use an upper bar to denote quantities evaluated in a frame co-rotating with $q_2$, e.g., displacement $\bar{\vect{r}}=\vect{R} \vect{r}$ and quadrupole moment 
\begin{equation}
    \bar{I}_{ij} = \int \bar{r}_i \bar{r}_j dM = \frac{a_i^2 }{5} \delta_{ij} \text{ (no summation)}.
    \label{eq:Ibar_ij}
\end{equation}

The equation of motion is obtained from variations of the energies. While the affine model is self-consistently satisfied only for an incompressible, homogeneous fluid with a polytropic index $n=0$ and adiabatic index $\Gamma=\infty$, we nonetheless follow \citep{Diener:95} (see also \citealt{Carter:83, Carter:85}) and allow the energies to be described by a generic pair of $(n, \Gamma)$, without requiring buoyancy neutral (i.e., $\Gamma$ can differ from $(1+1/n)$), while still assuming the perturbed fluid is described by Eq. (\ref{eq:r_vs_x}). The error due to this inconsistency will be assessed in future studies through direct comparisons with \citet{VanHoolst:94}. 

The kinetic energy can be computed from 
\begin{equation}
    T = {\rm Tr}[T_{ij}] = {\rm Tr}[ \frac{1}{2} \int \dot{r}_i\dot{r}_j dM ] = \frac{1}{2} \dot{q}_{ik} \dot{q}_{jk} \Ma = \frac{\Ma}{8}(50 \dot{q}_2\dot{q}_{-2} + 25\dot{q}_0^2 + 12\dot{q}_{\rm tr}^2),
    \label{eq:Delta_T}
\end{equation}
where $\Ma=\int x_ix_i dM/3$. Numerically, $\Ma=1/5$ for $n=0$ and $\Ma=0.1306910$ for $n=1$. 


\hang{The gas energy is $U=\int (\epsilon/\rho) dM$, where $\epsilon$ and $\rho$ are respectively the internal energy density and mass density. We assume that the gas follows a polytropic equation of state, where 
\begin{equation}
    P = (\Gamma-1) \epsilon \propto \rho^{\Gamma},
\end{equation}
with $P$ the pressure and $\Gamma=(\partial \log P/\partial \log \rho)_{\rm entropy}$ is the adiabatic index. Using further $\rho (\vect{r}) = \rho(\vect{x})/\det[q]$, we have 
\begin{align}
    \Pi = \int \frac{P}{\rho} dM = \frac{(\det[q])^{1-\Gamma}}{5-n}, \text{ and } U=\frac{\Pi}{\Gamma-1}. 
    \label{eq:Delta_U}
\end{align}
}
The explicit form of $U$ is readily obtained by plugging in $\det[\vect{q}] $ from Eq. (\ref{eq:det_qij}), and its derivative with respect to $q_{ij}$ is $\partial U/\partial q_{ij} = -\Pi (\vect{q}^{-1})_{ji}$. This will be used to derive the equation of motion in terms of $q_{ij}$. 

To assist comparisons with \citet{VanHoolst:94}, we also provide a series expansion in terms of $q_m$'s, as
\hang{
\begin{align}
    U&\simeq \frac{1}{(5-n)(\Gamma-1)}-\frac{3}{5-n} q_{\rm tr}
    +\frac{25 q_0^2+50 q_2 q_{-2}+12 (3 \Gamma -2) q_{\rm tr}^2}{8 (5-n)} \nonumber \\
    &\quad -\frac{125 \sqrt{6} q_0^3-750 \sqrt{6} q_0 q_2 q_{-2}+450 (3 \Gamma -1) q_0^2 q_{\rm tr} + 36 (3\Gamma-1)[25 q_2q_{-2} + 2 q_{\rm tr}^2(3\Gamma-2)]q_{\rm tr}}{144(5-n)} +...
\end{align}
}
We have verified that the series matches exactly eq. (19) of \citet{VanHoolst:94} for $n=0$, in which case $\vect{\xi}_m$ are given by Eq. (\ref{eq:vxi}). 

We then consider the self-gravity. For this, we introduce the self-gravitational potential energy tensor
\begin{equation}
    \Omega_{ij} = -\frac{1}{2}\int\int\frac{(r_i-r'_i)(r_j-r_j') }{|\vect{r}-\vect{r}'|^3} dM dM'.
\end{equation}
From the definition, it is easy to see that the total gravitational binding energy $\Omega$ is given by the trace of $\Omega_{ij}$, $\Omega = {\rm Tr}[\Omega_{ij}]$. 
Because the equations of motion depends on $\partial \Omega/\partial q_{ij}= - \Omega_{ik}(\vect{q}^{-1})_{jk}$, we compute the full $\Omega_{ij}$ matrix here. 
While \citet{Carter:83} gave a formula for $\Omega_{ij}$ in their eq. (6.15), it involves solving a complicated integral in their eq. (6.4) whose explicit solution is non-trivial to obtain. We instead follow \citet{Chandrasekhar:62} to obtain a closed-form solution. For this, it is convenient to first consider the potentials in the barred frame obtained by rotating the system azimuthally by an angle $\phi_{q}$ via the $\vect{R}$ matrix. In this system, the potential energy tensor is diagonal and given by 
\begin{equation}
    \bar{\Omega}_{ij} = -\frac{1}{2}\int\int\frac{(\bar{r}_i-\bar{r}'_i)(\bar{r}_j-\bar{r}_j') }{|\vect{r}-\vect{r}'|^3} dM dM' = -\frac{3}{2} \frac{5}{5-n} A_i \bar{I}_{ii} \delta_{ij} \text{ (no summation)},
    \label{eq:Omegabar_ij}
\end{equation}
where $\bar{I}_{ij}$ are still given by Eq. (\ref{eq:Ibar_ij}), and we explicitly include the $5/(5-n)$ factor to generalize the result to arbitrary values of $n$ (with the caveat that for $n>0$, $\vect{r}$ is not strictly affine). The one-index symbols $A_i$ are defined in eq. (5) of \citet{Chandrasekhar:62} and computed explicitly in terms of $a_i$ in their eqs. (15)-(17). For completeness, we reproduce them below. Define
\begin{equation}
    \sin^2 \theta_e=\frac{a_1^2 - a_3^2}{a_1^2-a_2^2}, \text{ and } \cos\phi_e=\frac{a_2}{a_1}, 
\end{equation}
and denote incomplete elliptic integrals of the first and second kinds, respectively, as 
\begin{align*}
    &F_e=F(\phi_e, m=\sin^2\theta_e) = \int^{\phi_e}_0 (1-m \sin^2 \varphi)^{-1/2} d\varphi, \\
    &E_e=E(\phi_e, m=\sin^2\theta_e) = \int^{\phi_e}_0 (1-m \sin^2 \varphi)^{+1/2} d\varphi,
\end{align*}
we can then write
\begin{subequations}
\begin{align}
    A_1 &= \frac{2}{a_1^3 \sin^3\phi_e} \frac{1}{\sin^2\theta_e}\left(F_e - E_e\right),  \nonumber \\
    &= \frac{2}{3} -2 \sqrt{q_2 q_{-2}} + \frac{\sqrt{6}}{3} q_0 - 2q_{\rm tr} + ... \\
    A_2 &= \frac{2}{a_1^3 \sin^3\phi_e} \frac{1}{\cos^2\theta_e}\left(\frac{a_3}{a_2} \sin\phi_e -E_e \right), \nonumber \\
    &= \frac{2}{3} + 2 \sqrt{q_2 q_{-2}} + \frac{\sqrt{6}}{3} q_0 - 2q_{\rm tr} + ...\\
    A_3&=\frac{2}{a_1^3 \sin^3\phi_e} \frac{1}{\sin^2\theta_e\cos^2\theta_e}\left(E_e-F_e\cos\theta_e^2 -\frac{a_2}{a_3}\sin^2\theta_e \sin\phi_e\right), \nonumber \\
    &=\frac{2}{3} - \frac{2\sqrt{6}}{3}q_0 - 2 q_{\rm tr} + ...
\end{align}
\label{eq:A1A2A3}
\end{subequations}
Note that the results are obtained by swapping the subscripts $(2,3)$ in the corresponding equations in \citet{Chandrasekhar:62}. This is because in our coordinate system, $a_1>a_3 >a_2$ most of the time, though occasionally, it is possible to have $a_2>a_3$. In this case, Eq. (\ref{eq:A1A2A3}) still holds as long as one lets $\sin^2 \theta_e>1$ and uses the reciprocal modulus transformation of elliptical integrals to handle $m>1$ \citep{Byrd:54}
\begin{subequations}
\begin{align}
    &F_e(\phi_e, m) = \frac{1}{\sqrt{m}} F_e[\arcsin (\sqrt{m} \sin \phi_e ), \frac{1}{m}], \nonumber \\
    &E_e(\phi_e, m) = \frac{1}{\sqrt{m}}  \left( m E_e[\arcsin (\sqrt{m} \sin \phi_e ), \frac{1}{m}] + (1-m) F_e[\arcsin (\sqrt{m} \sin \phi_e ), \frac{1}{m}] \right) \nonumber
\end{align}
\end{subequations}
In the limiting case where $a_2=a_3$ (achieved in the adiabatic tide limit), we have
\begin{subequations}
\begin{align}
    &A_1 = -\frac{2 \left[\sqrt{1-a_{21}^2}-\log \left(\sqrt{1-a_{21}^2}+1\right)+\log (a_{21})\right]}{a_1^3 \left(1-a_{21}^2\right)^{3/2}}, \\
    &A_2=\frac{\frac{1}{a_{21}^2-a_{21}^4}-\frac{{\rm arctanh}\left(\sqrt{1-a_{21}^2}\right)}{\left(1-a_{21}^2\right)^{3/2}}}{a_1^3}, \\
    &A_3=\frac{\sqrt{1-a_{21}^2}+a_{21}^2 \left(2 \log \left(\frac{a_{21}}{\sqrt{1-a_{21}^2}+1}\right)+{\rm arctanh}\left(\sqrt{1-a_{21}^2}\right)\right)}{a_1^3 a_{21}^2 \left(1-a_{21}^2\right)^{3/2}}
\end{align}
\end{subequations}
where $a_{21}=a_2/a_1$. 

Once $\bar{\Omega}_{ij}$ is obtained, $\Omega_{ij}$ can be readily computed by rotating back to the original, unbarred frame, as 
\begin{equation}
    \Omega_{ij} = R_{ai}\bar{\Omega}_{ab} R_{bj}
\end{equation}
and the total gravitational energy is 
\begin{align}
    \Omega&={\rm Tr}[\Omega_{ij}] = {\rm Tr}[\bar{\Omega}_{ij}]
    =-\frac{3}{10} \frac{5}{5-n}\sum_i A_i a_i^2 \nonumber \\
    &\simeq \frac{5}{5-n}\left(
    -\frac{3}{5} + \frac{3}{5} q_{\rm tr} - \frac{q_0^2}{8} - \frac{q_2q_{-2}}{4} - \frac{3}{5}q_{\rm tr}^2 \right. \nonumber \\
    &\quad\quad \left.
    -\frac{5 q_0^3}{56 \sqrt{6}}+\frac{5}{28} \sqrt{\frac{3}{2}} q_0 q_2q_{-2}+\frac{3 q_0^2 q_{\rm tr}}{8}+\frac{3 q_2 q_{-2} q_{\rm tr}}{4}+\frac{3 q_{\rm tr}^3}{5}
    \right)+...
    \label{eq:Delta_Omega}
\end{align}
where we again expanded the exact solution in powers of $q_m$ in the last equality to assist with a direct comparison with \citet{VanHoolst:94}. For a homogeneous ellipsoid, discontinuities in the derivatives of the gravitational potentials at the surface are non-trivial to evaluate, and we present an explicit evaluation of those terms in Appendix \ref{appx:grav_E}.

We consider both the tide and rotation as external perturbations to the system. For the tide, the interaction energy truncated to the quadrupolar order can be written as 
\begin{equation}
    \mathcal{E} = \frac{1}{2}\int \mathcal{E}_{ij} r_i r_j dM = \frac{1}{2} \mathcal{E}_{ij} q_{ik} q_{jk} \Ma, 
\end{equation}
where 
\begin{equation}
    \mathcal{E}_{ij} 
    = - \frac{\partial }{\partial D_i}\frac{\partial}{\partial D_j} \frac{M'}{|\vect{D} |} 
    =-\epsilon 
    \begin{bmatrix}
        \frac{1 + 3\cos\Phi_C}{2}, & 3\sin\Phi \cos\Phi_C, & 0 \\
        3 \sin\Phi\cos\Phi_C, & \frac{1-3\cos\Phi_C}{2}, & 0\\
        0, &0, & -1
    \end{bmatrix}.
\end{equation}
Note that we work in the frame corotating with the donor and use $\Phi_C=\Phi-\omega t$ to account for the Doppler effect. 
In comparison to, e.g., eq. (B1) of \cite{Diener:95}, we additionally included the centrifugal force due to a rotation $\omega$ (assumed to be along the $z$ axis, also the direction of the orbital angular momentum), giving rise to an energy 
\begin{equation}
    \mathcal{R}= -\frac{\omega^2}{2}\int (r^2-r_3^2) dM = -\frac{\omega^2}{2} (q_{ik} q_{ik} - q_{i3} q_{i3})\Ma.
\end{equation}

By varying the energies, the equations of motion are
\begin{equation}
    \Ma \ddot{q}_{ik}=(\Pi \delta_{ij} + \Omega_{ij}) (\vect{q}^{-1})_{kj} - \Ma \mathcal{E}_{ik} q_{kj} + \Ma \omega^2 (q_{ik} - \delta_{3i}q_{3k}),
    \label{eq:qij}
\end{equation}
By further contracting Eq. (\ref{eq:qij}) with the symmetric-tracefree tensor $\mathcal{Y}_{lm}^{ij}\hang{=(\mathcal{Y}_{l,-m}^{ij})^\ast}$ \citep{Poisson:14}, with
\begin{align}
    &\mathcal{Y}_{22}^{11} = -\mathcal{Y}_{22}^{22}=\sqrt{\frac{15}{32\pi}}, \quad \mathcal{Y}_{22}^{12} =  - i \sqrt{\frac{15}{32\pi}}, \nonumber \\
    & \mathcal{Y}_{20}^{11} = \mathcal{Y}_{20}^{22} = -\sqrt{\frac{5}{16\pi}}, \quad
    \mathcal{Y}_{20}^{33} = 2\sqrt{\frac{5}{16\pi}},
\end{align}
we obtain the equations of motion for each spherical harmonic mode. 
\begin{equation}
    \ddot{q}_{m} = \Pi_m + \Omega_m + \mathcal{E}_m + \mathcal{R}_m, 
\end{equation}
where on the right-hand side, the driving terms are due to variations of, respectively, the internal energy, the gravitational energy, the tidal interaction, and the centrifugal energy. Explicit expressions are provided as follows. By varying the internal energy, we have  
\begin{subequations}
\begin{align}
    \Pi_2 = \frac{2^{4 \Gamma -1} 3^{2 \Gamma -1} q_2 \left(125 \sqrt{6} q_0^3-450 q_0^2 (q_{\rm tr}+1)-750 \sqrt{6} q_0 q_2 q_{-2}+36 (q_{\rm tr}+1) \left(4 (q_{\rm tr}+1)^2-25 q_2 q_{-2}\right)\right)^{1-\Gamma }}{(5-n)\Ma \left(-25 q_0^2+20 \sqrt{6} q_0 (q_{\rm tr}+1)+6 \left(25 q_2 q_{-2}-4 (q_{\rm tr}+1)^2\right)\right)},
    \label{eq:Pi2_full}
\end{align}
\begin{align}
    \Pi_0 &=
    \left\{-125 q_0^4+150 \sqrt{6} q_0^3 (q_{\rm tr}+1)+40 q_0^2 \left[25 q_2 q_{-2}-9 (q_{\rm tr}+1)^2\right]\right.\nonumber \\
    &\left.\quad
    +4 \sqrt{6} q_0 (q_{\rm tr}+1) \left[12 (q_{\rm tr}+1)^2-125 q_2 q_{-2}\right]-60 q_2 q_{-2} \left[25 q_2 q_{-2}-4 (q_{\rm tr}+1)^2\right]\right\} \nonumber \\
    &\times\left\{-\frac{25}{8} q_{\rm tr} \left(q_0^2+2 q_2 q_{-2}\right)+\frac{25}{144} \left[5 \sqrt{6} q_0 \left(q_0^2-6 q_2 q_{-2}\right)-18 \left(q_0^2+2 q_2 q_{-2}\right)\right]+q_{\rm tr}^3+3 q_{\rm tr}^2+3 q_{\rm tr}+1\right\}^{-\Gamma }
    \nonumber \\
    &\times
    \left\{2 \sqrt{6} (5-n)\Ma \left(-25 q_0^2+20 \sqrt{6} q_0 (q_{\rm tr}+1)+6 \left[25 q_2 q_{-2}-4 (q_{\rm tr}+1)^2\right]\right)\right\}^{-1},
    \label{eq:Pi0_full}
\end{align}
\begin{align}
    \Pi_{\rm tr} &= (24 (5-n)\Ma)^{-1}
    \left(-25 q_0^2-50 q_2 q_{-2}+24 (q_{\rm tr}+1)^2\right) \nonumber \\
    &\times\left(-\frac{25}{8} q_{\rm tr} \left(q_0^2+2 q_2 q_{-2}\right)+\frac{25}{144} \left(5 \sqrt{6} q_0 \left(q_0^2-6 q_2 q_{-2}\right)-18 \left(q_0^2+2 q_2 q_{-2}\right)\right)+q_{\rm tr}^3+3 q_{\rm tr}^2+3 q_{\rm tr}+1\right)^{-\Gamma }.
    \label{eq:Pitr_full}
\end{align}
\label{eq:Pim}
\end{subequations}
And from the gravity,
\begin{subequations}
\begin{align}
    \Omega_2 = \frac{ (A_1-A_2)e^{2 i \phi_{q}} \left(5 \sqrt{6} q_0+30 |q_2|-12 (q_{\rm tr}+1)\right)-60 A_1 q_2}{40 (5-n)\Ma},
    \label{eq:Omega2_full}
\end{align}
\begin{align}
\Omega_0 & =  
    6 \left\{a_1^2 A_1 \left[25 \sqrt{6} q_0^2+30 q_0 (5 |q_2|-q_{\rm tr}-1)+6 \sqrt{6} (q_{\rm tr}+1) (5 |q_2|-2 q_{\rm tr}-2)\right] \right. \nonumber \\
    & 
    \quad\quad+a_2^2 A_2 \left[25 \sqrt{6} q_0^2-30 q_0 (5 |q_2|+q_{\rm tr}+1)-6 \sqrt{6} (q_{\rm tr}+1) (5 |q_2|+2 q_{\rm tr}+2)\right] \nonumber \\
    & \left.
    \quad\quad+a_3^2 A_3 \left[25 \sqrt{6} q_0^2-120 q_0 (q_{\rm tr}+1)+6 \sqrt{6} \left(4 (q_{\rm tr}+1)^2-25 |q_2|^2\right)\right]\right\}
     \nonumber \\
    & \times
    \left\{
    5 (5-n)\Ma \left(5 \sqrt{6} q_0+6 q_{\rm tr}+6\right) \left(-25 q_0^2+20 \sqrt{6} q_0 (q_{\rm tr}+1)+6 \left[25 |q_2|^2-4 (q_{\rm tr}+1)^2\right]\right)
    \right\}^{-1},
    \label{eq:Omega0_full}
\end{align}
\begin{align}
    \Omega_{\rm tr} = \frac{5 \sqrt{6} q_0 (A_1+A_2-2 A_3)
    +6 \left[-2 (q_{\rm tr} +1)(A_1+A_2+A_3)+5 |q_2| (A_2-A_1)\right]}{24 (5-n)\Ma}.
    \label{eq:Omegatr_full}
\end{align}
\label{eq:Omegam}
\end{subequations}

Note that the equations above are exact up to the affine approximation, Eq. (\ref{eq:r_vs_x}). We also present expanded forms of the driving terms that are complete to the four-wave interaction level. We will further write $q_{\rm tr}=q_{\rm tr}^{(1)} + q_{\rm tr}^{(2)}$ where $q_{\rm tr}^{(1)}\sim \omega^2\sim \epsilon$ will be sourced by the ($l=0$ part of the ) centrifugal potential, while $q_{\rm tr}^{(2)}\sim \epsilon^2$ from nonlinear mode interactions. 
From the internal pressure,
\begin{subequations}
\begin{align}
    &\Pi_{2}^{(1)} = -\frac{q_2}{(5-n)\Ma}, \quad
    \Pi_{2}^{(2)} = -\frac{q_2[5 \sqrt{6}q_0 - 6 \qtr (3\Gamma - 1) ]}{6(5-n)\Ma}, \nonumber \\
    &\Pi_{2}^{(3)} = - \frac{q_2[\Gamma(50 q_2 q_{-2} + 25 q_0^2 - 20\sqrt{6} q_0 \qtr ) + 4 (3\Gamma -1) (3\Gamma [\qtr]^2 - 2\qtrr) ]}{8 (5-n) \Ma},\\
    &\Pi_{0}^{(1)} = -\frac{q_0}{(5-n)\Ma}, \quad
    \Pi_{0}^{(2)} = \frac{-10\sqrt{6} q_2 q_{-2} + 5\sqrt{6} q_0^2 + 12 (3 \Gamma -1 ) q_0 \qtr }{12  (5-n)\Ma}, \nonumber \\
    &\Pi_{0}^{(3)} = -\frac{25 \Gamma  q_0 \left(2 q_2q_{-2} + q_0^2\right) - 10 \sqrt{ 6} \Gamma  \qtr \left(2 q_{2}q_{-2} - q_0^2\right)+4(3 \Gamma -1) q_0 \left(3 \Gamma  [\qtr]^2-2 \qtrr\right)}{8 (5-n) \Ma}, \\
    &\Pi_{\rm tr}^{(0)} = \frac{1}{(5-n)\Ma},\quad 
    \Pi_{\rm tr}^{(1)} = -\frac{(3\Gamma -2) \qtr}{(5-n) \Ma}, \nonumber \\
    &\Pi_{\rm tr}^{(2)} = \frac{ 25 (3 \Gamma -1) (2 q_2 q_{-2} + q_0^2) + 12 (3 \Gamma -2) \left((3 \Gamma -1) [\qtr]^2-2 \qtrr\right)}{24   (5-n) \Ma}.
\end{align}
\end{subequations}
And from self-gravity,
\begin{subequations}
\begin{align}
    &\Omega_2^{(1)} = \frac{q_2}{5 (5-n)\Ma}, \quad
    \Omega_2^{(2)} = -\frac{q_2(5\sqrt{6} q_0 + 42 q_{\rm tr}^{(1)}) }{70(5-n) \Ma}, \nonumber \\
    &\Omega_2^{(3)}= \frac{q_2 [50 q_2q_{-2} + 25q_0^2 + 40 \sqrt{6} q_0q_{\rm tr}^{(1)} + 84(2 [q_{\rm tr}^{(1)}]^2- q_{\rm tr}^{(2)})]}{140 (5-n)\Ma}, \\   
    &\Omega_{0}^{(1)} = \frac{q_0}{5(5-n)\Ma}, \quad
    \Omega_{0}^{(2)} = -\frac{10 \sqrt{6} q_2q_{-2} - 5\sqrt{6} q_0^2 + 84 q_0 q_{\rm tr}^{(1)} }{140(5-n)\Ma  },\nonumber \\
    &\Omega_{0}^{(3)} = \frac{q_0 [50 q_2 q_{-2} + 25 q_0^2 - 20\sqrt{6} q_0\qtr + 84 (2{[\qtr]}^2 - \qtrr)] + 40\sqrt{6}q_2q_{-2}\qtr }{140 (5-n) \Ma}, \\
    &\Omega_{\rm tr}^{(0)} = -\frac{1}{(5-n)\Ma}, \quad
    \Omega_{\rm tr}^{(1)} = \frac{2\qtr}{(5-n)\Ma}, \quad
    \Omega_{\rm tr2}^{(2)} = -\frac{10 q_2q_{-2} + 5 q_0^2 + 24[\qtr]^2 - 16 \qtrr}{8(5-n)\Ma}.
\end{align}
\end{subequations}

We now turn to external perturbations. For the tidal interaction, we have
\begin{subequations}
\begin{align}
     &\mathcal{E}_{2}^{(1)} = \frac{3}{5} \epsilon e^{-2i\Phi_C}, \quad 
    \mathcal{E}_{2}^{(2)} = \frac{ \epsilon  \left(10 q_2 - 5 \sqrt{6} q_0 e^{-2 i \Phi_C }  + 12 \qtr e^{-2 i \Phi_C }\right)}{20}, \\
    &\mathcal{E}_{0}^{(1)} = -\frac{\sqrt{6}}{5} \epsilon, \quad 
    \mathcal{E}_{0}^{(2)} = -\frac{\epsilon  \left[ 5 \left( \sqrt{6} q_2 e^{2 i \Phi_C } + \sqrt{6} q_{-2} e^{-2 i \Phi_C }  + 2 q_0\right) + 4 \sqrt{6} \qtr \right]}{20}, \\
    &\mathcal{E}_{\rm tr}^{(1)} = 0, \quad
    \mathcal{E}_{\rm tr}^{(2)} = \frac{5[3 (q_2 e^{-2 i\Phi_C} + q_{-2} e^{2i \Phi_C} ) - \sqrt{6} q_0]}{12} \epsilon.
\end{align}
\label{eq:Em}
\end{subequations}
Similarly, the centrifugal drive is decomposed as
\begin{subequations}
\begin{align}
    &\mathcal{R}_2^{(1)} = 0, \quad 
    \mathcal{R}_2^{(2)} = q_2 \omega^2, \\
    & \mathcal{R}_0^{(1)} = - \frac{2 \sqrt{6} \omega^2}{15}, \quad 
    \mathcal{R}_0^{(2)} = \frac{5 q_0 - 2\sqrt{6} \qtr}{15}\omega^2,\\
    & \mathcal{R}_{\rm tr}^{(1)} = \frac{2}{3}\omega^2, \quad 
    \mathcal{R}_{\rm tr}^{(2)} = \frac{-5 \sqrt{6} q_0 + 12 \qtr}{18}  \omega^2. 
\end{align}
\label{eq:Rm}
\end{subequations}

Since the affine model is self-consistent only for homogeneous ellipsoids with $(n=0, \Gamma=\infty)$, we discuss the equations of motion for this case with further details. First note that the natural frequency of the $l=0$ mode is $\omega_{\rm tr}^2=(3\Gamma-4)/(5-n)\hang{/\Ma}$, which approaches infinity as $\Gamma\to\infty$. Therefore, we can approximate $\ddot{q}_{\rm tr}=0$, allowing us to solve for the trace part as
\begin{align}
    &\qtr = \frac{2}{9 \Gamma} \omega^2, \nonumber \\
    &\qtrr=\frac{25 (2 q_2 q_{-2} + q_0^2) }{24} + 
    \frac{1}{\Gamma}\left[\frac{5}{6}(2q_2q_{-2} + q_0^2) + \frac{5\epsilon (3q_2e^{2i\Phi} + 3q_{-2} e^{-2i\Phi} - \sqrt{6} q_0 )}{36} -\frac{5 \sqrt{6}\omega^2 q_0}{54} + \frac{2\omega^4}{27} \right].
    \label{eq:qtr_incomp}
\end{align}
We have expanded the results in $1/\Gamma$ and kept only the leading-order terms. 
Substituting them back into the equations of motion for quadrupolar modes, we obtain the complete description of the dynamics for $(n, \Gamma)=(0, \infty)$ at the four-wave order up to the ignorance of the Coriolis force (which can be included at the leading order following Eqs. \ref{eq:qm_decom_into_cmpm}-\ref{eq:omega_f_pm}; \citealt{Yu:25a}) which reads
\begin{equation}
    \ddot{q}_m = - \omega_f^2 q_m + \omega_f^2K_m,
\end{equation}
where,
\begin{subequations}
\begin{align}
    K_2 =& \frac{3}{4} \epsilon e^{-2i\Phi} + q_2\left( -\frac{95\sqrt{6}}{84}q_0 + \frac{5}{8} \epsilon + \frac{5}{4} \omega^2 \right) - \frac{5\sqrt{6}}{16} \epsilon q_0 e^{-2i\Phi} \nonumber \\
    &+q_2\left[\frac{425}{336} (2q_2 q_{-2} + q_0^2) + \frac{25}{48} \epsilon (3 q_2  e^{2i\Phi}  + 3q_{-2}  e^{-2i\Phi} - \sqrt{6} q_0 )
    + \frac{25 \sqrt{6}}{72}\omega^2 q_0 
    \right], 
    \label{eq:K2_4m}\\
    K_0=&-\frac{\sqrt{6}}{4} \epsilon - \frac{\sqrt{6}}{6} \omega^2- \frac{95\sqrt{6}}{168}( 2q_2 q_{-2}- q_0^2)  - \frac{5\epsilon}{16} (\sqrt{6}q_2 e^{2i\Phi} +  \sqrt{6}q_{-2}e^{-2i\Phi} + 2q_0) + \frac{5}{4}\omega^2 q_0 \nonumber \\
    &+ q_0\left[\frac{425}{336}(2q_2q_{-2} +q_0^2)
     +\frac{25}{48}\epsilon(3q_2 e^{2i\Phi} + 3q_{-2} e^{-2i\Phi} - \sqrt{6}q_0)
     \right]
     + \frac{25\sqrt{6}}{36} \omega^2 (q_2q_{-2} - q_0^2).
     \label{eq:K0_4m}
\end{align}
\end{subequations}
Note that terms proportional to $\Gamma$ cancel out exactly in $\Pi_m^{(3)}$, making the result regular when $\Gamma\to \infty$.

\section{Surface terms in the third order perturbations of the gravitational potential energy}
\label{appx:grav_E}
At the formulation level, it is clear that the affine model agrees with that in \cite{VanHoolst:94} when applied to incompressible stars. However, it is not trivial to show the agreement, especially for terms of $\mathcal{O}(\xi^3)$ and above. This is because these terms in \cite{VanHoolst:94} often involve high-order derivatives of the quantities from the background or the lower perturbation levels, which produce surface terms that require integrating the Dirac delta functions. In this appendix, we explicitly calculate the $\mathcal{O}(\xi^3)$ terms in the gravitational potential energy and show their agreement with the affine model under careful treatment of the surface terms.

We first write the $\mathcal{O}(\xi^3)$ terms as
\begin{align}
    \Delta \Omega^{(3)} = \Delta \Omega^{(3a)} + \Delta \Omega^{(3b)},
\end{align}
where
\begin{align}
    \Delta \Omega^{(3a)} =& \frac{1}{6}\int \rho d^3x \xi^i \xi^j \xi^k \nabla_i \nabla_j \nabla_k \Omega_0, \\
    \Delta \Omega^{(3b)} =& \frac{1}{2}\int \rho d^3x \xi^i \xi^j \nabla_i \nabla_j \Omega^\prime,
\end{align}
where the integral is carried within the unperturbed volume. The Lagrangian displacement is expressed in spherical coordinates
\begin{align}
    \vect{\xi} = \xi^r \vect{e}_r+\xi^\theta \vect{e}_\theta+\xi^\phi \vect{e}_\phi.
\end{align}

The unperturbed potential $\Omega_0$ in all space is given by
\begin{align}
    \Omega_0 = \frac{1}{2}\left(r^2-3 R^2\right) \Theta(R-r) - \Theta(r-R),
\end{align}
where $\Theta(x)$ is the Heaviside step function\footnote{The one-half convention of the Heaviside step function is the result from treating both $\delta(x)$ and $\Theta(x)$ as finite distribution functions (with even and odd parity respectively) and taking the zero width limit for both of them. This gives $\int_{-\infty}^{\infty} dx \delta(x) \Theta(x) = 1/2$.} with $\Theta(0) = 1/2$. We also replace $\rho$ by $\rho\Theta(R-r)$ since the surface terms depend on the boundary value. The triple derivative contributes only to a Dirac delta term within the domain of integration:
\begin{align}
    \Delta \Omega^{(3a)} =& \frac{1}{6}\int d^3x \rho \Theta(R-r) (\xi^r)^3 \partial_r^3 U_0 \nonumber\\
    =& -\frac{1}{2}\int d^3x \rho \Theta(R-r) (\xi^r)^3 \delta(r-R) \nonumber\\
    =& - \frac{\rho}{4} \int \sin\theta d\theta d\phi (\xi^r|_{r=R})^3.
\end{align}
The Lagrangian displacement expanded in terms of the quadrupolar f-mode eigenfunctions is given by Eq.~\eqref{eq:vxi_expand}.
The integral of three spherical harmonics can be evaluated using the Wigner 3-j symbols (see, e.g., \cite{Weinberg:12}):
\begin{align}
    \int Y_{l_1 m_1} Y_{l_2 m_2} Y_{l_3 m_3} \sin\theta d\theta d\phi = \sqrt{\frac{(2l_1+1)(2l_2+1)(2l_3+1)}{4\pi}} 
    \begin{pmatrix}
    l_1 & l_2 & l_3\\
    m_1 & m_2 & m_3
    \end{pmatrix}
    \begin{pmatrix}
    l_1 & l_2 & l_3\\
    0 & 0 & 0
    \end{pmatrix}. \label{eq:gaunt_coef}
\end{align}
The non-zero coefficients for $l=2$, $m=-2,0,2$ are
\begin{align}
    \begin{pmatrix}
    2 & 2 & 2\\
    0 & 0 & 0
    \end{pmatrix}
    = -\sqrt{\frac{2}{35}}, \;\;\;\;
        \begin{pmatrix}
    2 & 2 & 2\\
    -2 & 2 & 0
    \end{pmatrix}
    = \sqrt{\frac{2}{35}}.
\end{align}
Focusing on the $l=2$, $m=-2,0,2$ modes, we have 
\begin{align}
    \Delta \Omega^{(3a)} = \frac{25}{28\sqrt{6}} q_0(6q_2 q_{-2}-q_0^2).\label{eq:grav_energy_3a}
\end{align}
The integral in $\Delta \Omega^{(3b)}$ contains both a surface integral term from the Dirac delta function and a bulk term.
\begin{align}
    \Delta \Omega^{(3b)} = \Delta \Omega^{(3b)}_\text{s} + \Delta \Omega^{(3b)}_\text{b},
\end{align}
where the subscripts s and b denote the surface term and bulk terms, respectively. To see the delta function arising from the derivatives of the Eulerian potential, we write the quadrupolar f-mode $\Omega^\prime$ as
\begin{align}
    \Omega^\prime(t,\boldsymbol{x}) = - \sum_{m} \sqrt{\frac{6\pi}{5}} \left[r^2 \Theta(R-r) + r^{-3} \Theta(r-R)\right] q_{m} Y_{2 m}.
\end{align}
Hence, $\partial_r^2 \Omega^{\prime}$ gives a term with the Dirac delta in the integrand.
\begin{align}
    \frac{1}{2}(\xi^r)^2\partial_r^2 \Omega^\prime =  (\xi^r)^2 \delta(r-R) \sum_{m} 3\sqrt{\frac{5\pi}{6}} q_{m} Y_{2m} + \text{bulk terms}.
\end{align}
For $l=2$, $m=-2,0,2$ modes, and using Eq.~\eqref{eq:gaunt_coef}, we have
\begin{align}
    \Delta \Omega^{(3b)}_\text{s} = -3\Delta \Omega^{(3a)} = -3\left(\frac{25}{28\sqrt{6}}\right) q_0(6q_2 q_{-2}-q_0^2). \label{eq:grav_energy_3b_surface}
\end{align}
Note that the delta term within $\partial_r^2 \Omega^{\prime}$ is equivalent to $-\xi^r\partial_r^3 \Omega_0$. Therefore, the factor of three in Eq.~\eqref{eq:grav_energy_3b_surface} comes from the ratio of the prefactors $1/6$ and $1/2$ of the potential terms.

The remaining terms without the Dirac delta function can be integrated to give
\begin{align}
    \Delta \Omega^{(3b)}_\text{b} = \frac{5}{8}\sqrt{\frac{3}{2}} q_0(6q_2 q_{-2}-q_0^2). \label{eq:grav_energy_3b_bulk}
\end{align}
Collecting Eqs.~\eqref{eq:grav_energy_3a}, \eqref{eq:grav_energy_3b_surface} and \eqref{eq:grav_energy_3b_bulk}, we obtain
\begin{align}
    \Delta \Omega^{(3)} = \Delta \Omega^{(3a)} + \Delta \Omega^{(3b)}_s + \Delta \Omega^{(3b)}_b =\frac{5}{56\sqrt{6}} q_0(6q_2 q_{-2}-q_0^2),
\end{align}
which agrees with Eq.~\eqref{eq:Delta_Omega}. 

\section{Coefficients of polynomials}
\label{appx:coeff_poly}
We present here the explicit expressions for the numerical coefficients of Eqs. (\ref{eq:v2_quad_form}), (\ref{eq:v2_cubic_form}), and (\ref{eq:z2_quad_form}). 

For Eq. (\ref{eq:v2_quad_form}), we have
\begin{subequations}
\begin{align}
    X_{22} &= -\frac{38 a_{31}}{7 (a_{31}+2)}, \\
    X_{21} &= \frac{35 \epsilon  [a_{31} (p-1)+2 p+1]-104 a_{31}+20}{35 (a_{31}+2)}, \\
    X_{20}&= \frac{3}{10} \epsilon  \left(\frac{6}{a_{31}+2}+5 p \epsilon \right),
\end{align}
\end{subequations}
where $a_{31}=a_3/a_1$, and
\begin{subequations}
\begin{align}
     X_{02} &= \frac{38 \sqrt{\frac{2}{3}} [(a_{31}-2) a_{31}-2]}{7 (a_{31}+2)^2}, \\
     X_{01} &= \frac{\sqrt{6} \{35 (a_{31}+2) \epsilon  [a_{31} (p-1)-1]+12 a_{31} (4 a_{31}-11)\}}{35 (a_{31}+2)^2}, \\
     X_{00} &= \frac{\sqrt{\frac{2}{3}} \{35 (a_{31}+2) \epsilon  [a_{31} (4 p-6)-10 p-3]+12 (a_{31}-1) (5 a_{31}-47)\}}{175 (a_{31}+2)^2}.
\end{align}
\end{subequations}
Note Eq. (\ref{eq:dcm_ad}) has been used to write $C_2^{(\rm st)}$. 

For Eq. (\ref{eq:v2_cubic_form}), we have
\begin{subequations}
\begin{align}
    W_{23} &= \frac{170 \left(a_{31}^2+a_{31}+1\right)}{21 (a_{31}+2)^2}, \\
    W_{22} &= \frac{5 \epsilon }{a_{31}+2}+\frac{2 a_{31} (8 a_{31}-137)}{21 (a_{31}+2)^2}, \\
    W_{21} &= \frac{105\epsilon  (a_{31}+2)  [a_{31} (p-2)+2 (p+1)]-2 a_{31} (77 a_{31}+356)+110}{105 (a_{31}+2)^2}, \\
    W_{20} &= \frac{3}{10} \epsilon  \left(\frac{6}{a_{31}+2}+5 p \epsilon \right)
\end{align}
\end{subequations}
and 
\begin{subequations}
\begin{align}
    W_{03} &= \frac{170 \sqrt{\frac{2}{3}} a_{31} \left(a_{31}^2+a_{31}+1\right)}{7 (a_{31}+2)^3} \\
    W_{02} &= \frac{\sqrt{\frac{2}{3}} \{315 a_{31} (a_{31}+2) \epsilon +2 a_{31} [a_{31} (242 a_{31}-237)-174]-436\}}{21 (a_{31}+2)^3} \\
    W_{01} &= \frac{\sqrt{\frac{2}{3}} \left\{105\epsilon (a_{31}+2)   \left[a_{31}^2 (p-2)+2 a_{31} p-4\right]+2 a_{31} [a_{31} (113 a_{31}-472)-19]\right\}}{35 (a_{31}+2)^3}\\
    W_{00} &=\frac{\sqrt{\frac{2}{3}} \left\{35 \epsilon (a_{31}+2)  \left[2 (a_{31}+2) (2 a_{31}-5) p-3 \left(4 a_{31}^2+a_{31}+4\right)\right]+4 (a_{31}-1) [a_{31} (10 a_{31}-269)-119]\right\}}{175 (a_{31}+2)^3}. 
\end{align}
\end{subequations}

Lastly, for Eq. (\ref{eq:z2_quad_form}), the coefficients are
\begin{subequations}
\begin{align}
    Z_{22} &= -\frac{38 a_{31} [q_2^{(0)}]^2}{7 (a_{31}+2)},  \\
    Z_{21} &= \frac{35 {q_2^{(0)}} \epsilon  [a_{31} (p-1)+2 p+1]-4 {q_2^{(0)}} [a_{31} (95 {q_2^{(0)}}+26)-5]}{35 (a_{31}+2)}-{[\ddot{q}_2^{(0)} e^{2i\Phi_c}] }, \\
    Z_{20} &= \Big(
    -70 (a_{31}+2) {[\ddot{q}_2^{(0)} e^{2i\Phi_c}] }+14 \epsilon  \{5 {q_2^{(0)}} [a_{31} (p-1)+2 p+1]+9\} \nonumber \\
    &\quad\quad +105 (a_{31}+2) p \epsilon ^2-4 a_{31} {q_2^{(0)}} (95 {q_2^{(0)}}+52)+40 {q_2^{(0)}}
    \Big) \nonumber \\
    &\times [70 (a_{31}+2)]^{-1}, 
\end{align}
\end{subequations}
and 
\begin{subequations}
\begin{align}
    Z_{02} &= \frac{38 \sqrt{\frac{2}{3}} [(a_{31}-2) a_{31}-2] [q_2^{(0)}]^2}{7 (a_{31}+2)^2}, \\
    Z_{01} &= 
    \sqrt{\frac{2}{3}} {q_2^{(0)}} \left\{
    -105 a_{31}^2 {\ddot{q}_0^{(0)}}+380 a_{31}^2 {q_0^{(0)}} {q_2^{(0)}}+144 a_{31}^2 {q_0^{(0)}}-210 a_{31} {\ddot{q}_0^{(0)}}
     \right.\nonumber \\
     &\left. \quad\quad\quad\quad +105 \epsilon (a_{31}+2) {q_0^{(0)}}   [a_{31} (p-1)-1] -760 a_{31} {q_0^{(0)}} {q_2^{(0)}}-396 a_{31} {q_0^{(0)}}-760 {q_0^{(0)}} {q_2^{(0)}}
    \right\}
     \nonumber \\
    &\times [35 (a_{31}+2)^2 {q_0^{(0)}}]^{-1}, \\
    Z_{00} &= 
    \sqrt{\frac{2}{3}} \left(
    -105 {\ddot{q}_0^{(0)}} (a_{31}+2)  [a_{31} (5 {q_2^{(0)}}+2)-2]+35 \epsilon {q_0^{(0)}}(a_{31}+2)    \{15 {q_2^{(0)}} [a_{31} (p-1)-1] +4 a_{31} p-6 a_{31}-10 p-3\}\right. \nonumber \\
    &\left.
    \quad\quad\quad\quad
    +2 {q_0^{(0)}} \left\{475 [(a_{31}-2) a_{31}-2] [q_2^{(0)}]^2+90 a_{31} (4 a_{31}-11) {q_2^{(0)}}+6 (a_{31}-1) (5 a_{31}-47)\right\}\right)
     \nonumber \\
    &\times [175 (a_{31}+2)^2 {q_0^{(0)}}]^{-1}.
\end{align}
\end{subequations}
Again, Eq. (\ref{eq:dcm_ad}) has been used to eliminate the Coriolis term, $C_2^{(\rm st)}$.


\bibliography{ref}{}

@ARTICLE{Lau:25,
       author = {{Lau}, Shu Yan and {Yu}, Hang},
        title = "{Gravitational Radiation-Driven Chaotic Tide in a White Dwarf-Massive Black Hole Binary as a Source of Repeating X-ray Transients}",
      journal = {arXiv e-prints},
         year = 2025,
        month = jun,
          eid = {arXiv:2506.10163},
        pages = {arXiv:2506.10163},
          doi = {10.48550/arXiv.2506.10163},
archivePrefix = {arXiv},
       eprint = {2506.10163},
 primaryClass = {astro-ph.HE},
       adsurl = {https://ui.adsabs.harvard.edu/abs/2025arXiv250610163L}
}

@misc{Lau:26,
  author =  {{Lau}, Shu Yan and {Yu}, Hang},
  title = {Diffusive-tide-powered repeated partial tidal disruption events with irregular recurrences},
  year = {2026},
  howpublished = {in prep}
}

@ARTICLE{Yu:21,
       author = {{Yu}, Hang and {Weinberg}, Nevin N. and {Arras}, Phil},
        title = "{Tides in the High-eccentricity Migration of Hot Jupiters: Triggering Diffusive Growth by Nonlinear Mode Interactions}",
      journal = {\apj},
         year = 2021,
        month = aug,
       volume = {917},
       number = {1},
          eid = {31},
        pages = {31},
          doi = {10.3847/1538-4357/ac0a79},
archivePrefix = {arXiv},
       eprint = {2104.04929},
 primaryClass = {astro-ph.EP},
       adsurl = {https://ui.adsabs.harvard.edu/abs/2021ApJ...917...31Y}
}

@ARTICLE{Yu:22a,
       author = {{Yu}, Hang and {Weinberg}, Nevin N. and {Arras}, Phil},
        title = "{Tidal Evolution and Diffusive Growth During High-eccentricity Planet Migration: Revisiting the Eccentricity Distribution of Hot Jupiters}",
      journal = {\apj},
         year = 2022,
        month = apr,
       volume = {928},
       number = {2},
          eid = {140},
        pages = {140},
          doi = {10.3847/1538-4357/ac5627},
archivePrefix = {arXiv},
       eprint = {2111.04649},
 primaryClass = {astro-ph.EP},
       adsurl = {https://ui.adsabs.harvard.edu/abs/2022ApJ...928..140Y}
}

@ARTICLE{Yu:23a,
       author = {{Yu}, Hang and {Weinberg}, Nevin N. and {Arras}, Phil and {Kwon}, James and {Venumadhav}, Tejaswi},
        title = "{Beyond the linear tide: impact of the non-linear tidal response of neutron stars on gravitational waveforms from binary inspirals}",
      journal = {\mnras},
         year = 2023,
        month = mar,
       volume = {519},
       number = {3},
        pages = {4325-4343},
          doi = {10.1093/mnras/stac3614},
archivePrefix = {arXiv},
       eprint = {2211.07002},
 primaryClass = {gr-qc},
       adsurl = {https://ui.adsabs.harvard.edu/abs/2023MNRAS.519.4325Y}
}

@ARTICLE{Yu:24a,
       author = {{Yu}, Hang and {Arras}, Phil and {Weinberg}, Nevin N.},
        title = "{Dynamical tides during the inspiral of rapidly spinning neutron stars: Solutions beyond mode resonance}",
      journal = {\prd},
         year = 2024,
        month = jul,
       volume = {110},
       number = {2},
          eid = {024039},
        pages = {024039},
          doi = {10.1103/PhysRevD.110.024039},
archivePrefix = {arXiv},
       eprint = {2404.00147},
 primaryClass = {gr-qc},
       adsurl = {https://ui.adsabs.harvard.edu/abs/2024PhRvD.110b4039Y}
}

@ARTICLE{Yu:24b,
       author = {{Yu}, Hang and {Dai}, Fei},
        title = "{Are WASP-107-like Systems Consistent with High-eccentricity Migration?}",
      journal = {\apj},
         year = 2024,
        month = sep,
       volume = {972},
       number = {2},
          eid = {159},
        pages = {159},
          doi = {10.3847/1538-4357/ad5ffb},
archivePrefix = {arXiv},
       eprint = {2406.00187},
 primaryClass = {astro-ph.EP},
       adsurl = {https://ui.adsabs.harvard.edu/abs/2024ApJ...972..159Y}
}

@ARTICLE{Yu:25a,
       author = {{Yu}, Hang and {Lau}, Shu Yan},
        title = "{Effective-one-body model for coalescing binary neutron stars: Incorporating tidal spin and enhanced radiation from dynamical tides}",
      journal = {\prd},
         year = 2025,
        month = apr,
       volume = {111},
       number = {8},
          eid = {084029},
        pages = {084029},
          doi = {10.1103/PhysRevD.111.084029},
       adsurl = {https://ui.adsabs.harvard.edu/abs/2025PhRvD.111h4029Y}
}

@ARTICLE{Kwon:24,
       author = {{Kwon}, K.~J. and {Yu}, Hang and {Venumadhav}, Tejaswi},
        title = "{Resonance Locking of Anharmonic $g$-Modes in Coalescing Neutron Star Binaries}",
      journal = {arXiv e-prints},
         year = 2024,
        month = oct,
          eid = {arXiv:2410.03831},
        pages = {arXiv:2410.03831},
          doi = {10.48550/arXiv.2410.03831},
archivePrefix = {arXiv},
       eprint = {2410.03831},
 primaryClass = {gr-qc},
       adsurl = {https://ui.adsabs.harvard.edu/abs/2024arXiv241003831K}
}

@ARTICLE{Kwon:25,
       author = {{Kwon}, K.~J. and {Yu}, Hang and {Venumadhav}, Tejaswi},
        title = "{Resonance locking: radian-level phase shifts due to nonlinear hydrodynamics of $g$-modes in merging neutron star binaries}",
      journal = {arXiv e-prints},
         year = 2025,
        month = mar,
          eid = {arXiv:2503.11837},
        pages = {arXiv:2503.11837},
          doi = {10.48550/arXiv.2503.11837},
archivePrefix = {arXiv},
       eprint = {2503.11837},
 primaryClass = {gr-qc},
       adsurl = {https://ui.adsabs.harvard.edu/abs/2025arXiv250311837K}
}

@ARTICLE{Schenk:02,
       author = {{Schenk}, A.~K. and {Arras}, P. and {Flanagan}, {\'E}. {\'E}. and
         {Teukolsky}, S.~A. and {Wasserman}, I.},
        title = "{Nonlinear mode coupling in rotating stars and the r-mode instability in neutron stars}",
      journal = {\prd},
         year = 2002,
        month = jan,
       volume = {65},
       number = {2},
          eid = {024001},
        pages = {024001},
          doi = {10.1103/PhysRevD.65.024001},
archivePrefix = {arXiv},
       eprint = {gr-qc/0101092},
 primaryClass = {gr-qc},
       adsurl = {https://ui.adsabs.harvard.edu/abs/2002PhRvD..65b4001S}
}

@ARTICLE{Weinberg:12,
       author = {{Weinberg}, Nevin N. and {Arras}, Phil and {Quataert}, Eliot and
         {Burkart}, Josh},
        title = "{Nonlinear Tides in Close Binary Systems}",
      journal = {\apj},
         year = "2012",
        month = "Jun",
       volume = {751},
       number = {2},
          eid = {136},
        pages = {136},
          doi = {10.1088/0004-637X/751/2/136},
archivePrefix = {arXiv},
       eprint = {1107.0946},
 primaryClass = {astro-ph.SR},
       adsurl = {https://ui.adsabs.harvard.edu/abs/2012ApJ...751..136W}
}

@ARTICLE{Weinberg:13,
       author = {{Weinberg}, Nevin N. and {Arras}, Phil and {Burkart}, Joshua},
        title = "{An Instability due to the Nonlinear Coupling of p-modes to g-modes: Implications for Coalescing Neutron Star Binaries}",
      journal = {\apj},
         year = 2013,
        month = jun,
       volume = {769},
       number = {2},
          eid = {121},
        pages = {121},
          doi = {10.1088/0004-637X/769/2/121},
archivePrefix = {arXiv},
       eprint = {1302.2292},
 primaryClass = {astro-ph.SR},
       adsurl = {https://ui.adsabs.harvard.edu/abs/2013ApJ...769..121W}
}

@ARTICLE{Weinberg:16,
   author = {{Weinberg}, N.~N.},
    title = "{Growth Rate of the Tidal p-Mode g-Mode Instability in Coalescing Binary Neutron Stars}",
  journal = {ApJ},
archivePrefix = "arXiv",
   eprint = {1509.06975},
 primaryClass = "astro-ph.SR",
     year = 2016,
    month = mar,
   volume = 819,
      eid = {109},
    pages = {109},
      doi = {10.3847/0004-637X/819/2/109},
   adsurl = {http://adsabs.harvard.edu/abs/2016ApJ...819..109W}
}

@ARTICLE{Venumadhav:14,
   author = {{Venumadhav}, T. and {Zimmerman}, A. and {Hirata}, C.~M.},
    title = "{The Stability of Tidally Deformed Neutron Stars to Three- and Four-mode Coupling}",
  journal = {ApJ},
archivePrefix = "arXiv",
   eprint = {1307.2890},
 primaryClass = "astro-ph.HE",
     year = 2014,
    month = jan,
   volume = 781,
      eid = {23},
    pages = {23},
      doi = {10.1088/0004-637X/781/1/23},
   adsurl = {http://adsabs.harvard.edu/abs/2014ApJ...781...23V}
}

@ARTICLE{Arras:23,
       author = {{Arras}, Phil and {Yu}, Hang and {Weinberg}, Nevin N.},
        title = "{Large Dynamical Tide Amplitudes from Small Kicks at Pericenter}",
      journal = {arXiv e-prints},
         year = 2023,
        month = jun,
          eid = {arXiv:2306.02382},
        pages = {arXiv:2306.02382},
          doi = {10.48550/arXiv.2306.02382},
archivePrefix = {arXiv},
       eprint = {2306.02382},
 primaryClass = {astro-ph.EP},
       adsurl = {https://ui.adsabs.harvard.edu/abs/2023arXiv230602382A}
}

@BOOK{Poisson:14,
       author = {{Poisson}, Eric and {Will}, Clifford M.},
        title = "{Gravity}",
         year = "2014",
    publisher = "Cambridge University Press",
       adsurl = {https://ui.adsabs.harvard.edu/abs/2014grav.book.....P}
}

@ARTICLE{Chandrasekhar:62,
       author = {{Chandrasekhar}, S. and {Lebovitz}, Norman R.},
        title = "{The Potentials and the Superpotentials of Homogeneous Ellipsoids.}",
      journal = {\apj},
         year = 1962,
        month = nov,
       volume = {136},
        pages = {1037},
          doi = {10.1086/147456},
       adsurl = {https://ui.adsabs.harvard.edu/abs/1962ApJ...136.1037C}
}

@ARTICLE{Chandrasekhar:63,
       author = {{Chandrasekhar}, S.},
        title = "{The Equilibrium and the Stability of the Roche Ellipsoids.}",
      journal = {\apj},
         year = 1963,
        month = nov,
       volume = {138},
        pages = {1182},
          doi = {10.1086/147716},
       adsurl = {https://ui.adsabs.harvard.edu/abs/1963ApJ...138.1182C}
}

@BOOK{Chandrasekhar:87,
       author = {{Chandrasekhar}, Subrahmanyan},
        title = "{Ellipsoidal figures of equilibrium}",
         year = 1987,
       adsurl = {https://ui.adsabs.harvard.edu/abs/1987efe..book.....C}
}

@ARTICLE{VanHoolst:94,
       author = {{Van Hoolst}, T.},
        title = "{Nonlinear, nonradial, isentropic oscillations of stars: Hamiltonian formalism}",
      journal = {\aap},
         year = 1994,
        month = jun,
       volume = {286},
        pages = {879-889},
       adsurl = {https://ui.adsabs.harvard.edu/abs/1994A&A...286..879V}
}

@ARTICLE{Hut:81,
       author = {{Hut}, P.},
        title = "{Tidal evolution in close binary systems.}",
      journal = {\aap},
         year = 1981,
        month = jun,
       volume = {99},
        pages = {126-140},
       adsurl = {https://ui.adsabs.harvard.edu/abs/1981A&A....99..126H}
}

@ARTICLE{Vick:18,
       author = {{Vick}, Michelle and {Lai}, Dong},
        title = "{Dynamical tides in highly eccentric binaries: chaos, dissipation, and quasi-steady state}",
      journal = {\mnras},
         year = 2018,
        month = may,
       volume = {476},
       number = {1},
        pages = {482-495},
          doi = {10.1093/mnras/sty225},
archivePrefix = {arXiv},
       eprint = {1708.09392},
 primaryClass = {astro-ph.SR},
       adsurl = {https://ui.adsabs.harvard.edu/abs/2018MNRAS.476..482V}
}

@ARTICLE{Vick:19,
       author = {{Vick}, Michelle and {Lai}, Dong and {Anderson}, Kassandra R.},
        title = "{Chaotic tides in migrating gas giants: forming hot and transient warm Jupiters via Lidov-Kozai migration}",
      journal = {\mnras},
         year = 2019,
        month = apr,
       volume = {484},
       number = {4},
        pages = {5645-5668},
          doi = {10.1093/mnras/stz354},
archivePrefix = {arXiv},
       eprint = {1812.05618},
 primaryClass = {astro-ph.EP},
       adsurl = {https://ui.adsabs.harvard.edu/abs/2019MNRAS.484.5645V}
}

@ARTICLE{Wu:18,
       author = {{Wu}, Yanqin},
        title = "{Diffusive Tidal Evolution for Migrating Hot Jupiters}",
      journal = {\aj},
         year = 2018,
        month = mar,
       volume = {155},
       number = {3},
          eid = {118},
        pages = {118},
          doi = {10.3847/1538-3881/aaa970},
archivePrefix = {arXiv},
       eprint = {1710.02542},
 primaryClass = {astro-ph.EP},
       adsurl = {https://ui.adsabs.harvard.edu/abs/2018AJ....155..118W}
}

@ARTICLE{Wu:05,
       author = {{Wu}, Yanqin},
        title = "{Origin of Tidal Dissipation in Jupiter. II. The Value of Q}",
      journal = {\apj},
         year = 2005,
        month = dec,
       volume = {635},
       number = {1},
        pages = {688-710},
          doi = {10.1086/497355},
archivePrefix = {arXiv},
       eprint = {astro-ph/0407628},
 primaryClass = {astro-ph},
       adsurl = {https://ui.adsabs.harvard.edu/abs/2005ApJ...635..688W}
}

@ARTICLE{Anderson:16,
       author = {{Anderson}, Kassandra R. and {Storch}, Natalia I. and {Lai}, Dong},
        title = "{Formation and stellar spin-orbit misalignment of hot Jupiters from Lidov-Kozai oscillations in stellar binaries}",
      journal = {\mnras},
         year = 2016,
        month = mar,
       volume = {456},
       number = {4},
        pages = {3671-3701},
          doi = {10.1093/mnras/stv2906},
archivePrefix = {arXiv},
       eprint = {1510.08918},
 primaryClass = {astro-ph.EP},
       adsurl = {https://ui.adsabs.harvard.edu/abs/2016MNRAS.456.3671A}
}

@ARTICLE{Guillochon:11,
       author = {{Guillochon}, James and {Ramirez-Ruiz}, Enrico and {Lin}, Douglas},
        title = "{Consequences of the Ejection and Disruption of Giant Planets}",
      journal = {\apj},
         year = 2011,
        month = may,
       volume = {732},
       number = {2},
          eid = {74},
        pages = {74},
          doi = {10.1088/0004-637X/732/2/74},
archivePrefix = {arXiv},
       eprint = {1012.2382},
 primaryClass = {astro-ph.EP},
       adsurl = {https://ui.adsabs.harvard.edu/abs/2011ApJ...732...74G}
}

@ARTICLE{Guillochon:13,
       author = {{Guillochon}, James and {Ramirez-Ruiz}, Enrico},
        title = "{Hydrodynamical Simulations to Determine the Feeding Rate of Black Holes by the Tidal Disruption of Stars: The Importance of the Impact Parameter and Stellar Structure}",
      journal = {\apj},
         year = 2013,
        month = apr,
       volume = {767},
       number = {1},
          eid = {25},
        pages = {25},
          doi = {10.1088/0004-637X/767/1/25},
archivePrefix = {arXiv},
       eprint = {1206.2350},
 primaryClass = {astro-ph.HE},
       adsurl = {https://ui.adsabs.harvard.edu/abs/2013ApJ...767...25G}
}

@ARTICLE{Sepinsky:07a,
       author = {{Sepinsky}, J.~F. and {Willems}, B. and {Kalogera}, V.},
        title = "{Equipotential Surfaces and Lagrangian Points in Nonsynchronous, Eccentric Binary and Planetary Systems}",
      journal = {\apj},
         year = 2007,
        month = may,
       volume = {660},
       number = {2},
        pages = {1624-1635},
          doi = {10.1086/513736},
archivePrefix = {arXiv},
       eprint = {astro-ph/0612508},
 primaryClass = {astro-ph},
       adsurl = {https://ui.adsabs.harvard.edu/abs/2007ApJ...660.1624S}
}

@ARTICLE{Sepinsky:09,
       author = {{Sepinsky}, J.~F. and {Willems}, B. and {Kalogera}, V. and {Rasio}, F.~A.},
        title = "{Interacting Binaries with Eccentric Orbits. II. Secular Orbital Evolution due to Non-conservative Mass Transfer}",
      journal = {\apj},
         year = 2009,
        month = sep,
       volume = {702},
       number = {2},
        pages = {1387-1392},
          doi = {10.1088/0004-637X/702/2/1387},
archivePrefix = {arXiv},
       eprint = {0903.0621},
 primaryClass = {astro-ph.SR},
       adsurl = {https://ui.adsabs.harvard.edu/abs/2009ApJ...702.1387S}
}

@ARTICLE{Dawson:18,
       author = {{Dawson}, Rebekah I. and {Johnson}, John Asher},
        title = "{Origins of Hot Jupiters}",
      journal = {\araa},
         year = 2018,
        month = sep,
       volume = {56},
        pages = {175-221},
          doi = {10.1146/annurev-astro-081817-051853},
archivePrefix = {arXiv},
       eprint = {1801.06117},
 primaryClass = {astro-ph.EP},
       adsurl = {https://ui.adsabs.harvard.edu/abs/2018ARA&A..56..175D}
}

@ARTICLE{Mayor:95,
       author = {{Mayor}, Michel and {Queloz}, Didier},
        title = "{A Jupiter-mass companion to a solar-type star}",
      journal = {\nat},
         year = 1995,
        month = nov,
       volume = {378},
       number = {6555},
        pages = {355-359},
          doi = {10.1038/378355a0},
       adsurl = {https://ui.adsabs.harvard.edu/abs/1995Natur.378..355M}
}

@ARTICLE{Eggleton:83,
       author = {{Eggleton}, P.~P.},
        title = "{Aproximations to the radii of Roche lobes.}",
      journal = {\apj},
         year = 1983,
        month = may,
       volume = {268},
        pages = {368-369},
          doi = {10.1086/160960},
       adsurl = {https://ui.adsabs.harvard.edu/abs/1983ApJ...268..368E}
}

@ARTICLE{Fabrycky:07,
       author = {{Fabrycky}, Daniel and {Tremaine}, Scott},
        title = "{Shrinking Binary and Planetary Orbits by Kozai Cycles with Tidal Friction}",
      journal = {\apj},
         year = 2007,
        month = nov,
       volume = {669},
       number = {2},
        pages = {1298-1315},
          doi = {10.1086/521702},
archivePrefix = {arXiv},
       eprint = {0705.4285},
 primaryClass = {astro-ph},
       adsurl = {https://ui.adsabs.harvard.edu/abs/2007ApJ...669.1298F}
}

@ARTICLE{Hills:88,
       author = {{Hills}, J.~G.},
        title = "{Hyper-velocity and tidal stars from binaries disrupted by a massive Galactic black hole}",
      journal = {\nat},
         year = 1988,
        month = feb,
       volume = {331},
       number = {6158},
        pages = {687-689},
          doi = {10.1038/331687a0},
       adsurl = {https://ui.adsabs.harvard.edu/abs/1988Natur.331..687H}
}

@ARTICLE{Wang:22,
       author = {{Wang}, Mengye and {Yin}, Jinjing and {Ma}, Yiqiu and {Wu}, Qingwen},
        title = "{A Model for the Possible Connection Between a Tidal Disruption Event and Quasi-periodic Eruption in GSN 069}",
      journal = {\apj},
         year = 2022,
        month = jul,
       volume = {933},
       number = {2},
          eid = {225},
        pages = {225},
          doi = {10.3847/1538-4357/ac75e6},
archivePrefix = {arXiv},
       eprint = {2206.03092},
 primaryClass = {astro-ph.HE},
       adsurl = {https://ui.adsabs.harvard.edu/abs/2022ApJ...933..225W}
}

@ARTICLE{Miniutti:19,
       author = {{Miniutti}, G. and {Saxton}, R.~D. and {Giustini}, M. and {Alexander}, K.~D. and {Fender}, R.~P. and {Heywood}, I. and {Monageng}, I. and {Coriat}, M. and {Tzioumis}, A.~K. and {Read}, A.~M. and {Knigge}, C. and {Gandhi}, P. and {Pretorius}, M.~L. and {Ag{\'\i}s-Gonz{\'a}lez}, B.},
        title = "{Nine-hour X-ray quasi-periodic eruptions from a low-mass black hole galactic nucleus}",
      journal = {\nat},
         year = 2019,
        month = sep,
       volume = {573},
       number = {7774},
        pages = {381-384},
          doi = {10.1038/s41586-019-1556-x},
archivePrefix = {arXiv},
       eprint = {1909.04693},
 primaryClass = {astro-ph.HE},
       adsurl = {https://ui.adsabs.harvard.edu/abs/2019Natur.573..381M}
}

@ARTICLE{Giustini:20,
       author = {{Giustini}, Margherita and {Miniutti}, Giovanni and {Saxton}, Richard D.},
        title = "{X-ray quasi-periodic eruptions from the galactic nucleus of RX J1301.9+2747}",
      journal = {\aap},
         year = 2020,
        month = apr,
       volume = {636},
          eid = {L2},
        pages = {L2},
          doi = {10.1051/0004-6361/202037610},
archivePrefix = {arXiv},
       eprint = {2002.08967},
 primaryClass = {astro-ph.HE},
       adsurl = {https://ui.adsabs.harvard.edu/abs/2020A&A...636L...2G}
}

@ARTICLE{Arcodia:22,
       author = {{Arcodia}, R. and {Miniutti}, G. and {Ponti}, G. and {Buchner}, J. and {Giustini}, M. and {Merloni}, A. and {Nandra}, K. and {Vincentelli}, F. and {Kara}, E. and {Salvato}, M. and {Pasham}, D.},
        title = "{The complex time and energy evolution of quasi-periodic eruptions in eRO-QPE1}",
      journal = {\aap},
         year = 2022,
        month = jun,
       volume = {662},
          eid = {A49},
        pages = {A49},
          doi = {10.1051/0004-6361/202243259},
archivePrefix = {arXiv},
       eprint = {2203.11939},
 primaryClass = {astro-ph.HE},
       adsurl = {https://ui.adsabs.harvard.edu/abs/2022A&A...662A..49A}
}

@ARTICLE{King:20,
       author = {{King}, Andrew},
        title = "{GSN 069 - A tidal disruption near miss}",
      journal = {\mnras},
         year = 2020,
        month = mar,
       volume = {493},
       number = {1},
        pages = {L120-L123},
          doi = {10.1093/mnrasl/slaa020},
archivePrefix = {arXiv},
       eprint = {2002.00970},
 primaryClass = {astro-ph.HE},
       adsurl = {https://ui.adsabs.harvard.edu/abs/2020MNRAS.493L.120K}
}

@ARTICLE{King:22,
       author = {{King}, Andrew},
        title = "{Quasi-periodic eruptions from galaxy nuclei}",
      journal = {\mnras},
         year = 2022,
        month = sep,
       volume = {515},
       number = {3},
        pages = {4344-4349},
          doi = {10.1093/mnras/stac1641},
archivePrefix = {arXiv},
       eprint = {2206.04698},
 primaryClass = {astro-ph.GA},
       adsurl = {https://ui.adsabs.harvard.edu/abs/2022MNRAS.515.4344K}
}

@ARTICLE{Linial:23,
       author = {{Linial}, Itai and {Sari}, Re'em},
        title = "{Unstable Mass Transfer from a Main-sequence Star to a Supermassive Black Hole and Quasiperiodic Eruptions}",
      journal = {\apj},
         year = 2023,
        month = mar,
       volume = {945},
       number = {2},
          eid = {86},
        pages = {86},
          doi = {10.3847/1538-4357/acbd3d},
archivePrefix = {arXiv},
       eprint = {2211.09851},
 primaryClass = {astro-ph.HE},
       adsurl = {https://ui.adsabs.harvard.edu/abs/2023ApJ...945...86L}
}

@ARTICLE{Rees:88,
       author = {{Rees}, Martin J.},
        title = "{Tidal disruption of stars by black holes of {}10$^{6}$-{}10$^{8}$ solar masses in nearby galaxies}",
      journal = {\nat},
         year = 1988,
        month = jun,
       volume = {333},
       number = {6173},
        pages = {523-528},
          doi = {10.1038/333523a0},
       adsurl = {https://ui.adsabs.harvard.edu/abs/1988Natur.333..523R}
}

@ARTICLE{Komossa:15,
       author = {{Komossa}, S.},
        title = "{Tidal disruption of stars by supermassive black holes: Status of observations}",
      journal = {Journal of High Energy Astrophysics},
         year = 2015,
        month = sep,
       volume = {7},
        pages = {148-157},
          doi = {10.1016/j.jheap.2015.04.006},
archivePrefix = {arXiv},
       eprint = {1505.01093},
 primaryClass = {astro-ph.HE},
       adsurl = {https://ui.adsabs.harvard.edu/abs/2015JHEAp...7..148K}
}

@ARTICLE{Hamers:19,
       author = {{Hamers}, Adrian S. and {Dosopoulou}, Fani},
        title = "{An Analytic Model for Mass Transfer in Binaries with Arbitrary Eccentricity, with Applications to Triple-star Systems}",
      journal = {\apj},
         year = 2019,
        month = feb,
       volume = {872},
       number = {2},
          eid = {119},
        pages = {119},
          doi = {10.3847/1538-4357/ab001d},
archivePrefix = {arXiv},
       eprint = {1812.05624},
 primaryClass = {astro-ph.SR},
       adsurl = {https://ui.adsabs.harvard.edu/abs/2019ApJ...872..119H}
}

@ARTICLE{Maguire:20,
       author = {{Maguire}, Kate and {Eracleous}, Michael and {Jonker}, Peter G. and {MacLeod}, Morgan and {Rosswog}, Stephan},
        title = "{Tidal Disruptions of White Dwarfs: Theoretical Models and Observational Prospects}",
      journal = {\ssr},
         year = 2020,
        month = mar,
       volume = {216},
       number = {3},
          eid = {39},
        pages = {39},
          doi = {10.1007/s11214-020-00661-2},
archivePrefix = {arXiv},
       eprint = {2004.00146},
 primaryClass = {astro-ph.HE},
       adsurl = {https://ui.adsabs.harvard.edu/abs/2020SSRv..216...39M}
}

@ARTICLE{Press:77,
       author = {{Press}, W.~H. and {Teukolsky}, S.~A.},
        title = "{On formation of close binaries by two-body tidal capture.}",
      journal = {\apj},
         year = 1977,
        month = apr,
       volume = {213},
        pages = {183-192},
          doi = {10.1086/155143},
       adsurl = {https://ui.adsabs.harvard.edu/abs/1977ApJ...213..183P}
}

@ARTICLE{Mardling:95,
       author = {{Mardling}, Rosemary A.},
        title = "{The Role of Chaos in the Circularization of Tidal Capture Binaries. I. The Chaos Boundary}",
      journal = {\apj},
         year = 1995,
        month = sep,
       volume = {450},
        pages = {722},
          doi = {10.1086/176178},
       adsurl = {https://ui.adsabs.harvard.edu/abs/1995ApJ...450..722M}
}

@ARTICLE{Ivanov:04,
       author = {{Ivanov}, P.~B. and {Papaloizou}, J.~C.~B.},
        title = "{On the tidal interaction of massive extrasolar planets on highly eccentric orbits}",
      journal = {\mnras},
         year = 2004,
        month = jan,
       volume = {347},
       number = {2},
        pages = {437-453},
          doi = {10.1111/j.1365-2966.2004.07238.x},
archivePrefix = {arXiv},
       eprint = {astro-ph/0303669},
 primaryClass = {astro-ph},
       adsurl = {https://ui.adsabs.harvard.edu/abs/2004MNRAS.347..437I}
}

@ARTICLE{Ivanov:07,
       author = {{Ivanov}, P.~B. and {Papaloizou}, J.~C.~B.},
        title = "{Orbital circularisation of white dwarfs and the formation of gravitational radiation sources in star clusters containing an intermediate mass black hole}",
      journal = {\aap},
         year = 2007,
        month = dec,
       volume = {476},
       number = {1},
        pages = {121-135},
          doi = {10.1051/0004-6361:20077105},
archivePrefix = {arXiv},
       eprint = {0709.0480},
 primaryClass = {astro-ph},
       adsurl = {https://ui.adsabs.harvard.edu/abs/2007A&A...476..121I}
}

@ARTICLE{Dawson:15,
       author = {{Dawson}, Rebekah I. and {Murray-Clay}, Ruth A. and {Johnson}, John Asher},
        title = "{The Photoeccentric Effect and Proto-hot Jupiters. III. A Paucity of Proto-hot Jupiters on Super-eccentric Orbits}",
      journal = {\apj},
         year = 2015,
        month = jan,
       volume = {798},
       number = {2},
          eid = {66},
        pages = {66},
          doi = {10.1088/0004-637X/798/2/66},
archivePrefix = {arXiv},
       eprint = {1211.0554},
 primaryClass = {astro-ph.EP},
       adsurl = {https://ui.adsabs.harvard.edu/abs/2015ApJ...798...66D}
}

@ARTICLE{Socrates:12,
       author = {{Socrates}, Aristotle and {Katz}, Boaz and {Dong}, Subo and {Tremaine}, Scott},
        title = "{Super-eccentric Migrating Jupiters}",
      journal = {\apj},
         year = 2012,
        month = may,
       volume = {750},
       number = {2},
          eid = {106},
        pages = {106},
          doi = {10.1088/0004-637X/750/2/106},
archivePrefix = {arXiv},
       eprint = {1110.1644},
 primaryClass = {astro-ph.EP},
       adsurl = {https://ui.adsabs.harvard.edu/abs/2012ApJ...750..106S}
}

@ARTICLE{Goldreich:77,
       author = {{Goldreich}, P. and {Nicholson}, P.~D.},
        title = "{Turbulent Viscosity and Jupiter's Tidal Q}",
      journal = {\icarus},
         year = 1977,
        month = feb,
       volume = {30},
       number = {2},
        pages = {301-304},
          doi = {10.1016/0019-1035(77)90163-4},
       adsurl = {https://ui.adsabs.harvard.edu/abs/1977Icar...30..301G}
}

@ARTICLE{Armstrong:20,
       author = {{Armstrong}, David J. and {Lopez}, Th{\'e}o A. and {Adibekyan}, Vardan and {Booth}, Richard A. and {Bryant}, Edward M. and {Collins}, Karen A. and {Deleuil}, Magali and {Emsenhuber}, Alexandre and {Huang}, Chelsea X. and {King}, George W. and {Lillo-Box}, Jorge and {Lissauer}, Jack J. and {Matthews}, Elisabeth and {Mousis}, Olivier and {Nielsen}, Louise D. and {Osborn}, Hugh and {Otegi}, Jon and {Santos}, Nuno C. and {Sousa}, S{\'e}rgio G. and {Stassun}, Keivan G. and {Veras}, Dimitri and {Ziegler}, Carl and {Acton}, Jack S. and {Almenara}, Jose M. and {Anderson}, David R. and {Barrado}, David and {Barros}, Susana C.~C. and {Bayliss}, Daniel and {Belardi}, Claudia and {Bouchy}, Francois and {Brice{\~n}o}, C{\'e}sar and {Brogi}, Matteo and {Brown}, David J.~A. and {Burleigh}, Matthew R. and {Casewell}, Sarah L. and {Chaushev}, Alexander and {Ciardi}, David R. and {Collins}, Kevin I. and {Col{\'o}n}, Knicole D. and {Cooke}, Benjamin F. and {Crossfield}, Ian J.~M. and {D{\'\i}az}, Rodrigo F. and {Delgado Mena}, Elisa and {Demangeon}, Olivier D.~S. and {Dorn}, Caroline and {Dumusque}, Xavier and {Eigm{\"u}ller}, Philipp and {Fausnaugh}, Michael and {Figueira}, Pedro and {Gan}, Tianjun and {Gandhi}, Siddharth and {Gill}, Samuel and {Gonzales}, Erica J. and {Goad}, Michael R. and {G{\"u}nther}, Maximilian N. and {Helled}, Ravit and {Hojjatpanah}, Saeed and {Howell}, Steve B. and {Jackman}, James and {Jenkins}, James S. and {Jenkins}, Jon M. and {Jensen}, Eric L.~N. and {Kennedy}, Grant M. and {Latham}, David W. and {Law}, Nicholas and {Lendl}, Monika and {Lozovsky}, Michael and {Mann}, Andrew W. and {Moyano}, Maximiliano and {McCormac}, James and {Meru}, Farzana and {Mordasini}, Christoph and {Osborn}, Ares and {Pollacco}, Don and {Queloz}, Didier and {Raynard}, Liam and {Ricker}, George R. and {Rowden}, Pamela and {Santerne}, Alexandre and {Schlieder}, Joshua E. and {Seager}, Sara and {Sha}, Lizhou and {Tan}, Thiam-Guan and {Tilbrook}, Rosanna H. and {Ting}, Eric and {Udry}, St{\'e}phane and {Vanderspek}, Roland and {Watson}, Christopher A. and {West}, Richard G. and {Wilson}, Paul A. and {Winn}, Joshua N. and {Wheatley}, Peter and {Villasenor}, Jesus Noel and {Vines}, Jose I. and {Zhan}, Zhuchang},
        title = "{A remnant planetary core in the hot-Neptune desert}",
      journal = {\nat},
         year = 2020,
        month = jul,
       volume = {583},
       number = {7814},
        pages = {39-42},
          doi = {10.1038/s41586-020-2421-7},
archivePrefix = {arXiv},
       eprint = {2003.10314},
 primaryClass = {astro-ph.EP},
       adsurl = {https://ui.adsabs.harvard.edu/abs/2020Natur.583...39A}
}

@ARTICLE{Liu:13,
       author = {{Liu}, Shang-Fei and {Guillochon}, James and {Lin}, Douglas N.~C. and {Ramirez-Ruiz}, Enrico},
        title = "{On the Survivability and Metamorphism of Tidally Disrupted Giant Planets: The Role of Dense Cores}",
      journal = {\apj},
         year = 2013,
        month = jan,
       volume = {762},
       number = {1},
          eid = {37},
        pages = {37},
          doi = {10.1088/0004-637X/762/1/37},
archivePrefix = {arXiv},
       eprint = {1211.1971},
 primaryClass = {astro-ph.EP},
       adsurl = {https://ui.adsabs.harvard.edu/abs/2013ApJ...762...37L}
}

@article{Coughlin:2019,
  title = {Partial Stellar Disruption by a Supermassive Black Hole: Is the Light Curve Really Proportional to t−9/4?},
  volume = {883},
  ISSN = {2041-8213},
  url = {http://dx.doi.org/10.3847/2041-8213/ab412d},
  DOI = {10.3847/2041-8213/ab412d},
  number = {1},
  journal = {\apjl},
  publisher = {American Astronomical Society},
  author = {Coughlin,  Eric R. and Nixon,  C. J.},
  year = {2019},
  month = sep,
  pages = {L17}
}

@ARTICLE{Bandopadhyay:24,
       author = {{Bandopadhyay}, Ananya and {Coughlin}, Eric R. and {Nixon}, C.~J. and {Pasham}, Dheeraj R.},
        title = "{Repeating Nuclear Transients from Repeating Partial Tidal Disruption Events: Reproducing ASASSN-14ko and AT2020vdq}",
      journal = {\apj},
         year = 2024,
        month = oct,
       volume = {974},
       number = {1},
          eid = {80},
        pages = {80},
          doi = {10.3847/1538-4357/ad6a5a},
archivePrefix = {arXiv},
       eprint = {2406.03675},
 primaryClass = {astro-ph.HE},
       adsurl = {https://ui.adsabs.harvard.edu/abs/2024ApJ...974...80B}
}

@ARTICLE{Bandopadhyay:25,
       author = {{Bandopadhyay}, Ananya and {Coughlin}, Eric R. and {Nixon}, C.~J.},
        title = "{Repeated Tidal Interactions Between Stars and Supermassive Black Holes: Mass Transfer, Stability, and Implications for Repeating Partial Tidal Disruption Events}",
      journal = {\apj},
         year = 2025,
        month = jul,
       volume = {987},
       number = {1},
          eid = {16},
        pages = {16},
          doi = {10.3847/1538-4357/add9a5},
archivePrefix = {arXiv},
       eprint = {2504.18614},
 primaryClass = {astro-ph.HE},
       adsurl = {https://ui.adsabs.harvard.edu/abs/2025ApJ...987...16B}
}

@article{Ryu:2020,
  title = {Tidal Disruptions of Main-sequence Stars. II. Simulation Methodology and Stellar Mass Dependence of the Character of Full Tidal Disruptions},
  volume = {904},
  ISSN = {1538-4357},
  url = {http://dx.doi.org/10.3847/1538-4357/abb3cd},
  DOI = {10.3847/1538-4357/abb3cd},
  number = {2},
  journal = {\apj},
  publisher = {American Astronomical Society},
  author = {Ryu,  Taeho and Krolik,  Julian and Piran,  Tsvi and Noble,  Scott C.},
  year = {2020},
  month = nov,
  pages = {99}
}

@article{Rosswog:2009,
  title = {TIDAL DISRUPTION AND IGNITION OF WHITE DWARFS BY MODERATELY MASSIVE BLACK HOLES},
  volume = {695},
  ISSN = {1538-4357},
  url = {http://dx.doi.org/10.1088/0004-637X/695/1/404},
  DOI = {10.1088/0004-637x/695/1/404},
  number = {1},
  journal = {\apj},
  publisher = {American Astronomical Society},
  author = {Rosswog,  S. and Ramirez-Ruiz,  E. and Hix,  W. R.},
  year = {2009},
  month = mar,
  pages = {404–419}
}

@ARTICLE{Yao:25,
       author = {{Yao}, Philippe Z. and {Quataert}, Eliot},
        title = "{Mass Transfer in Tidally Heated Stars Orbiting Massive Black Holes and Implications for Repeating Nuclear Transients}",
      journal = {arXiv e-prints},
         year = 2025,
        month = may,
          eid = {arXiv:2505.10611},
        pages = {arXiv:2505.10611},
          doi = {10.48550/arXiv.2505.10611},
archivePrefix = {arXiv},
       eprint = {2505.10611},
 primaryClass = {astro-ph.HE},
       adsurl = {https://ui.adsabs.harvard.edu/abs/2025arXiv250510611Y}
}

@ARTICLE{Lai:97,
       author = {{Lai}, Dong},
        title = "{Dynamical Tides in Rotating Binary Stars}",
      journal = {\apj},
         year = 1997,
        month = dec,
       volume = {490},
       number = {2},
        pages = {847-862},
          doi = {10.1086/304899},
archivePrefix = {arXiv},
       eprint = {astro-ph/9704132},
 primaryClass = {astro-ph},
       adsurl = {https://ui.adsabs.harvard.edu/abs/1997ApJ...490..847L}
}

@ARTICLE{Diener:95,
       author = {{Diener}, P. and {Kosovichev}, A.~G. and {Kotok}, E.~V. and {Novikov}, I.~D. and {Pethick}, C.~J.},
        title = "{Non-linear effects at tidal capture of stars by a massive black hole - II. Compressible affine models and tidal interaction after capture}",
      journal = {\mnras},
         year = 1995,
        month = jul,
       volume = {275},
       number = {2},
        pages = {498-506},
          doi = {10.1093/mnras/275.2.498},
       adsurl = {https://ui.adsabs.harvard.edu/abs/1995MNRAS.275..498D}
}

@ARTICLE{Merritt:13,
       author = {{Merritt}, David},
        title = "{Loss-cone dynamics}",
      journal = {Classical and Quantum Gravity},
         year = 2013,
        month = dec,
       volume = {30},
       number = {24},
          eid = {244005},
        pages = {244005},
          doi = {10.1088/0264-9381/30/24/244005},
archivePrefix = {arXiv},
       eprint = {1307.3268},
 primaryClass = {astro-ph.GA},
       adsurl = {https://ui.adsabs.harvard.edu/abs/2013CQGra..30x4005M}
}

@ARTICLE{Carter:83,
       author = {{Carter}, B. and {Luminet}, J. -P.},
        title = "{Tidal compression of a star by a large black hole. I Mechanical evolution and nuclear energy release by proton capture}",
      journal = {\aap},
         year = 1983,
        month = may,
       volume = {121},
       number = {1},
        pages = {97-113},
       adsurl = {https://ui.adsabs.harvard.edu/abs/1983A&A...121...97C}
}

@ARTICLE{Carter:85,
       author = {{Carter}, B. and {Luminet}, J.~P.},
        title = "{Mechanics of the affine star model}",
      journal = {\mnras},
         year = 1985,
        month = jan,
       volume = {212},
        pages = {23-55},
          doi = {10.1093/mnras/212.1.23},
       adsurl = {https://ui.adsabs.harvard.edu/abs/1985MNRAS.212...23C}
}

@phdthesis{Wu:98,
    author = "Wu, Yanqin",
    title = "{Excitation and saturation of white dwarf pulsations}",
    school = "California Institute of Technology",
    year = "1998", 
    doi = "10.7907/nc60-8v65", 
    url = "https://thesis.library.caltech.edu/3736/"
}

@ARTICLE{MacLeod:22,
       author = {{MacLeod}, Morgan and {Vick}, Michelle and {Loeb}, Abraham},
        title = "{Tidal Wave Breaking in the Eccentric Lead-in to Mass Transfer and Common Envelope Phases}",
      journal = {\apj},
         year = 2022,
        month = sep,
       volume = {937},
       number = {1},
          eid = {37},
        pages = {37},
          doi = {10.3847/1538-4357/ac8aff},
archivePrefix = {arXiv},
       eprint = {2203.01947},
 primaryClass = {astro-ph.SR},
       adsurl = {https://ui.adsabs.harvard.edu/abs/2022ApJ...937...37M}
}

@ARTICLE{Pitre:25,
       author = {{Pitre}, Tristan and {Poisson}, Eric},
        title = "{Impact of nonlinearities on relativistic dynamical tides in compact binary inspirals}",
      journal = {arXiv e-prints},
         year = 2025,
        month = jun,
          eid = {arXiv:2506.08722},
        pages = {arXiv:2506.08722},
archivePrefix = {arXiv},
       eprint = {2506.08722},
 primaryClass = {gr-qc},
       adsurl = {https://ui.adsabs.harvard.edu/abs/2025arXiv250608722P}
}

@ARTICLE{Luminet:89,
       author = {{Luminet}, J. -P. and {Pichon}, B.},
        title = "{Tidal pinching of white dwarfs}",
      journal = {\aap},
         year = 1989,
        month = jan,
       volume = {209},
       number = {1-2},
        pages = {103-110},
       adsurl = {https://ui.adsabs.harvard.edu/abs/1989A&A...209..103L}
}

@article{Rosswog:09,
  title = {TIDAL DISRUPTION AND IGNITION OF WHITE DWARFS BY MODERATELY MASSIVE BLACK HOLES},
  volume = {695},
  ISSN = {1538-4357},
  url = {http://dx.doi.org/10.1088/0004-637X/695/1/404},
  DOI = {10.1088/0004-637x/695/1/404},
  number = {1},
  journal = {\apj},
  publisher = {American Astronomical Society},
  author = {Rosswog,  S. and Ramirez-Ruiz,  E. and Hix,  W. R.},
  year = {2009},
  month = mar,
  pages = {404–419}
}

@article{MacLeod:16,
  title = {OPTICAL THERMONUCLEAR TRANSIENTS FROM TIDAL COMPRESSION OF WHITE DWARFS AS TRACERS OF THE LOW END OF THE MASSIVE BLACK HOLE MASS FUNCTION},
  volume = {819},
  ISSN = {1538-4357},
  url = {http://dx.doi.org/10.3847/0004-637X/819/1/3},
  DOI = {10.3847/0004-637x/819/1/3},
  number = {1},
  journal = {\apj},
  publisher = {American Astronomical Society},
  author = {MacLeod,  Morgan and Guillochon,  James and Ramirez-Ruiz,  Enrico and Kasen,  Daniel and Rosswog,  Stephan},
  year = {2016},
  month = feb,
  pages = {3}
}

@article{Tanikawa:18,
  title = {High-resolution Hydrodynamic Simulation of Tidal Detonation of a Helium White Dwarf by an Intermediate Mass Black Hole},
  volume = {858},
  ISSN = {1538-4357},
  url = {http://dx.doi.org/10.3847/1538-4357/aaba79},
  DOI = {10.3847/1538-4357/aaba79},
  number = {1},
  journal = {\apj},
  publisher = {American Astronomical Society},
  author = {Tanikawa,  Ataru},
  year = {2018},
  month = apr,
  pages = {26}
}

@ARTICLE{Byrd:54,
       author = {{Byrd}, P.~F. and {Friedman}, M.},
        title = "{Handbook of Elliptic Integrals for Engineers and Physicists (Ref. U. Wegner)}",
      journal = {Mitteilungen der Astronomischen Gesellschaft Hamburg},
         year = 1954,
        month = jan,
       volume = {5},
        pages = {99},
       adsurl = {https://ui.adsabs.harvard.edu/abs/1954MitAG...5...99B}
}

@misc{MathWorld:Resultant,
      author = {{Weisstein}, Eric~W.},
      title = "{Resultant. From \textit{MathWorld}--A Wolfram Resource.}",
      howpublished = {\url{https://mathworld.wolfram.com/Resultant.html}},
      year = 2025, 
      note = {Last visited on 07/31/2025}
    }

@book{Hairer:93,
author = {Hairer, Ernst and Norsett, Syvert and Wanner, Gerhard},
year = {1993},
month = {01},
pages = {},
title = {Solving Ordinary Differential Equations I: Nonstiff Problems},
volume = {8},
isbn = {978-3-540-56670-0},
doi = {10.1007/978-3-540-78862-1}, 
publisher = {Springer Berlin, Heidelberg}
}

@ARTICLE{Lai:06,
       author = {{Lai}, Dong and {Wu}, Yanqin},
        title = "{Resonant tidal excitations of inertial modes in coalescing neutron star binaries}",
      journal = {\prd},
         year = 2006,
        month = jul,
       volume = {74},
       number = {2},
          eid = {024007},
        pages = {024007},
          doi = {10.1103/PhysRevD.74.024007},
archivePrefix = {arXiv},
       eprint = {astro-ph/0604163},
 primaryClass = {astro-ph},
       adsurl = {https://ui.adsabs.harvard.edu/abs/2006PhRvD..74b4007L}
}

@ARTICLE{Lai:21,
       author = {{Lai}, Dong},
        title = "{Jupiter's Dynamical Love Number}",
      journal = {Planet. sci. j.},
         year = 2021,
        month = aug,
       volume = {2},
       number = {4},
          eid = {122},
        pages = {122},
          doi = {10.3847/PSJ/ac013b},
archivePrefix = {arXiv},
       eprint = {2103.06186},
 primaryClass = {astro-ph.EP},
       adsurl = {https://ui.adsabs.harvard.edu/abs/2021PSJ.....2..122L}
}

@ARTICLE{Dewberry:22,
       author = {{Dewberry}, Janosz W. and {Lai}, Dong},
        title = "{Dynamical Tidal Love Numbers of Rapidly Rotating Planets and Stars}",
      journal = {\apj},
         year = 2022,
        month = feb,
       volume = {925},
       number = {2},
          eid = {124},
        pages = {124},
          doi = {10.3847/1538-4357/ac3ede},
archivePrefix = {arXiv},
       eprint = {2110.12129},
 primaryClass = {astro-ph.EP},
       adsurl = {https://ui.adsabs.harvard.edu/abs/2022ApJ...925..124D}
}

@ARTICLE{Ho:99,
       author = {{Ho}, Wynn C.~G. and {Lai}, Dong},
        title = "{Resonant tidal excitations of rotating neutron stars in coalescing binaries}",
      journal = {\mnras},
         year = 1999,
        month = sep,
       volume = {308},
       number = {1},
        pages = {153-166},
          doi = {10.1046/j.1365-8711.1999.02703.x},
archivePrefix = {arXiv},
       eprint = {astro-ph/9812116},
 primaryClass = {astro-ph},
       adsurl = {https://ui.adsabs.harvard.edu/abs/1999MNRAS.308..153H}
}

@ARTICLE{Braviner:15,
       author = {{Braviner}, Harry J. and {Ogilvie}, Gordon I.},
        title = "{Tidal interactions of a Maclaurin spheroid - II. Resonant excitation of modes by a close, misaligned orbit}",
      journal = {\mnras},
         year = 2015,
        month = feb,
       volume = {447},
       number = {2},
        pages = {1141-1153},
          doi = {10.1093/mnras/stu2521},
archivePrefix = {arXiv},
       eprint = {1412.2514},
 primaryClass = {astro-ph.SR},
       adsurl = {https://ui.adsabs.harvard.edu/abs/2015MNRAS.447.1141B}
}

@ARTICLE{Xu:17,
       author = {{Xu}, Wenrui and {Lai}, Dong},
        title = "{Resonant tidal excitation of oscillation modes in merging binary neutron stars: Inertial-gravity modes}",
      journal = {\prd},
         year = 2017,
        month = oct,
       volume = {96},
       number = {8},
          eid = {083005},
        pages = {083005},
          doi = {10.1103/PhysRevD.96.083005},
       adsurl = {https://ui.adsabs.harvard.edu/abs/2017PhRvD..96h3005X}
}

@ARTICLE{Huang:63,
       author = {{Huang}, Su-Shu},
        title = "{Modes of Mass Ejection by Binary Stars and the Effect on Their Orbital Periods.}",
      journal = {\apj},
         year = 1963,
        month = aug,
       volume = {138},
        pages = {471},
          doi = {10.1086/147659},
       adsurl = {https://ui.adsabs.harvard.edu/abs/1963ApJ...138..471H}
}

@ARTICLE{Shu:81,
       author = {{Shu}, F.~H. and {Lubow}, S.~H.},
        title = "{Mass, angular momentum, and energy transfer in close binary stars}",
      journal = {\araa},
         year = 1981,
        month = jan,
       volume = {19},
        pages = {277-293},
          doi = {10.1146/annurev.aa.19.090181.001425},
       adsurl = {https://ui.adsabs.harvard.edu/abs/1981ARA&A..19..277S}
}

@ARTICLE{Ogilvie:14,
       author = {{Ogilvie}, Gordon I.},
        title = "{Tidal Dissipation in Stars and Giant Planets}",
      journal = {\araa},
         year = 2014,
        month = aug,
       volume = {52},
        pages = {171-210},
          doi = {10.1146/annurev-astro-081913-035941},
archivePrefix = {arXiv},
       eprint = {1406.2207},
 primaryClass = {astro-ph.SR},
       adsurl = {https://ui.adsabs.harvard.edu/abs/2014ARA&A..52..171O}
}

@article{Hayasaki:13,
  title = {Finite,  intense accretion bursts from tidal disruption of stars on bound orbits},
  volume = {434},
  ISSN = {1365-2966},
  url = {http://dx.doi.org/10.1093/mnras/stt871},
  DOI = {10.1093/mnras/stt871},
  number = {2},
  journal = {\mnras},
  publisher = {Oxford University Press (OUP)},
  author = {Hayasaki,  Kimitake and Stone,  Nicholas and Loeb,  Abraham},
  year = {2013},
  month = jul,
  pages = {909–924}
}

@article{Cufari:22,
  title = {Using the Hills Mechanism to Generate Repeating Partial Tidal Disruption Events and ASASSN-14ko},
  volume = {929},
  ISSN = {2041-8213},
  url = {http://dx.doi.org/10.3847/2041-8213/ac6021},
  DOI = {10.3847/2041-8213/ac6021},
  number = {2},
  journal = {\apjl},
  publisher = {American Astronomical Society},
  author = {Cufari,  M. and Coughlin,  Eric R. and Nixon,  C. J.},
  year = {2022},
  month = apr,
  pages = {L20}
}

@article{Liu:24,
  title = {Rapid evolution of the recurrence time in the repeating partial tidal disruption event eRASSt J045650.3−203750},
  volume = {683},
  ISSN = {1432-0746},
  url = {http://dx.doi.org/10.1051/0004-6361/202348682},
  DOI = {10.1051/0004-6361/202348682},
  journal = {\aap},
  publisher = {EDP Sciences},
  author = {Liu,  Zhu and Ryu,  Taeho and Goodwin,  A. J. and Rau,  A. and Homan,  D. and Krumpe,  M. and Merloni,  A. and Grotova,  I. and Anderson,  G. E. and Malyali,  A. and Miller-Jones,  J. C. A.},
  year = {2024},
  month = mar,
  pages = {L13}
}

@ARTICLE{Zhou:24,
       author = {{Zhou}, Cong and {Zhong}, Binyu and {Zeng}, Yuhe and {Huang}, Lei and {Pan}, Zhen},
        title = "{Probing orbits of stellar mass objects deep in galactic nuclei with quasiperiodic eruptions. II. Population analysis}",
      journal = {\prd},
         year = 2024,
        month = oct,
       volume = {110},
       number = {8},
          eid = {083019},
        pages = {083019},
          doi = {10.1103/PhysRevD.110.083019},
archivePrefix = {arXiv},
       eprint = {2405.06429},
 primaryClass = {astro-ph.HE},
       adsurl = {https://ui.adsabs.harvard.edu/abs/2024PhRvD.110h3019Z}
}

@ARTICLE{Godet:14,
       author = {{Godet}, O. and {Lombardi}, J.~C. and {Antonini}, F. and {Barret}, D. and {Webb}, N.~A. and {Vingless}, J. and {Thomas}, M.},
        title = "{Implications of the Delayed 2013 Outburst of ESO 243-49 HLX-1}",
      journal = {\apj},
         year = 2014,
        month = oct,
       volume = {793},
       number = {2},
          eid = {105},
        pages = {105},
          doi = {10.1088/0004-637X/793/2/105},
archivePrefix = {arXiv},
       eprint = {1408.1819},
 primaryClass = {astro-ph.HE},
       adsurl = {https://ui.adsabs.harvard.edu/abs/2014ApJ...793..105G}
}

@ARTICLE{Webb:23,
       author = {{Webb}, Natalie A. and {Barret}, Didier and {Godet}, Olivier and {Gupta}, Maitrayee and {Lin}, Dacheng and {Quintin}, Erwan and {Tranin}, Hugo},
        title = "{Tidal disruption events and quasi‑periodic eruptions}",
      journal = {Astronomische Nachrichten},
         year = 2023,
        month = may,
       volume = {344},
       number = {4},
          eid = {e20230051},
        pages = {e20230051},
          doi = {10.1002/asna.20230051},
archivePrefix = {arXiv},
       eprint = {2304.08828},
 primaryClass = {astro-ph.HE},
       adsurl = {https://ui.adsabs.harvard.edu/abs/2023AN....34430051W}
}

@ARTICLE{Bortolas:23,
       author = {{Bortolas}, Elisa and {Ryu}, Taeho and {Broggi}, Luca and {Sesana}, Alberto},
        title = "{Partial stellar tidal disruption events and their rates}",
      journal = {\mnras},
         year = 2023,
        month = sep,
       volume = {524},
       number = {2},
        pages = {3026-3038},
          doi = {10.1093/mnras/stad2024},
archivePrefix = {arXiv},
       eprint = {2303.03408},
 primaryClass = {astro-ph.HE},
       adsurl = {https://ui.adsabs.harvard.edu/abs/2023MNRAS.524.3026B}
}

@article{Campana:15,
  title = {Multiple tidal disruption flares in the active galaxy IC 3599},
  volume = {581},
  ISSN = {1432-0746},
  url = {http://dx.doi.org/10.1051/0004-6361/201525965},
  DOI = {10.1051/0004-6361/201525965},
  journal = {\aap},
  publisher = {EDP Sciences},
  author = {Campana,  S. and Mainetti,  D. and Colpi,  M. and Lodato,  G. and D’Avanzo,  P. and Evans,  P. A. and Moretti,  A.},
  year = {2015},
  month = aug,
  pages = {A17}
}

@article{Wevers:23,
  title = {Live to Die Another Day: The Rebrightening of AT 2018fyk as a Repeating Partial Tidal Disruption Event},
  volume = {942},
  ISSN = {2041-8213},
  url = {http://dx.doi.org/10.3847/2041-8213/ac9f36},
  DOI = {10.3847/2041-8213/ac9f36},
  number = {2},
  journal = {\apjl},
  publisher = {American Astronomical Society},
  author = {Wevers,  T. and Coughlin,  E. R. and Pasham,  D. R. and Guolo,  M. and Sun,  Y. and Wen,  S. and Jonker,  P. G. and Zabludoff,  A. and Malyali,  A. and Arcodia,  R. and Liu,  Z. and Merloni,  A. and Rau,  A. and Grotova,  I. and Short,  P. and Cao,  Z.},
  year = {2023},
  month = jan,
  pages = {L33}
}

@article{Miniutti:23b,
  title = {Repeating tidal disruptions in GSN 069: Long-term evolution and constraints on quasi-periodic eruptions’ models},
  volume = {670},
  ISSN = {1432-0746},
  url = {http://dx.doi.org/10.1051/0004-6361/202244512},
  DOI = {10.1051/0004-6361/202244512},
  journal = {\aap},
  publisher = {EDP Sciences},
  author = {Miniutti,  G. and Giustini,  M. and Arcodia,  R. and Saxton,  R. D. and Read,  A. M. and Bianchi,  S. and Alexander,  K. D.},
  year = {2023},
  month = feb,
  pages = {A93}
}

@article{Somalwar:25,
  title = {The First Systematically Identified Repeating Partial Tidal Disruption Event},
  volume = {985},
  ISSN = {1538-4357},
  url = {http://dx.doi.org/10.3847/1538-4357/adcc19},
  DOI = {10.3847/1538-4357/adcc19},
  number = {2},
  journal = {\apj},
  publisher = {American Astronomical Society},
  author = {Somalwar,  Jean J. and Ravi,  Vikram and Yao,  Yuhan and Guolo,  Muryel and Graham,  Matthew and Hammerstein,  Erica and Lu,  Wenbin and Nicholl,  Matt and Sharma,  Yashvi and Stein,  Robert and van Velzen,  Sjoert and Bellm,  Eric C. and Coughlin,  Michael W. and Groom,  Steven L. and Masci,  Frank J. and Riddle,  Reed},
  year = {2025},
  month = may,
  pages = {175}
}

@ARTICLE{Weldon:25,
       author = {{Weldon}, Grant C. and {Hansen}, Bradley M.~S. and {Naoz}, Smadar},
        title = "{Saving Doomed Planets: Mass Loss and Angular Momentum Return Boost Hot Jupiter Survival Rates}",
      journal = {arXiv e-prints},
         year = 2025,
        month = oct,
          eid = {arXiv:2510.26882},
        pages = {arXiv:2510.26882},
          doi = {10.48550/arXiv.2510.26882},
archivePrefix = {arXiv},
       eprint = {2510.26882},
 primaryClass = {astro-ph.EP},
       adsurl = {https://ui.adsabs.harvard.edu/abs/2025arXiv251026882W}
}

@ARTICLE{Kumar:96,
   author = {{Kumar}, P. and {Goodman}, J.},
    title = "{Nonlinear Damping of Oscillations in Tidal-Capture Binaries}",
  journal = {\apj},
   eprint = {astro-ph/9509112},
     year = 1996,
    month = aug,
   volume = 466,
    pages = {946},
      doi = {10.1086/177565},
   adsurl = {https://ui.adsabs.harvard.edu/abs/1996ApJ...466..946K}
}

@ARTICLE{Scipy,
  author  = {Virtanen, Pauli and Gommers, Ralf and Oliphant, Travis E. and
            Haberland, Matt and Reddy, Tyler and Cournapeau, David and
            Burovski, Evgeni and Peterson, Pearu and Weckesser, Warren and
            Bright, Jonathan and {van der Walt}, St{\'e}fan J. and
            Brett, Matthew and Wilson, Joshua and Millman, K. Jarrod and
            Mayorov, Nikolay and Nelson, Andrew R. J. and Jones, Eric and
            Kern, Robert and Larson, Eric and Carey, C J and
            Polat, {\.I}lhan and Feng, Yu and Moore, Eric W. and
            {VanderPlas}, Jake and Laxalde, Denis and Perktold, Josef and
            Cimrman, Robert and Henriksen, Ian and Quintero, E. A. and
            Harris, Charles R. and Archibald, Anne M. and
            Ribeiro, Ant{\^o}nio H. and Pedregosa, Fabian and
            {van Mulbregt}, Paul and {SciPy 1.0 Contributors}},
  title   = {{{SciPy} 1.0: Fundamental Algorithms for Scientific
            Computing in Python}},
  journal = {Nature Methods},
  year    = {2020},
  volume  = {17},
  pages   = {261--272},
  adsurl  = {https://rdcu.be/b08Wh},
  doi     = {10.1038/s41592-019-0686-2},
}

@Article{Numpy,
 title         = {Array programming with {NumPy}},
 author        = {Charles R. Harris and K. Jarrod Millman and St{\'{e}}fan J.
                 van der Walt and Ralf Gommers and Pauli Virtanen and David
                 Cournapeau and Eric Wieser and Julian Taylor and Sebastian
                 Berg and Nathaniel J. Smith and Robert Kern and Matti Picus
                 and Stephan Hoyer and Marten H. van Kerkwijk and Matthew
                 Brett and Allan Haldane and Jaime Fern{\'{a}}ndez del
                 R{\'{i}}o and Mark Wiebe and Pearu Peterson and Pierre
                 G{\'{e}}rard-Marchant and Kevin Sheppard and Tyler Reddy and
                 Warren Weckesser and Hameer Abbasi and Christoph Gohlke and
                 Travis E. Oliphant},
 year          = {2020},
 month         = sep,
 journal       = {Nature},
 volume        = {585},
 number        = {7825},
 pages         = {357--362},
 doi           = {10.1038/s41586-020-2649-2},
 publisher     = {Springer Science and Business Media {LLC}},
 url           = {https://doi.org/10.1038/s41586-020-2649-2}
}

@Article{Matplotlib,
  Author    = {Hunter, J. D.},
  Title     = {Matplotlib: A 2D graphics environment},
  Journal   = {Computing in Science \& Engineering},
  Volume    = {9},
  Number    = {3},
  Pages     = {90--95},
  publisher = {IEEE COMPUTER SOC},
  doi       = {10.1109/MCSE.2007.55},
  year      = 2007
}

@inproceedings{Numba,
  title={Numba: A llvm-based python jit compiler},
  author={Lam, Siu Kwan and Pitrou, Antoine and Seibert, Stanley},
  booktitle={Proceedings of the Second Workshop on the LLVM Compiler Infrastructure in HPC},
  pages={1--6},
  year={2015}
}

@misc{Mathematica,
  author = {{Wolfram Research, Inc.}},
  title = {Mathematica, {V}ersion 14.1},
  url = {https://www.wolfram.com/mathematica},
  year = {2025},
  note = {Champaign, IL, 2025}
}
\bibliographystyle{aasjournalv7}



\end{document}